\documentclass[twocolumn,tighten,times]{aastex631}

\usepackage[T1]{fontenc}
\usepackage{amsmath}
\let\tablenum\relax
\usepackage{siunitx}
\DeclareSIUnit\Mpc{Mpc}
\usepackage{xspace}
\usepackage{acronym}
\usepackage{booktabs}
\usepackage{xcolor}
\usepackage{xfp}
\usepackage{orcidlink}
\usepackage[normalem]{ulem}

\newcommand{\Msun}{\ensuremath{\mathit{M_\odot}}}

\newcommand{\DL}{\ensuremath{D_\mathrm{L}}}

\newcommand\PEpdfp{\ensuremath{p}}
\newcommand\PEdata{\ensuremath{d}}
\newcommand{\PEparameter}{\ensuremath{\boldsymbol{\theta}}}%

\newcommand\PEhyparameter{\ensuremath{\boldsymbol\Lambda}}
\newcommand\PEpdf[2][?]{\ensuremath{\PEpdfp({#2}\ifx#1?\else | {#1}\fi)}}

\newcommand\PEpriorpdfpi{\ensuremath{\pi}}
\newcommand\PEpdfprior[1]{\ensuremath{\PEpriorpdfpi({#1})}}
\newcommand\PEprior[1][\PEparameter]{\PEpdfprior{#1}}
\newcommand\PEpriorpe[1][\PEparameter]{{\let\keepPEpriorpdfpi\PEpriorpdfpi\def\PEpriorpdfpi{\keepPEpriorpdfpi_{\text{PE}}}\PEprior[#1]\let\PEpriorpdfpi\keepPEpriorpdfpi}}

\newcommand{\soft}[1]{\textsc{#1}}

\newcommand{\GSTLAL}{\soft{GstLAL}\xspace}

\newcommand{\CWB}{\soft{cWB}\xspace}

\newcommand{\PYCBC}{\soft{PyCBC}\xspace}
\newcommand{\MBTA}{\soft{MBTA}\xspace}

\newcommand{\BAYESWAVE}{\soft{BayesWave}\xspace}
\newcommand{\BILBY}{\soft{Bilby}\xspace}

\newcommand{\LALSUITE}{\soft{LALSuite}\xspace}

\newcommand{\ASIMOV}{\soft{Asimov}\xspace}
\newcommand{\PESUMMARY}{\soft{PESummary}\xspace}

\newcommand{\NUMPY}{\soft{NumPy}\xspace}
\newcommand{\SCIPY}{\soft{SciPy}\xspace}
\newcommand{\MATPLOTLIB}{\soft{Matplotlib}\xspace}
\newcommand{\SEABORN}{\soft{seaborn}\xspace}
\newcommand{\GWPY}{\soft{GWpy}\xspace}
\newcommand{\DYNESTY}{\soft{Dynesty}\xspace}

\newcommand{\IMRPhenomXPHM}{\soft{IMRPhenomXPHM}\xspace}
\newcommand{\IMRPhenomXPHMST}{\soft{IMRPhenomXPHM\_SpinTaylor}\xspace}

\newcommand{\IMRPhenomPTWONRTidal}{\soft{IMRPhenomPv2\_NRTidal}\xspace}

\usepackage{amstext}

\makeatletter
\@ifpackageloaded{cleveref}{
  \AtBeginDocument{
    \def\ltx@label#1{\cref@label{#1}}%
    \def\label@in@display@noarg#1{\cref@old@label@in@display{#1}}%
    \def\label@in@mmeasure@noarg#1{%
      \begingroup%
        \measuring@false%
        \cref@old@label@in@display{#1}%
      \endgroup}%
  }
}{}
\makeatother

\protected\def\protectedacused{\acused}

\acrodef{LIGO}[LIGO]{Laser Interferometer Gravitational-Wave Observatory}
\acrodef{LHO}[LHO]{\ac{LIGO} Hanford Observatory}
\acrodef{LLO}[LLO]{\ac{LIGO} Livingston Observatory}
\acrodef{KAGRA}[KAGRA]{KAGRA}\acused{KAGRA}
\acrodef{iKAGRA}[iKAGRA]{initial-phase \ac{KAGRA}}
\acrodef{bKAGRA}[bKAGRA]{baseline-design \ac{KAGRA}}
\acrodef{GEO}[GEO]{GEO\,600 \ac{GW} detector}
\acrodef{aLIGO}{Advanced \ac{LIGO}}
\acrodef{A+}{Advanced+ \ac{LIGO}}
\acrodef{Asharp}[\ensuremath{\text{A}^\sharp}]{\ac{LIGO} \acs{Asharp}}
\acrodef{AdV}{Advanced \acl{Virgo}}
\acrodef{AdV+}{Advanced \acl{Virgo}+}
\acrodef{Virgo}{Virgo}\acused{Virgo}
\acrodef{VirgoNEXT}[Virgo\_nEXT]{Virgo\_nEXT}\acused{VirgoNEXT}

\acrodef{LSC}[LSC]{\acs{LIGO} Scientific Collaboration}
\acrodef{LV}[LV]{\acs{LIGO}--\acs{Virgo} Collaboration\protect\protectedacused{LVC}}
\acrodef{LVC}[LV]{\acs{LIGO}--\acs{Virgo} Collaboration\protect\protectedacused{LV}}
\acrodef{LVK}[LVK]{\acs{LIGO}--\ac{Virgo}--\ac{KAGRA} Collaboration}
\acrodef{IGWN}[IGWN]{International \ac{GWH} Observatory Network}

\acrodef{O1}[O1]{first observing run}
\acrodef{O2}[O2]{second observing run}
\acrodef{O3}[O3]{third observing run}
\acrodef{O3a}[O3a]{first half of the third observing run}
\acrodef{O3b}[O3b]{second half of the third observing run}
\acrodef{O3GK}[O3GK]{observing run}
\acrodef{O4}[O4]{fourth observing run}
\acrodef{O4a}[O4a]{first part of the fourth observing run}
\acrodef{O4b}[O4b]{second part of the fourth observing run}
\acrodef{O4c}[O4c]{third part of the fourth observing run}
\acrodef{O5}[O5]{fifth observing run}

\acrodef{BH}[BH]{black hole}
\acrodef{BBH}[BBH]{binary black hole}
\acrodef{BNS}[BNS]{binary neutron star}
\acrodef{IMBH}[IMBH]{intermediate-mass black hole}
\acrodef{NS}[NS]{neutron star}
\acrodef{BHNS}[BHNS]{black hole--neutron star binary}
\acrodefplural{BHNS}[BHNSs]{black hole--neutron star binaries}
\acrodef{NSBH}[NSBH]{neutron star--black hole binary}
\acrodefplural{NSBH}[NSBHs]{neutron star--black hole binaries}
\acrodef{PBH}[PBH]{primordial \ac{BH}}
\acrodef{CBC}[CBC]{compact binary coalescence}

\acrodef{GW}[GW]{gravitational wave\protect\protectedacused{GWH}}
\acrodef{GWH}[GW]{gravitational-wave\protect\protectedacused{GW}}
\acrodef{IFO}[IFO]{interferometer}
\acrodef{SNR}[SNR]{signal-to-noise ratio}
\acrodef{FAR}[FAR]{false alarm rate}
\acrodef{IFAR}[IFAR]{inverse false alarm rate}
\acrodef{FAP}[FAP]{false alarm probability}
\acrodef{PSD}[PSD]{power spectral density}
\acrodef{GR}[GR]{general relativity}
\acrodef{NR}[NR]{numerical relativity}
\acrodef{PN}[PN]{post-Newtonian}
\acrodef{EOB}[EOB]{effective-one-body}
\acrodef{ROM}[ROM]{reduced-order model}
\acrodef{IMR}[IMR]{inspiral--merger--ringdown}
\acrodef{PDF}[PDF]{probability density function}
\acrodef{PE}[PE]{parameter estimation}
\acrodef{CI}[CI]{credible interval}
\acrodef{CL}[CL]{credible level}
\acrodef{EOS}[EoS]{equation of state}
\acrodef{KLD}[KLD]{Kullback--Leibler divergence}
\acrodef{JSD}[JSD]{Jensen--Shannon divergence}
\acrodef{GCN}[GCN]{General Coordinates Network}
\acrodef{GWTC}[GWTC]{Gravitational-Wave Transient Catalog}
\acrodef{GWOSC}[GWOSC]{Gravitational Wave Open Science Center}
\acrodef{WDM}[WDM]{Wilson--Debauchies--Meyer}

\acrodef{CWB}[cWB]{coherent WaveBurst}
\acrodef{LAL}[LAL]{\ac{LIGO} algorithm library}

\acrodef{CHRoCC}{central heating radius of curvature correction}
\acrodef{NonSENS}{non-stationary estimation and noise subtraction}

\acrodef{PTA}{Pulsar Timing Array}

\newcommand\gwtc[1][?]{\mbox{GWTC\if#1?\else-#1\fi}}
\newcommand\thisgwtcversionmajor{5}
\newcommand\thisgwtcversionminor{0}
\newcommand\thisgwtcversionfull{\thisgwtcversionmajor.\thisgwtcversionminor}
\newcommand\thisgwtcversion\thisgwtcversionfull
\newcommand\thisgwtc{\gwtc[\thisgwtcversion]}

\acrodef{LSC}[LSC]{LIGO Scientific Collaboration}
\acrodef{LVC}[LVC]{LIGO Scientific and Virgo Collaboration}
\acrodef{aLIGO}{Advanced Laser Interferometer Gravitational-Wave Observatory}
\acrodef{aVirgo}{Advanced Virgo}
\acrodef{LIGO}[LIGO]{Laser Interferometer Gravitational-Wave Observatory}
\acrodef{IFO}[IFO]{interferometer}
\acrodef{LHO}[LHO]{LIGO-Hanford}
\acrodef{LLO}[LLO]{LIGO-Livingston}
\acrodef{O2}[O2]{second observing run}
\acrodef{O1}[O1]{first observing run}
\acrodef{O3}[O3]{third observing run}
\acrodef{O3a}[O3a]{first half of the third observing run}
\acrodef{O3b}[O3b]{second half of the third observing run}

\acrodef{BH}[BH]{black hole}
\acrodef{BBH}[BBH]{binary black hole}
\acrodef{BNS}[BNS]{binary neutron star}
\acrodef{IMBH}[IMBH]{intermediate-mass black hole}
\acrodef{NS}[NS]{neutron star}
\acrodef{BHNS}[BHNS]{black hole--neutron star binaries}
\acrodef{NSBH}[NSBH]{neutron star--black hole binary}
\acrodef{PBH}[PBH]{primordial black hole binaries}
\acrodef{CBC}[CBC]{compact binary coalescence}
\acrodef{GW}[GW]{gravitational wave}
\acrodef{GWH}[GW]{gravitational-wave}
\acrodef{EM}[EM]{electromagnetic}

\acrodef{CWB}[cWB]{coherent WaveBurst}
\acrodef{SNR}[SNR]{signal-to-noise ratio}
\acrodef{IFAR}[IFAR]{inverse false alarm rate}
\acrodef{FAP}[FAP]{false alarm probability}
\acrodef{PSD}[PSD]{power spectral density}

\acrodef{LCDM}[$\Lambda$CDM]{Lambda cold dark matter}
\acrodef{FLRW}[FLRW]{Friedmann\textendash Lema\^itre\textendash Robertson\textendash Walker}
\acrodef{LSS}[LSS]{large scale structure}

\acrodef{GR}[GR]{general relativity}
\acrodef{NR}[NR]{numerical relativity}
\acrodef{PN}[PN]{post-Newtonian}
\acrodef{EOB}[EOB]{effective-one-body}
\acrodef{ROM}[ROM]{reduced-order model}
\acrodef{IMR}[IMR]{inspiral--merger--ringdown}

\acrodef{PDF}[PDF]{probability density function}
\acrodef{PE}[PE]{parameter estimation}
\acrodef{CI}[CI]{credible interval}
\acrodef{CR}[CR]{credible region}

\acrodef{EOS}[EoS]{equation of state}

\acrodef{LAL}[LAL]{LIGO Algorithm Library}
\acrodef{MC}[MC]{Monte Carlo}
\acrodef{KLD}[KLD]{Kullback--Leibler divergence}
\acrodef{JSD}[JSD]{Jensen--Shannon divergence}

\newcommand{\DLGW}{\ensuremath{D_\mathrm{L}^\mathrm{GW}}}
\newcommand{\DLEM}{\ensuremath{D_\mathrm{L}^{\mathrm{EM}}}}
\newcommand{\skyareasymbol}{\ensuremath{\Delta \Omega}\xspace}
\newcommand{\Hunit}{\ensuremath{\text{km\,s}^{-1}\,\text{Mpc}^{-1}}}

\acrodef{MLTP}[MLTP]{\textsc{Multi Peak}\xspace}
\newcommand{\fullpop}[0]{\textsc{FullPop}-4.0\xspace}
\newcommand{\fullpopthreepeak}[0]{\textsc{FullPop 3 Peaks}\xspace}
\acrodef{PLP}[PLP]{\textsc{PowerLaw+Peak}\xspace}

\acrodef{EFT}[EFT]{effective field theory}
\acrodef{CMB}[CMB]{cosmic microwave background}
\acrodef{ISW}[ISW]{integrated Sachs--Wolfe}

\newcommand{\gwcosmo}{\texttt{gwcosmo}\xspace}
\newcommand{\icarogw}{\texttt{icarogw}\xspace}

\newcommand{\NbrCBCgwtcfour}{141\xspace}

\newcommand{\NbrCBCtotgwtcfive}{{\ensuremath{236}}}
\newcommand{\NbrCBCgwtcfive}{{\ensuremath{235}}}
\newcommand{\NbrBBHgwtcfive}{{\ensuremath{231}}}
\newcommand{\NbrBBHBrightgwtcfive}{{\ensuremath{232}}}
\newcommand{\NbrNSgwtcfive}{{\ensuremath{5}}}

\newcommand{\FARcutgwtcfive}{{\ensuremath{0.25}}}

\newcommand{\Hzerofiducial}{{\ensuremath{71.7}_{-7.5}^{+9.4}}}

\newcommand{\Hzerobrightonlysixtygwtcfive}{{\ensuremath{79.1}_{-12.4}^{+27.6}}}

\newcommand{\HzeroImprovementspectralsixtyfromgwtcfour}{{\ensuremath{14.6\%}}}

\newcommand{\HzeroCatFullpopsixtygwtcfive}{{\ensuremath{68.8}_{-13.2}^{+14.2}}}
\newcommand{\HzeroCatFullpopninetygwtcfive}{{\ensuremath{68.8}_{-21.8}^{+24.7}}}

\newcommand{\HzeroCatFullpopCombinedsixtygwtcfive}{{\ensuremath{71.7}_{-7.5}^{+9.4}}}
\newcommand{\HzeroCatFullpopCombinedninetygwtcfive}{{\ensuremath{71.7}_{-11.4}^{+17.7}}}

\newcommand{\HzeroImprovementCatAndEmptyfiducialsixty}{{\ensuremath{3.5\%}}}

\newcommand{\HzeroImprovementGladePepsOneSixty}{{\ensuremath{2.8\%}}}

\newcommand{\HzeroImprovementCombinedsixtyfromgwtcfour}{{\ensuremath{22.0\%}}}

\newcommand{\HzeroImprovementCombinedsixtyfromgwtcfourgladep}{{\ensuremath{13.2\%}}}

\newcommand{\HzeroEmptycatFullpopsixtygwtcfive}{{\ensuremath{70.1}_{-13.3}^{+15.1}}}
\newcommand{\HzeroEmptycatFullpopninetygwtcfive}{{\ensuremath{70.1}_{-21.5}^{+26.8}}}

\newcommand{\HzeroFullpopspectralbrightsixtygwtcfive}{{\ensuremath{72.2}_{-7.8}^{+10.2}}}
\newcommand{\HzeroFullpopspectralbrightninetygwtcfive}{{\ensuremath{72.2}_{-11.8}^{+19.1}}}

\newcommand{\HzeroMLTPspectralsixtygwtcfive}{{\ensuremath{71.0}_{-17.5}^{+21.0}}}
\newcommand{\HzeroMLTPspectralninetygwtcfive}{{\ensuremath{71.0}_{-26.5}^{+36.0}}}

\newcommand{\HzeroCatMLTPsixtygwtcfive}{{\ensuremath{70.6}_{-16.7}^{+20.3}}}
\newcommand{\HzeroCatMLTPninetygwtcfive}{{\ensuremath{70.6}_{-27.1}^{+36.0}}}

\newcommand{\HzeroCatMLTPbrightsixtygwtcfive}{{\ensuremath{73.1}_{-8.6}^{+12.3}}}
\newcommand{\HzeroCatMLTPbrightninetygwtcfive}{{\ensuremath{73.1}_{-12.8}^{+23.5}}}

\newcommand{\HzeroMLTPspectralbrightsixtygwtcfive}{{\ensuremath{73.4}_{-8.8}^{+12.5}}}
\newcommand{\HzeroMLTPspectralbrightninetygwtcfive}{{\ensuremath{73.4}_{-13.0}^{+24.1}}}

\newcommand{\MGxidarksixtygwtcfive}{{\ensuremath{1.1}_{-0.3}^{+0.6}}}
\newcommand{\MGxidarkninetygwtcfive}{{\ensuremath{1.1}_{-0.5}^{+1.7}}}
\newcommand{\MGndarksixtygwtcfive}{{\ensuremath{3.4}_{-2.6}^{+4.2}}}
\newcommand{\MGndarkninetygwtcfive}{{\ensuremath{3.4}_{-3.1}^{+5.9}}}
\newcommand{\MGxidarknarrowsixtygwtcfive}{{\ensuremath{1.0}_{-0.2}^{+0.3}}}
\newcommand{\MGxidarknarrowninetygwtcfive}{{\ensuremath{1.0}_{-0.3}^{+0.6}}}
\newcommand{\MGndarknarrowsixtygwtcfive}{{\ensuremath{3.8}_{-2.8}^{+3.8}}}
\newcommand{\MGndarknarrowninetygwtcfive}{{\ensuremath{3.8}_{-3.4}^{+5.4}}}

\newcommand{\MGcMdarksixtygwtcfive}{{\ensuremath{-0.4}_{-1.3}^{+1.6}}} 
\newcommand{\MGcMdarkninetygwtcfive}{{\ensuremath{-0.4}_{-2.0}^{+2.7}}}
\newcommand{\MGcMdarknarrowsixtygwtcfive}{{\ensuremath{-0.1}_{-0.8}^{+1.0}}} 
\newcommand{\MGcMdarknarrowninetygwtcfive}{{\ensuremath{-0.1}_{-1.2}^{+1.9}}}

\newcommand{\HzeroFullpopDarkDESepsOnesixty}{{\ensuremath{68.8}_{-13.2}^{+14.2}}}

\newcommand{\HzeroBPLthreePspectralsixty}{{\ensuremath{73.6}_{-13.0}^{+14.5}}}
\newcommand{\HzeroBPLthreePspectralninety}{{\ensuremath{73.6}_{-20.9}^{+25.0}}}
\newcommand{\HzeroBPLthreePspectralbrightsixty}{{\ensuremath{73.6}_{-8.0}^{+10.5}}}
\newcommand{\HzeroBPLthreePspectralbrightninety}{{\ensuremath{73.6}_{-12.1}^{+19.1}}}
\newcommand{\HzeroBPLthreePdarksixty}{{\ensuremath{71.1}_{-12.1}^{+12.9}}}
\newcommand{\HzeroBPLthreePdarkninety}{{\ensuremath{71.1}_{-20.1}^{+22.9}}}
\newcommand{\HzeroBPLthreePdarkbrightsixty}{{\ensuremath{72.3}_{-7.4}^{+9.0}}}
\newcommand{\HzeroBPLthreePdarkbrightninety}{{\ensuremath{72.3}_{-11.3}^{+17.3}}}

\newcommand{\MugLowSpectralsixtygwtcfive}{{\ensuremath{9.1}_{-0.5}^{+0.4}}}

\newcommand{\MugHighSpectralsixtygwtcfive}{{\ensuremath{27.1}_{-3.0}^{+2.4}}}

\newcommand{\LeftDipSpectralsixtygwtcfive}{{\ensuremath{2.3}_{-0.5}^{+0.4}}}

\newcommand{\RightDipSpectralsixtygwtcfive}{{\ensuremath{6.9}_{-1.5}^{+1.7}}}

\newcommand{\HzeroMLTPspectralGaussianSpinsixty}{{\ensuremath{72.4}_{-18.3}^{+22.5}}}

\newcommand{\HzeroMLTPspectralTransitionSpinsixty}{{\ensuremath{67.0}_{-17.6}^{+21.9}}}

\newcommand{\HzeroImprovementspectraMLTPGaussspinvsNospinsixtyRaw}{-8.8}

\newcommand{\HzeroImprovementspectraMLTPTransspinvsNospinsixtyRaw}{-5.3}

\newcommand{\HzeroImprovementspectraMLTPGaussspinvsNospinsixtyAbs}{\ensuremath{\fpeval{abs(\HzeroImprovementspectraMLTPGaussspinvsNospinsixtyRaw)}\%}}

\newcommand{\HzeroImprovementspectraMLTPTransspinvsNospinsixtyAbs}{\ensuremath{\fpeval{abs(\HzeroImprovementspectraMLTPTransspinvsNospinsixtyRaw)}\%}}

\newcommand{\BFHzeroMassFullPopBPLThreeP}{{\ensuremath{0.61 \pm 0.14}}}
\newcommand{\BFHzeroMassFullPop}{{\ensuremath{0.00 \pm 0.14}}}

\newcommand{\BFHzeroSpinGaussian}{{\ensuremath{0.00 \pm 0.15}}}
\newcommand{\BFHzeroSpinTransition}{{\ensuremath{3.16 \pm 0.10}}}

\newcommand{\BFHzeroDarkGladePlusLum}{{\ensuremath{-0.06 \pm 0.14}}}
\newcommand{\BFHzeroDarkGladePlusUni}{{\ensuremath{0.52 \pm 0.14}}}
\newcommand{\BFHzeroDarkDESLum}{{\ensuremath{0.19 \pm 0.14}}}
\newcommand{\BFHzeroDarkDESUni}{{\ensuremath{0.01 \pm 0.14}}}
\newcommand{\BFHzeroDarkDESLumthreeP}{\ensuremath{0.48 \pm 0.14}}
\newcommand{\BFHzeroDarkDESUnithreeP}{\ensuremath{0.79 \pm 0.14}}
\newcommand{\BFHzeroDarkGladePlusLumthreeP}{\ensuremath{0.50 \pm 0.14}}
\newcommand{\BFHzeroDarkGladePlusUnithreeP}{\ensuremath{0.56 \pm 0.14}}

\newcommand{\PearsonHcmCataloggwtcfive}{{\ensuremath{0.74}}}

\newcommand{\ImprovementcMwidesixty}{\ensuremath{\fpeval{abs(27.5)}\%}} 
 
\newcommand{\ImprovementcMnarrowsixty}{\ensuremath{\fpeval{abs(45.5)}\%}}

\newcommand{\ImprovementXizerowidesixty}{\ensuremath{\fpeval{abs(35.7)}\%}} 
 
\newcommand{\ImprovementXizeronarrowsixty}{\ensuremath{\fpeval{abs(50.0)}\%}}

\newcommand{\HzeroImprovementCatAndEmptygwcosmosixty}{\ensuremath{8\%}}
\newcommand{\HzeroImprovementCatAndEmptyicarogwsixty}{\ensuremath{2\%}}

\newcommand{\gladep}{\textsc{GLADE+ $K_s$-band}\xspace}
\newcommand{\des}{\textsc{DES $r$-band}\xspace}

\newcommand{\preferredmassmodel}{\textsc{FullPop}-4.0\xspace}

\begin{document}

\title{GWTC-5.0: Constraints on the Cosmic Expansion Rate and Modified Gravitational-wave Propagation}

\author[0000-0003-4786-2698]{A.~G.~Abac}
\affiliation{Max Planck Institute for Gravitational Physics (Albert Einstein Institute), D-14476 Potsdam, Germany}
\author{A.~Abe}
\affiliation{Department of Physics, Graduate School of Science, Osaka Metropolitan University, 3-3-138 Sugimoto-cho, Sumiyoshi-ku, Osaka City, Osaka 558-8585, Japan  }
\author{I.~Abouelfettouh}
\affiliation{LIGO Hanford Observatory, Richland, WA 99352, USA}
\author{F.~Acernese}
\affiliation{Dipartimento di Fisica ``E.R. Caianiello'', Universit\`a di Salerno, I-84084 Fisciano, Salerno, Italy}
\affiliation{INFN, Sezione di Napoli, I-80126 Napoli, Italy}
\author[0000-0002-8648-0767]{K.~Ackley}
\affiliation{University of Warwick, Coventry CV4 7AL, United Kingdom}
\author{A.~Adam}
\affiliation{OzGrav, University of Western Australia, Crawley, Western Australia 6009, Australia}
\author[0009-0004-2101-5428]{S.~Adhicary}
\affiliation{The Pennsylvania State University, University Park, PA 16802, USA}
\author{D.~Adhikari}
\affiliation{Max Planck Institute for Gravitational Physics (Albert Einstein Institute), D-30167 Hannover, Germany}
\affiliation{Leibniz Universit\"{a}t Hannover, D-30167 Hannover, Germany}
\author[0000-0002-5731-5076]{R.~X.~Adhikari}
\affiliation{LIGO Laboratory, California Institute of Technology, Pasadena, CA 91125, USA}
\author{V.~K.~Adkins}
\affiliation{Louisiana State University, Baton Rouge, LA 70803, USA}
% ...

\collaboration{9999}{The LIGO Scientific Collaboration, the Virgo Collaboration, and the KAGRA Collaboration}

\correspondingauthor{LSC P\&P Committee via LVK Publications as proxy}
\email{lvc.publications@ligo.org}

\begin{abstract}
We employ \NbrCBCtotgwtcfive{} gravitational-wave (GW) sources in the fifth LIGO--Virgo--KAGRA Collaboration (LVK) Gravitational-Wave Transient Catalog (GWTC-5.0) to estimate the Hubble constant $H_0$. 
We compare the luminosity distance measured from GWs to the redshift inferred i) using features in the mass spectrum, and ii) using statistical host galaxy association.
Probing the relationship between source luminosity distances and redshifts obtained in this way yields constraints on cosmological parameters. 
We estimate $H_0 = \Hzerofiducial \,\Hunit$ (median with 68\% symmetric credible interval).
This combines information from the source-frame mass distribution with the $H_0$ measurement from GW170817 and its electromagnetic counterpart as well as galaxy catalog information from Dark Energy Survey Year 6 (DES-Y6). 
We improve over the GWTC-4.0 measurement by using more GW sources, some with significantly smaller sky localization volumes, which leads to a reduction by \HzeroImprovementCombinedsixtyfromgwtcfour{} of the $H_0$ uncertainty and a reconstructed mass distribution with lower uncertainties.
We also constrain deviations from general relativity (GR) which affect GW propagation, specifically those that modify the luminosity distance inferred from the GW signal. We find no departures from GR in parameterized tests of GW propagation.
\end{abstract}

\keywords{Gravitational wave astronomy (675) -- Gravitational wave sources (677) -- Hubble constant (758) -- Observational cosmology (1146)}

\section{Introduction}
\label{sec:intro}

\noindent Obtaining independent measurements of the Hubble constant ($H_0$) is a major focus of gravitational-wave (\acsu{GW}) cosmology, driven by the existing discrepancy between early Universe measurements
from the \ac{CMB} radiation and local measurements from standardizable sources such as Type Ia supernovae (SNe Ia). 
Measurements of $H_0$ made by the Planck Collaboration in the Planck 2018 Data Release~\citep{Planck:2018vyg} and the Supernovae H0 for the Equation of State (SH0ES) project with the recalibration of supernovae by Large Magellanic Cloud Cepheids~\citep{Riess:2021jrx} have now reached an ${\sim}8\%$ discrepancy with $\gtrsim 5\sigma$ credibility, although other local measurements, including alternative methods of calibrating the distance ladder, suggest a smaller tension (e.g.,~\citealt{DiValentino:2024yew}).

The possibility of using \ac{GW} detections to infer cosmological parameters, such as $H_0$, was first proposed by~\citet{Schutz:1986gp}.
\acp{GW} from \acp{CBC} serve as \emph{standard sirens}~\citep{Holz:2005df}, providing a self-calibrated measure of luminosity distance that is independent of traditional methods such as the cosmic distance ladder. If combined with redshift information, \acp{GW}
can be used as probes of the luminosity distance-redshift relation, which depends on the cosmological model and its parameters. In this way GW sources may help to resolve the $H_0$ discrepancy, and can also provide insights into possible new physics beyond the standard \ac{LCDM} cosmological model~\citep{Bull:2015stt,Perivolaropoulos:2021jda, Abdalla:2022yfr, CosmoVerseNetwork:2025alb}. 

However, the  redshift of a CBC source cannot be determined from the \ac{GW} signal itself due to its degeneracy with the binary source masses~\citep{Krolak:1987ofj}. 
Several methods have been proposed to break this degeneracy. If a counterpart in the \ac{EM} spectrum can be uniquely associated to the \ac{GW} event, the redshift of the host galaxy can be determined via astronomical photometry or spectroscopy \citep{Holz:2005df,Dalal:2006qt, Nissanke:2009kt, Nissanke:2013fka,LIGOScientific:2017adf,Chen:2017rfc,Feeney:2018mkj} - referred to as a \emph{bright} siren. The only bright siren observed to date is the \ac{BNS} merger GW170817~\citep{LIGOScientific:2017vwq}, which, combined with coincident \ac{EM} transients associated with the host galaxy NGC 4993~\citep{LIGOScientific:2017ync}, provided the first bright standard siren measurement of $H_0$~\citep{LIGOScientific:2017adf}.
In the absence of further bright siren events, the steady increase of detections from \ac{BBH}, \ac{NSBH} and other \ac{BNS} candidates without confident \ac{EM} counterparts has driven forward other methods to measure $H_0$.

One approach relies on the presence of features in the mass spectrum of binary compact objects to break the mass--redshift degeneracy~\citep{Chernoff:1993th,Markovic:1993cr,Taylor:2011fs,Farr:2019twy,You:2020wju,Mastrogiovanni:2021wsd,Ezquiaga:2020tns,Ezquiaga:2022zkx}, referred to as the \emph{spectral} siren method (also sometimes called the \emph{population} method). 
The cosmological parameters are sampled together with a set of population parameters describing the intrinsic CBC parameter distribution (for instance the source-frame masses) and the \ac{CBC} merger rate. 

A second approach, referred to as the \emph{dark} siren method (also called \emph{galaxy catalog} method), consists of supplementing the spectral siren method with additional redshift information from galaxy surveys~\citep{Schutz:1986gp,MacLeod:2007jd,DelPozzo:2011vcw,Nishizawa:2016ood,LIGOScientific:2018gmd,DES:2019ccw,Gray:2019ksv,DES:2020nay,LIGOScientific:2019zcs,Finke:2021aom,LIGOScientific:2021aug,Gair:2022zsa, Borghi:2023opd, Bom:2024afj}.
Alternative approaches to infer the source redshift, which we will not consider in this work, take advantage of the cross-correlation between the spatial distribution of \acp{GW} and galaxies~\citep{Camera:2013xfa,Oguri:2016dgk,Mukherjee:2019wcg,Mukherjee:2020hyn,Mukherjee:2022afz,Afroz:2024joi,Fonseca:2023uay, Zazzera:2024agl,Ferri:2024amc,Pedrotti:2025tfg,Cheng:2026atn}, the adoption of theoretical priors on the merger-redshift distributions~\citep{Ding:2018zrk,Ye:2021klk}, and the use of tidal distortions of \acp{NS}~\citep{Messenger:2011gi,DelPozzo:2015bna,Chatterjee:2021xrm}.

Both spectral and dark siren approaches have been applied in~\cite{LIGOScientific:2025jau} to the merger candidates reported in the fourth \ac{GWTC}~\citep{LIGOScientific:2025slb} using the latest version of the codes used by the \ac{LVK}, \gwcosmo{}\footnote{\url{https://git.ligo.org/lscsoft/gwcosmo}}~\citep{Gray:2019ksv,Gray:2021sew,Gray:2023wgj,Papadopoulos:2026puy} and \icarogw\footnote{\url{https://github.com/simone-mastrogiovanni/icarogw}}~\citep{Mastrogiovanni:2023emh,Mastrogiovanni:2023zbw}.
Both codes implement the dark siren method allowing marginalization over the \ac{GW} population parameters, while incorporating galaxy catalog information.
This approach delivers cosmological constraints that are more robust to the systematic uncertainties introduced by fixing population parameters~\citep{Mastrogiovanni:2021wsd,LIGOScientific:2021aug,Pierra:2023deu,Agarwal:2024hld}.

The \ac{O4b} of the \ac{LVK} network of detectors began on 2024 April 10 at 15:00:00 UTC, following the end of the \ac{O4a} which ended on 2024 January 16 at 16:00:00 UTC.
\ac{O4b} included the two \acl{LIGO} \citep[\acsu{LIGO};][]{LIGOScientific:2014pky} detectors and the Virgo detector \citep{VIRGO:2014yos} in observing mode, while the KAGRA~\citep{KAGRA:2020tym} detector did not join the observing run.
\ac{O4b} ended on 2025 January 28 at 17:00:00 UTC, and the accompanying version of the \ac{GWTC}, hereafter referred to as \thisgwtc{}~\citep{GWTC:Introduction,GWTC:Methods,GWTC:Results}, contains all the candidates reported in previous observing runs, which include the \acl{O1}~\citep[\acsu{O1};][]{LIGOScientific:2016dsl},
the \acl{O2}~\citep[\acsu{O2};][]{LIGOScientific:2018mvr}, the \acl{O3}~\citep[\acsu{O3};][]{LIGOScientific:2020ibl,KAGRA:2021vkt,LIGOScientific:2021usb}, and \ac{O4a} \citep{LIGOScientific:2025slb} in addition to the latest observations from \ac{O4b}.
See~\citep{GWTC:Introduction} for a general introduction to \thisgwtc{}, and the
articles presented in the \thisgwtc{} Focus Issue~\citep{FocusIssue} for other aspects of this data set.

In this paper we present an updated estimate of $H_0$ using the full population of \ac{BNS}, \ac{NSBH}, and \ac{BBH} candidates reported in \thisgwtc{}. We select candidates for inclusion in the analysis based on a \ac{FAR} of less than \FARcutgwtcfive{} per year to reduce contamination from noise events. This allows us to combine the bright siren event GW170817 with an additional \NbrCBCgwtcfive{} \ac{GW} detections used as dark sirens to obtain our final estimate of $H_0$ with a total of \NbrCBCtotgwtcfive{} candidates.

In addition, we present constraints on deviations from \ac{GR} that affect the propagation of \acp{GW} and which can be parameterized in terms of a modified \ac{GW}--\ac{EM} luminosity-distance ratio~\citep{Belgacem:2018lbp,Ezquiaga:2021ayr,Mancarella:2021ecn,Leyde:2022orh,Mastrogiovanni:2023emh,Chen:2023wpj,LIGOScientific:2025jau}. These constraints test the hypothesis that gravity behaves differently from GR on cosmological scales, which is interpreted as the existence of a dark energy component (see~\citealt{Clifton:2011jh} for a comprehensive review of modified gravity models).

The remainder of this paper is organized as follows. In Section~\ref{sec:method} we summarize the spectral and dark siren statistical methods adopted in this study to jointly infer the cosmological and population parameters. In Section~\ref{sec:data} we detail the properties of the \ac{GW} candidates and the galaxy catalog used. In Section~\ref{sec:results} we present the results of our analysis and the tests made to check its robustness against systematic errors, while in Section~\ref{sec:discussion} we discuss how our results compare with the literature and the limitations of our analysis. In Section~\ref{sec:conclusion} we present our conclusions.

Throughout this paper, unless otherwise stated, we assume a flat-$\Lambda$CDM cosmology and the best-fit Planck-2015 value of $\Omega_{\rm m} = 0.3065$ for the fractional matter density in the current epoch~\citep{Planck:2015fie}. 
\section{Methods}\label{sec:method}

\subsection{Dark Siren Statistical Framework}\label{Dark Sirens Statistical Framework}
\noindent To jointly infer cosmology and population\textendash level properties of \ac{GW} sources from the observed event catalog, we employ a hierarchical Bayesian framework~\citep{Mandel:2018mve,Thrane:2018qnx,Vitale:2020aaz}. The observed sample is modeled as resulting from an inhomogeneous Poisson process in the presence of selection effects, assuming statistically independent and non-overlapping events. 
Each event in the catalog is described by detector\textendash frame parameters ${\PEparameter}^\mathrm{det}$, which include the detector\textendash frame masses and \ac{GW} luminosity distance, $\PEparameter^\mathrm{det} \ni \{  m_1^\mathrm{det}, m_2^\mathrm{det}, \DLGW \}$ (where $m_1^\mathrm{det} \geq m_2^\mathrm{det}$). 
For each event, labeled by the index $i$, individual parameter constraints are given in the form of samples from the posterior probability $p\big( {\PEparameter}^\mathrm{det}_{i} | \PEdata_i \big)$ for the parameters ${\PEparameter}^\mathrm{det}_{i}$ given the observed data $\PEdata_i$. These are assumed to be obtained with a parameter estimation prior that we denote $\pi_{\rm PE}(\PEparameter^\mathrm{det})$.
The event parameters are drawn from a distribution which is modeled as a function of source\textendash frame quantities ${\PEparameter}$, which include the source\textendash frame masses and redshift, $ \PEparameter \ni \{  m_1, m_2, z \}$.
The population distribution 
$p_{\mathrm{pop}}(\PEparameter | \PEhyparameter)$ is described parametrically by a set of \emph{hyperparameters} $\PEhyparameter$ (sometimes simply referred as \emph{parameters}) that includes the cosmological parameters $\PEhyparameter_{\rm c}$, and the parameters describing the astrophysical distribution of sources, such as the mass distribution and the redshift evolution of the merger rate.
The set of hyperparameters is inferred through the population-level likelihood, given by \citep{Loredo:2004nn,Mandel:2018mve,Vitale:2020aaz}
\begin{align}
p\left( \{\PEdata_i\}_{i=1}^{N_{\mathrm{det}}} | \PEhyparameter \right) &\propto \xi(\PEhyparameter)^{-N_{\mathrm{det}}} \, \prod_{i=1}^{N_{\mathrm{det}}} \int \mathrm{d}\PEparameter^\mathrm{det}_{i} \, \frac{p\big( {\PEparameter}^\mathrm{det}_{i} | \PEdata_i \big)}{\pi_{\rm PE}(\PEparameter^\mathrm{det}_{i})} \nonumber \\ 
&\times \left[ \Big|\frac{\mathrm{d}\PEparameter^\mathrm{det}_{i}(\theta, \PEhyparameter)}{\mathrm{d} \PEparameter_{ i}} \Big| ^ {-1}
 p_{\mathrm{pop}}(\PEparameter_i | \PEhyparameter)\right] \, , \label{eqn:sec2_marginal_lk}
\end{align}
where $\xi(\PEhyparameter) $ is the expected fraction of detected events in the population, accounting for selection effects, and the Jacobian term $\left|\mathrm{d}\PEparameter^\mathrm{det}_{i} / \mathrm{d} \PEparameter_{ i} \right|$ accounts for the transformation from source to detector frame. 
Finally, $N_{\mathrm{det}}$ is the total number of detected events in the catalog, and we assume marginalization over the overall total number of mergers in the observing time, $N$, with a scale--invariant prior $\propto 1/N$~\citep{Mandel:2018mve}.

A description of the implementation of the likelihood computation is provided in the O4a cosmology paper \citep{LIGOScientific:2025jau}.
Throughout, we use two pipelines to sample from the hyperparameter posterior: {\icarogw} and {\gwcosmo} \citep{Mastrogiovanni:2023zbw, Papadopoulos:2026puy} that adopt different strategies to evaluate the posterior in Eq.~\eqref{eqn:sec2_marginal_lk}.

\subsection{Construction of Redshift Priors}\label{subsec:redshift_prior}

\noindent In this section, we summarize the most relevant aspects of the construction of redshift distributions as a function of sky position. 
Since the methodological framework is virtually unchanged when compared to the GWTC-4 analysis \citep{LIGOScientific:2025jau}, we only focus on the most relevant equations and refer to \citet{Mastrogiovanni:2023emh, Gray:2023wgj} for further details.

The line-of-sight redshift prior for a specific element in the sky, $\Omega$, receives two contributions
\begin{equation}
    p(z | \Omega) \propto  \frac{\psi(z | \PEhyparameter)}{1+z} \left[\frac{\mathrm{d}N^{\rm eff}_{\rm gal,cat}}{\mathrm{d}z \mathrm{d}\Omega }+\frac{\mathrm{d}N^{\rm eff}_{\rm gal,out}}{\mathrm{d}z \mathrm{d}\Omega} \right]
    \,,
    \label{eq:total}
\end{equation}
where $\psi(z|\PEhyparameter)$ parametrizes the redshift dependence of the \ac{CBC} merger rate and the factor of $(1+z)^{-1}$ accounts for the conversion of time intervals from source to observer frame.
The terms in square brackets represent the contributions to the redshift prior from galaxies within the catalog (first term), and a model for unobserved `out-of-catalog' galaxies (second term).
We note that the two pipelines treat the galaxy catalog completeness correction in slightly different ways: one depends on all Schechter function parameters, including the galaxy number density per comoving volume (\texttt{icarogw}), while the other is independent of it (\texttt{gwcosmo}). 
This can result in a systematic effect, as seen for the DESY6 catalog (see Section~\ref{sec:results}).

\paragraph{In\textendash catalog part, $\mathrm{d}N^{\rm eff}_{\rm gal,cat}/(\mathrm{d}z \mathrm{d}\Omega)$} This term is built starting from the galaxies in the catalog. The sky is divided in equal\textendash size pixels, labeled with their central coordinates $\Omega$, 
with the \texttt{healpix} pixelization algorithm~\citep{Gorski:2004by,Zonca:2019vzt}. 
Inside each pixel, we select all galaxies with apparent magnitude smaller than the median inside the pixel, denoted as $m_{\rm thr}(\Omega)$.

For each pixel, a redshift prior is constructed as a weighted sum of the posterior distributions for the true redshift $z$ given observed redshifts $z^j_{\rm obs}$ for the selected galaxies $j=1, \dots, N_{\rm gal}(\Omega)$ in the pixel, each denoted by $p(z|z^j_{\rm obs},\sigma^j_{z,\mathrm{obs}}, \PEhyparameter_{\rm c})$, where $\sigma^j_{z,\mathrm{obs}}$ stands for redshift uncertainty (standard deviation of a Gaussian). 
The weight each galaxy contributes with is computed from the absolute magnitude of each galaxy, $M_j$ (or equivalently its absolute luminosity $L_j$),
\begin{equation}
    \label{eq:gal_weights}
    w_j(\epsilon, M_j) =  \left|\frac{L_j}{L_*}\right|^{\epsilon} = 
    10^{-0.4 \epsilon ( M_j-M_*)} \,,  
\end{equation}
 \citep{Gray:2019ksv}, where $L_*$ and $M_*$ are the reference luminosity and corresponding magnitude at the knee of the luminosity function, respectively. We assume the luminosity function to be given by the \cite{Schechter:1976iz} form, described in more detail in Appendix~B of \cite{LIGOScientific:2025jau}. 

In Eq.~\eqref{eq:gal_weights}, we weight each galaxy by the absolute luminosity in a specific band, $L_j$, raised to a power $\epsilon$ which we treat as a fixed parameter. 
In particular, we consider the cases $\epsilon=0$, corresponding to equal probability for all galaxies to host \acp{CBC}, which we will refer to as \emph{no\textendash weighting} case, and $\epsilon=1$ corresponding to a linear weight of galaxies by their luminosity, which we will refer to as \emph{luminosity\textendash weighting} case.
It is known that luminosity in specific magnitude bands correlates with galaxy properties such as stellar mass or star formation rate, for example \citep{Neijssel:2019irh,Adhikari:2020wpn,Santoliquido:2020axb,Broekgaarden:2021efa,Rauf:2023oiy,Srinivasan:2023vaa,Hanselman:2024hqy,Li:2025hrh}. Luminosity\textendash weighting reflects an assumption that such galaxy properties may also correlate with likelihood to host \ac{CBC}s, see e.g. \cite{OShaughnessy:2016nny,Artale:2019doq,Gray:2019ksv,Vijaykumar:2023bgs} for more extended discussions.

For the out-of-catalog contribution, see Appendix~B of \cite{LIGOScientific:2025jau} for details.

\subsection{Population Models}
\label{subsubsec:population_models}

\begin{figure*}[ht!]
    \centering
    \includegraphics[width=0.95\textwidth]{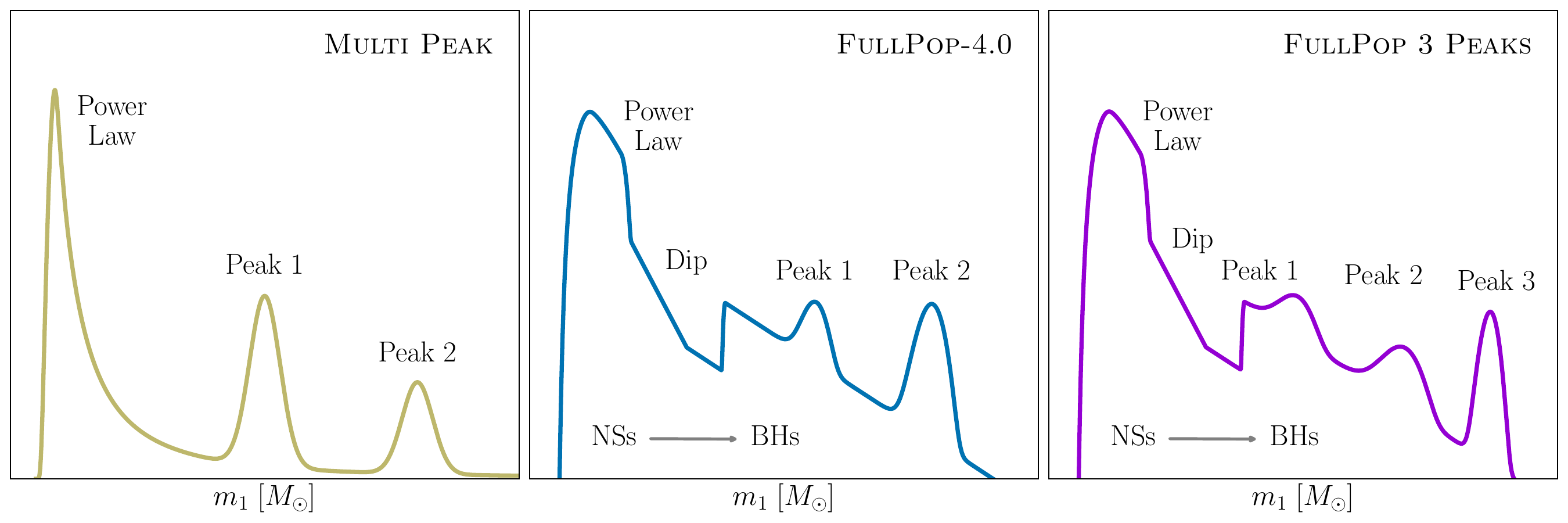} 
    \caption{Qualitative graphical representation of the three source-frame mass models considered in this paper and described in Section~\ref{subsubsec:population_models} and Appendix~\ref{sec:appendix_pop}.  
    The mass distribution model displayed in the first panel represent the mass ranges of \acp{BH}, while the second two panels include both \acp{BH} and \acp{NS}. The mass ranges shown are not to scale.
    }
    \label{fig:mass model sketch}
\end{figure*}

\noindent We construct \ac{CBC} rate models from independent redshift and source mass distributions, and while we assume in most analyses the \ac{CBC} spins to be isotropically distributed with uniform distribution in the spin magnitudes, in a subset of analyses we additionally model the spin population.

We consider three different models for the distribution of primary mass, $p\left(m_{1} | \PEhyparameter \right)$, which enters the term $p_{\mathrm{pop}}$ in Eq.~\eqref{eqn:sec2_marginal_lk}. These models are denoted as: \ac{MLTP}, \fullpop and \fullpopthreepeak.
These are phenomenological parametric models defined in terms of relatively simple functional forms that contain features motivated by either astrophysical expectations or previous \ac{GW} observations. These models are constructed as superpositions of truncated Gaussian and power\textendash law distributions with different parameters (described in Appendix~\ref{sec:appendix_pop}). In this work we consider these mass models as redshift-independent; see \citet{Mukherjee:2021rtw, Karathanasis:2022rtr, Rinaldi:2023bbd, Heinzel:2024hva,Lalleman:2025xcs,Gennari:2025nho,GWTC:AstroDist} for investigations into their possible evolution. We will comment upon this further in Section~\ref{sec:perspective}.

Figure~\ref{fig:mass model sketch} shows a sketch of the typical form of these models, with the different mass features that characterize them highlighted. We briefly describe these models next (see Appendix~\ref{sec:appendix_pop} and~\citealt{GWTC:AstroDist} for more details). The \ac{PLP} model used in \citet{LIGOScientific:2025jau} has not been continued in this work: it was slightly disfavored relative to \ac{MLTP} in \citet{LIGOScientific:2025jau}, and we find it more strongly disfavored with \thisgwtc{}.

The \ac{MLTP} mass model was originally introduced in \citet{LIGOScientific:2020kqk}. It features a power\textendash law distribution for the primary mass spectrum with a smooth low\textendash mass cutoff and includes a pair of Gaussian peaks to capture excesses at intermediate masses. This model is similar to the \textsc{Broken Power Law + 2 Peaks} model adopted in~\citet{LIGOScientific:2025pvj}, except our model has one power law instead of two.
This model is characterized by eleven population parameters.

In the \ac{MLTP} model the full mass distributions are factorized as
\begin{equation}
\begin{split}
p(m_{1},m_{2} | \PEhyparameter) = &\,\,p(m_{1} | \PEhyparameter)\,S_{\rm h}(m_1|\PEhyparameter)\\
&\times p(m_{2} | m_{1},\PEhyparameter)\,S_{\rm h}(m_2|\PEhyparameter),
    \label{massprior0}
\end{split}
\end{equation}
where $p\left(m_{2} | m_{1}, \PEhyparameter\right)$ is the distribution of the secondary mass component conditioned on the primary mass and $S_{\rm h}(m|\PEhyparameter)$ is a smoothing function defined in Appendix~C of \cite{LIGOScientific:2025jau}. 
This is modeled assuming that the mass ratio $q=m_{2} /m_{1}$ follows a power\textendash law distribution. 

The \fullpop model is a generalization of the \ac{MLTP}, extending the distribution to encompass the full mass spectrum of \acp{CBC}, including \ac{BNS}, \ac{NSBH}, and \ac{BBH} mergers. It is designed to cover a wide mass range, from a few to several hundred solar masses~\citep{Fishbach:2020ryj,Farah:2021qom,Mali:2024wpq}. 
The model combines a first power\textendash law component for the low\textendash mass region (representing \ac{NS}-containing events) with a smooth low\textendash mass cut\textendash off, and a second power\textendash law component for the \ac{BBH} mass distribution, which includes two Gaussian peaks. 
A dip function is introduced at the junction between the two power\textendash law regimes, aiming to model the apparent mass gap between \acp{NS} and \acp{BH}. The parameters governing this dip are treated as population parameters.
This model is characterized by nineteen parameters.

By modeling the full population of compact objects in a unified framework, the \fullpop model allows us to include a larger set of GW events in our analysis, offering greater sensitivity to features in the mass spectrum.
Another major distinction from the \ac{MLTP} mass model lies in the parametrization of the secondary mass. Instead of modeling $m_{2}$ as a power\textendash law conditioned on $m_{1}$, as in Eq.~\eqref{massprior0}, the \fullpop model assumes that the distribution of $m_{2}$ is  given by $p(m_{2}|\PEhyparameter)$ and employs a pairing function $f(m_1, m_2 | \PEhyparameter)$ enforcing the condition $m_{1} \geq m_{2}$ and allowing for further flexibility for the secondary mass~\citep{Fishbach:2019bbm}, with
\begin{equation}
\label{massprior1}
p(m_1, m_2|\PEhyparameter) \propto \;p_{\rm S}(m_1|\PEhyparameter)\, p_{\rm S}(m_2|\PEhyparameter) f(m_1, m_2| \PEhyparameter)\,,
\end{equation}
where $p_{\rm S}(m|\PEhyparameter)$ is defined in terms of $p(m|\PEhyparameter)$ and the smoothing functions defined in Appendix~C of \citep{LIGOScientific:2025jau}. 

Finally, we include a further extension of the \fullpop model, which we denote as \fullpopthreepeak, which was first used in~\citet{Pierra:2026hbh}. This model modifies the \fullpop mass distributions by the addition of a third Gaussian peak intended to model further structure in the BBH mass distribution. 

The equations which describe our three population models can be found in Appendix~\ref{sec:appendix_pop}. For more details, see also~\citet{GWTC:AstroDist}.
In Section~\ref{sec:results} we compare our analysis obtained using \textit{single\textendash population} models (the \ac{MLTP} model which is valid for \ac{BBH} candidates only) to that obtained using a \textit{multi-population} model (\ac{BNS} + \ac{NSBH} + \ac{BBH} candidates), i.e., the \fullpop and \fullpopthreepeak models.

For the merger rate evolution as a function of the redshift, the term $\psi(z | \PEhyparameter)$ in Eq.~\eqref{eq:total}, is modeled with a Madau\textendash Dickinson parametrization~\citep{Madau:2014bja}, which is characterized by parameters $\{ \gamma, \kappa, z_{\mathrm{p}}\} \in \PEhyparameter$, see Eq.~\eqref{eq:MDrate} in Appendix~\ref{sec:appendix_pop} for details.
In particular, the parameter $\gamma$ controls the low\textendash redshift slope of the merger rate, with $\psi(z) \propto (1+z)^{\gamma}$ for $z \ll z_{\mathrm{p}}$. 

In addition, rather than assuming that the spin magnitudes are uniformly distributed and that their orientations are isotropic, these properties can be jointly modeled and inferred alongside the mass and merger rate distribution, defined as an extra $p_{\rm pop}(\cdot)$ term in Eq.~\eqref{eqn:sec2_marginal_lk}. Such inclusion to the inference framework makes it more general, through a more complete modeling of the CBC population, and potentially could improve the constraints on the cosmology. In this work, we consider two parametrizations, to model the full spin distribution denoted by
\begin{equation}
    p(\vec{\chi},\cos\vec{\theta}|\boldsymbol{\Lambda})=p(\vec{\chi}|\boldsymbol{\Lambda})p(\cos\vec{\theta} | \boldsymbol{\Lambda}),
\end{equation}
factorized as the product between the PDF for the spin magnitude part $p(\vec{\chi}|\boldsymbol{\Lambda})$ and tilt angles part $p(\cos\vec{\theta}|\boldsymbol{\Lambda})$. The first parametrization, called the \textsc{Gaussian} model, assumes that the spin magnitudes for each component are given by truncated Gaussian distributions over $[0,1]$, while the tilt angles are given by a mixture model between a uniform distribution between $[-1,1]$ and a truncated Gaussian centered around unity. This model is directly taken from \cite{LIGOScientific:2025pvj}. 
The second spin model is taken from \cite{Pierra:2024fbl}, in particular we use the so-called \textsc{Transition} model which allows for a possible transition between two spin magnitude distributions via the mass, both described with truncated Gaussians between 0 and 1. The PDF for the spin magnitude now depends directly on the masses and reads as
\begin{eqnarray}
    \pi(\chi|m,\Lambda) &=& W(m)\mathcal{N}_{1,[0,1]}(\chi|\mu_{\chi_1},\sigma_{\chi_1}) +\nonumber\\
    & +& (1-W(m))\mathcal{N}_{2,[0,1]}(\chi|\mu_{\chi_2},\sigma_{\chi_2}) \, ,
\end{eqnarray}
where $W(m)$ is a mass dependent window function which can transition from an initial value $\lambda_{\rm f}$ to 0.
The window function is constructed as a logistic function such that
\begin{eqnarray}
    W(m;m_{\rm t},\delta_{\rm m_t},\lambda_{\rm f}) = \frac{\lambda_{\rm f}}{1+e^{\frac{m-m_{\rm t}}{\delta_{\rm m_t}}}} \, ,
\end{eqnarray}
where $m_{\rm t}$ is the mass at which the first spin magnitude distribution transitions to the second one and $\delta_{\rm m_t}$ controls the steepness of that transition. With the \textsc{Transition} prescription, the angle distribution of the spins is the same as for the \textsc{Gaussian} model.

Such a spin model is of interest given the recent evidence for correlations between the spin magnitudes and masses of BBHs \citep{Pierra:2024fbl,Li:2023yyt,LIGOScientific:2025pvj,Li:2025rhu}. 
These correlations could improve the resulting $H_0$ constraint through a more faithful reconstruction of a spin-dependent feature of the mass distribution.

\subsection{Cosmological Models}
\label{subsec: model params}

\subsubsection{Background Evolution}
\label{subsubsec: cosmological models}
\noindent Under the assumptions of homogeneity and isotropy, the luminosity distance can be computed based on the \ac{FLRW} metric as
\begin{equation}
\label{eq:dlz}
    \DL =\frac{c(1+z)}{H_0} \int_0^z \frac{\mathrm{d}z'}{E(z')}\,,
\end{equation}
where $E(z)=H(z)/H_0$ is the dimensionless expansion rate of the Universe. This depends on the cosmological model assumed and can be computed using the Friedmann equations.
In this paper, we restrict our focus to a flat\textendash $\Lambda$CDM model. 
Under this assumption, $E(z)$ is given by
\begin{equation}
\label{eq:Ez}
E(z)=\left[\Omega_{\rm m}(1+z)^3+\Omega_{\Lambda}\right]^{1/2}\,.
\end{equation}
Here, $\Omega_{\rm m}$ is the fractional energy density in  matter components today (cold dark matter + baryonic matter), 
and we have ignored the radiation energy density which is negligible at the redshifts of our interest. Under this approximation, the dark energy density fraction today is $\Omega_{\Lambda} = 1 - \Omega_{\rm m}$.

As our data currently have no constraining power on the equation\textendash of\textendash state parameter of dark energy $w_0$, our main results will be based on the flat\textendash $\Lambda$CDM model with $w_0 = -1$.

\subsubsection{Models of Modified \ac{GW} Propagation}
\label{subsubsec: gravity models}
\noindent We also analyze our data in the context of cosmological modified\textendash gravity models. Some of the analyses presented here are continued from \citep{LIGOScientific:2025jau}, where more extensive details can be found.
We focus here on a common feature of these models, sometimes referred to as \emph{GW friction}~\citep{Saltas:2014dha,Pettorino:2014bka,Nishizawa:2017nef,Amendola:2017ovw,Lagos:2019kds}.
Under this effect, new terms in the \ac{GW} propagation equation result in modifications to the \ac{GW} amplitude received at the observer. This effect is indistinguishable from a change in the luminosity distance to the \ac{GW} source. The result is that the luminosity distance $\DLGW$ inferred for a \ac{GW} source differs from the \ac{EM} luminosity distance $\DL$ given by Eq.~\eqref{eq:dlz}. Any measurement of the \ac{GW} source luminosity distance obtained using \ac{EM} observables would be unaffected, i.e., $\DLEM=\DL$.

Based on this, multiple studies~\citep{{Belgacem:2017ihm,Belgacem:2018lbp},LISACosmologyWorkingGroup:2019mwx,Belgacem:2019tbw,Mukherjee:2020mha,Finke:2021aom,Finke:2021znb,Finke:2021eio,Ezquiaga:2021ayr,Finke:2021aom,Mancarella:2021ecn,Kalogera:2021bya,Leyde:2022orh,Liu:2023onj,Branchesi:2023mws, Chen:2023wpj, Abac:2025saz} have considered the ratio $\DLGW / \DLEM$ as a convenient probe of departures from \ac{GR} on cosmological scales. This ratio, equal to one in \ac{GR}, can become a function of redshift in cosmological modified\textendash gravity models.
Here we consider two commonly used parameterized forms for the \ac{GW}\textendash \ac{EM} luminosity-distance ratio.

Two assumptions are relevant to both parameterized forms. Firstly, we assume the \ac{GW} propagation speed, $c_T$, is luminal due to the tight constraint on this parameter from GW170817~\citep{LIGOScientific:2017zic}.
Secondly, we treat departures from \ac{GR} impacting only the propagation phase of \ac{GW} signals.
That is, we do not consider modifications to the generation of \ac{GW}s, which would affect the waveform \textit{at source}.

\paragraph{$\Xi_0$\textendash $n$ Parametrization} 
In this parametrization, $\DLGW$ is described by \citep{Belgacem:2018lbp}
\begin{equation}
    \label{eq: def Xi parametrization dl}
    \DLGW = \DLEM \left(\Xi_0 + \frac{1-\Xi_0}{(1+z)^n}\right)\,,
\end{equation}
where both parameters $\Xi_0$ and $n$ are positive. The primary parameter of interest is $\Xi_0$, which controls the overall amplitude of departures from \ac{GR}. At low redshifts, $\DLGW/\DLEM \rightarrow 1$ (irrespective of $\Xi_0$).
At high redshifts, $\DLGW/\DLEM \rightarrow \Xi_0$
as changes to $\DLGW$ should saturate at redshifts where the fractional energy density of dark energy, 
$\Omega_{\Lambda}(z)$, is negligible. This holds under the assumption that deviations from \ac{GR} are associated to the late\textendash time emergence of dark energy. The power\textendash law index $n$ controls the rate of transition between these two regimes.

The specific form of the $\Xi_0$\textendash$n$ parametrization is an assumption, and Eq.~\eqref{eq: def Xi parametrization dl} was calibrated to cover a large spectrum of known luminal modified\textendash gravity theories \citep{LISACosmologyWorkingGroup:2019mwx}.
The \ac{GR} limit of the theory is $\Xi_0\rightarrow 1$ (for any value of $n$). However, the parametrization is imperfectly behaved, since $n\rightarrow 0$ also recovers the \ac{GR} behavior $\DLGW = \DLEM$.

\paragraph{$\alpha_M$ Parametrization} 
This parametrization is inspired by Horndeski gravity~\citep{Horndeski:1974wa, Deffayet:2011gz,Kobayashi:2011nu}, which is the most general family of scalar\textendash tensor gravity models with second-order equations of motion. See \citet{Kobayashi:2019hrl} and references therein for a review of Horndeski gravity and its phenomena. In the widespread basis of~\cite{Bellini:2014fua}, adopted for describing linear cosmological perturbations of Horndeski theories around a \ac{FLRW} solution, $\alpha_M(z)$ is the rate of change of the effective Planck mass, and hence the effective gravitational coupling strength  \citep{Bellini:2014fua, Gleyzes:2014rba}. This results in the following expression for $\DLGW$~\citep{Lagos:2019kds}:
\begin{equation}
    \begin{split}
         \DLGW &= 
        \DLEM \exp \left\{ \frac{1}{2} \int_0^z \frac{\mathrm{d} z^{\prime}}{1+z^{\prime}} \alpha_M\left(z^{\prime}\right)\right\},
        \label{eq:aM_param}
    \end{split}
\end{equation}
where in this work we will use the following ansatz for $\alpha_M(z)$ 
\begin{eqnarray}
    \alpha_M(z)=c_M \frac{\Omega_{\Lambda}(z)}{\Omega_{\Lambda}}= c_M \frac{1}{E^2(z)}\,,
    \label{eq:aM_ansatz}
\end{eqnarray}
where $ \Omega_{\Lambda}=\Omega_{\Lambda}(z=0)$ and $c_M$ is a constant of proportionality. For the dimensionless expansion rate, $E(z)$, we use Eq.~\eqref{eq:Ez} which assumes a flat\textendash $\Lambda$CDM model with constant dark energy density, as in this work we are not considering changes to the cosmological expansion history. 
In principle, $\alpha_M(z)$ also enters the background evolution equations; however, any resulting change can be absorbed into other functions such as the effective dark energy equation of state~\citep{Bellini:2014fua}.
The \ac{GR} limit of the model is obtained for $c_M=0$.

The redshift\textendash dependent form of $\alpha_M(z)$ is a choice.
The form of Eq.~\eqref{eq:aM_ansatz} has been widely adopted for \ac{LSS} constraints~\citep{Bellini:2015xja,Noller:2018wyv,Baker:2020apq,Seraille:2024beb, Ishak:2024jhs}.

\paragraph{Connection to gravitational coupling.}
We noted above that, in Horndeski gravity, the function $\alpha_M(z)$ controls the rate of change of an effective gravitational coupling strength. 
As detailed further in Appendix~\ref{sec:appendix_MG}, in Horndeski gravity the following relation holds:
\begin{align}
\label{eq:DGW_GGW}
\frac{G_{\rm GW}(z)}{G_{\rm GW}(0)}
=\left(\frac{D_{\rm L}^{\rm GW}}{D_{\rm L}^{\rm EM}}\right)^2
\end{align}
where $G_{\rm GW}(z)$ is an effective gravitational coupling associated to \ac{GW} propagation, which may be redshift-dependent. In GR this quantity would be Newton's constant, $G_{\rm N}$. A change in the value of $G_{\rm GW}(z)$ between $z_{\rm source}$ and $z=0$ is a common leading-order change to \ac{GW} propagation in Horndeski gravity.
By squaring our constraints on the GW-EM distance ratio, we can interpret our constraints in this light via Eq.~\eqref{eq:DGW_GGW}.

We note that a different gravitational coupling strength can also be constrained by large-scale structure surveys (LSS), sometimes denoted $G_{\rm eff}$, $G_{\rm matter}$ or $\mu$. The quantities $G_{\rm GW}$ and $G_{\rm eff}$ can be the same under some conditions, which allows for a comparison of GW and LSS results \citep{Bellini:2014fua,Gleyzes:2014rba,Perenon:2015sla,BeltranJimenez:2015sgd,Romano:2025pcs}.

\section{Data}\label{sec:data}

\subsection{GW Events}\label{subsec:gw_events}
\noindent The analyses presented here use \thisgwtc{} and are based on the detection of \ac{GW} candidates produced by merging compact binaries between \ac{O1} and the end of \ac{O4b}. To reduce the noise contamination of the datasets used in cosmological studies, we select a subset of GW events with the lowest \ac{FAR} among all search pipelines, ensuring all events have \ac{FAR}  $<\FARcutgwtcfive\,\text{yr}^{-1}$.

A total of \NbrCBCtotgwtcfive{} \ac{CBC} \ac{GW} candidates with \acp{FAR} below this threshold have been detected by our search pipelines: \textsc{cWB-BBH} \citep{Klimenko:2005xv, Klimenko:2008fu, Klimenko:2015ypf, Mishra:2024zzs}, \textsc{GstLAL} \citep{Messick:2016aqy, Sachdev:2019vvd, Tsukada:2023edh, Sakon:2022ibh, Joshi:2025nty}, \textsc{MBTA} \citep{Allene:2025saz}, and \textsc{PyCBC} \citep{Allen:2005fk, DalCanton:2014hxh, Usman:2015kfa, Nitz:2017svb}.
Following the \thisgwtc{} classification of candidates into potential \acp{BBH} or \ac{NS}-binaries~\citep{GWTC:Results}, \NbrBBHgwtcfive{} out of \NbrCBCtotgwtcfive{} events are believed to originate from the coalescence of \ac{BBH} candidates and \NbrNSgwtcfive{} from binaries where at least one component could have been a \ac{NS}.
This analysis includes the \NbrCBCgwtcfour{} dark siren events considered in our previous cosmological analysis of GWTC-4.0 \citep{LIGOScientific:2025jau}.
As in the GWTC-4.0 analysis, we exclude GW231123\_135430~\citep{LIGOScientific:2025rsn}, as some of its inferred properties, such as the binary masses or its luminosity distance, vary dramatically with different waveforms, and the reason for these differences is not well understood.
Added to the \NbrCBCgwtcfive{} dark sirens is the special case of the multi-messenger event GW170817, which is again treated differently from the others and will be used in the rest of the paper as a bright siren.

\begin{figure}[t!]
    \centering
    \includegraphics[width=\columnwidth]{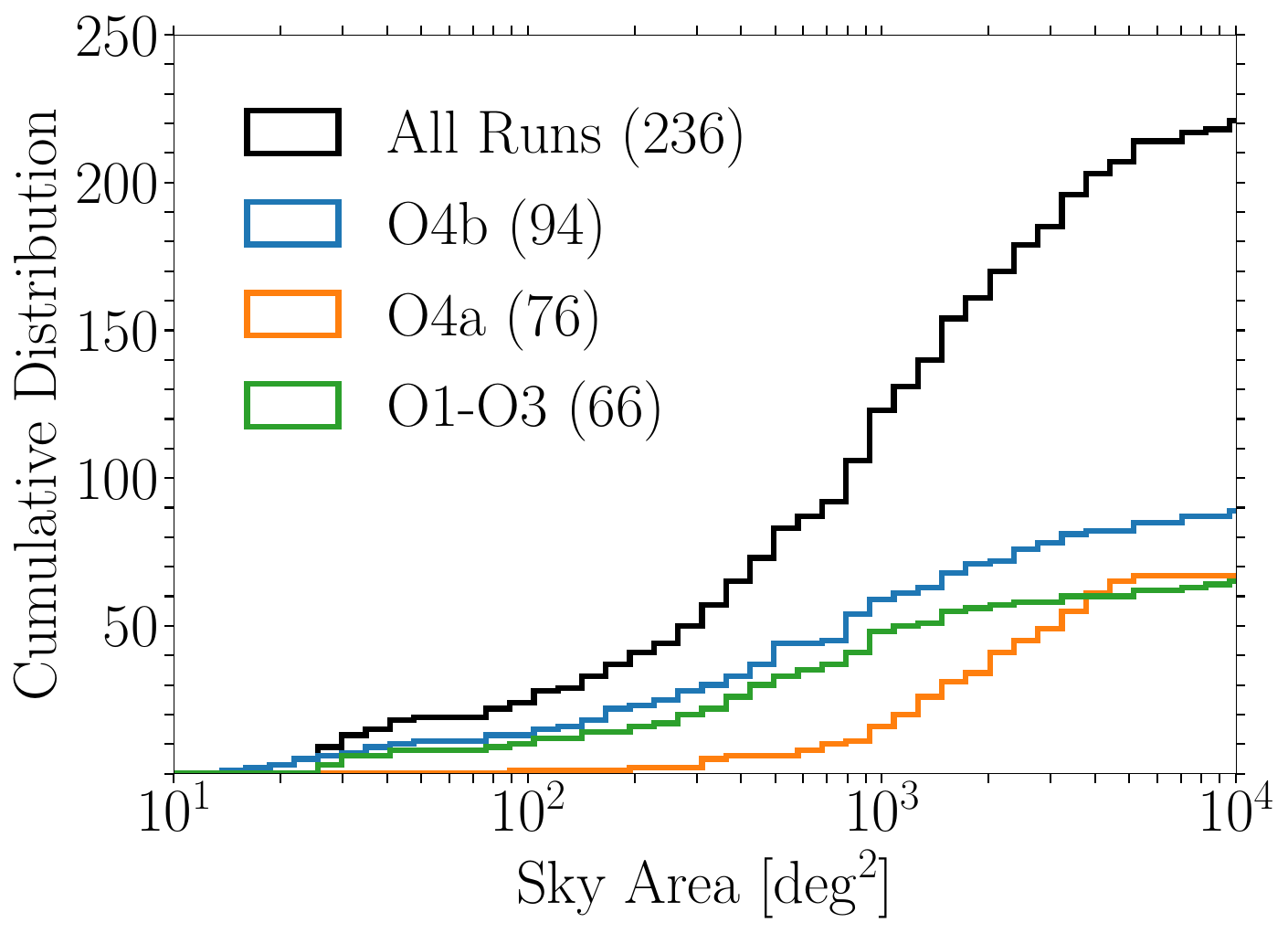}
    \caption{Cumulative distribution of the size of the 90$\%$ \ac{CR} of the sky localization of \ac{CBC} candidates observed during \ac{O1}+\ac{O2}+\ac{O3}+\ac{O4a}+\ac{O4b} in black (\NbrCBCtotgwtcfive{} total events including GW170817), \ac{O1}-\ac{O3} in green (66 events), \ac{O4a} in orange (76 events), and \ac{O4b} in blue (94 events). The exclusion of \ac{Virgo} in \ac{O4a} greatly increased the sky localization uncertainty.}
    \label{fig:sky_localization}
\end{figure}

Figure~\ref{fig:sky_localization} shows the cumulative distribution of the 90\% \ac{CR} of the sky localization of \ac{CBC} events observed in all \ac{LVK} observing runs since \ac{O1}, as well as that of the \ac{O4b} events only. 
The sky localization of the GW events detected during the \ac{O4b} observing run is, on average, better than previous events and markedly more precise than O4a events (see Figures \ref{fig:sky_localization} and \ref{fig:galaxy_catalog_comparison}).
This improvement can be attributed to the participation of \ac{Virgo} in \ac{O4b} (which was not observing in O4a), resulting in better triangulation from three detectors \citep{Essick:2014wwa,Ouzriat:2025ben}.
A full list of luminosity distances and sky uncertainties of the \ac{GW} candidates considered in our study can be found in Appendix~\ref{sec:appendix_event_list}.

Different waveform models have been used to perform the \ac{PE} for each \ac{GW} candidate across the observing runs~\citep{GWTC:Methods}. 
For our analysis, we use posterior samples produced with a single waveform approximant rather than a mixture of samples from different waveforms~\citep{GWTC:Methods,GWTC:Results} to mitigate potential systematics.
In particular, for candidates from the \ac{O1}, \ac{O2}, and \ac{O3} runs, we use the posterior samples based on the \IMRPhenomXPHM waveform model~\citep{Pratten:2020ceb}, where 
for GW200115\_042309~\citep{LIGOScientific:2021qlt} we use the large-spin magnitude prior posterior samples, while for the BNS mergers GW190425\_081805~\citep{LIGOScientific:2020aai} and GW170817~\citep{LIGOScientific:2019zcs} we use the large-spin magnitude prior posterior samples obtained with the \IMRPhenomPTWONRTidal~\citep{Dietrich:2017aum,Dietrich:2018uni} and a prior allowing for high-spin and low-spin magnitudes, respectively.
For events from both the \ac{O4a} and \ac{O4b} observing runs, we use the posterior samples produced with the \IMRPhenomXPHMST model~\citep{Colleoni:2024knd}, except for GW230529\_181500, for which we use posterior samples produced using the \IMRPhenomXPHM waveform model and released in~\citet{LIGOScientific:2024elc}.
In this study we do not consider the impact of waveform systematics, as they are expected to be relevant only in population analyses of \ac{GW} events with \ac{SNR} above 100~\citep{Kapil:2024zdn, Dhani:2024jja}.

Finally, we estimate the \ac{GW} detection probability in denominator of Eq.~\eqref{eqn:sec2_marginal_lk} by using a set of simulated \ac{GW} signals (called injections) \citep{GWTC4RnPInjMethod,GWTC:Methods}. More details on how the injections are used to compute this expression can be found in Appendix~A of \citet{LIGOScientific:2025jau}.

\subsection{Galaxy Catalogs}\label{subsec:gal_cat}

We use two galaxy catalogs for our dark siren analysis, each with unique properties for cosmological inference. We describe them in more detail below. As the approach uses pixelation to correlate galaxy catalog information with \ac{GW} localization \citep{Gray:2023wgj}, we adopt the \texttt{healpix} scheme~\citep{Gorski:2004by,Zonca:2019vzt} with $\mathtt{nside} = 128$ for all catalogs, chosen so that a minimum of 25 pixels are able to cover the 99.9\% credible localization region of every event.

The first galaxy catalog used in this analysis is the $K_s$-band selection from GLADE+, exactly the same as used previously with GWTC-3.0 \citep{LIGOScientific:2021aug} and GWTC-4.0 \citep{LIGOScientific:2025jau}. GLADE+ \citep{Dalya:2018cnd, Dalya:2021ewn} is combined from six previous catalogs: the Gravitational Wave Galaxy Catalog (GWGC, \citealt{White:2011qf}), HyperLEDA \citep{Makarov:2014txa}, the 2 Micron All-Sky Survey Extended Source Catalog (2MASS XSC, \citealt{2MASS:2006qir}), the 2MASS Photometric Redshift Catalog (2MPZ, \citealt{Bilicki:2013sza}), the WISExSCOS Photometric Redshift Catalog (WISExSCOSPZ, \citealt{Bilicki:2016irk}) and the Sloan Digital Sky Survey quasar catalog from the 16th data release (SDSS-DR16Q, \citealt{eBOSS:2020jck}). Before masking, this dataset covers almost the entire sky except for the Milky Way area, where dust and stars obscure background galaxies.

The catalog includes all galaxies in GLADE+ with measurements in the near-infrared $K_\mathrm{s}$-band, numbering $\sim1$ million galaxies with median redshift $\langle z \rangle \sim 0.08$ and covering 37,201 $\mathrm{deg}^2$. About $77\%$ of the objects have photometric redshift estimates (photo-$z$s), originating mainly from 2MPZ \citep{Bilicki:2013sza} with a typical accuracy of $\sigma_z / (1+z) < 1.5\%$. The remaining $\sim 23\%$ of the objects have more precise spectroscopic redshifts (spec-$z$s) from other input datasets with errors of order $\sigma_z < 10^{-3}$. Objects with $z\leq 0.05$ were also corrected for peculiar velocities using the framework of \citet{Mukherjee:2019qmm}, relying on the Bayesian Origin Reconstruction from Galaxies (BORG) approach \citep{Jasche:2012kq}. We use a coarse $\mathtt{nside}=32$ to compute the magnitude threshold $m_\mathrm{thr}$ and normalization maps. Further details of all choices and methodology for \gladep can be found in Sec.\ 3.2 of both \citet{LIGOScientific:2021aug} and \citet{LIGOScientific:2025jau}.

\begin{table*}[t!]
    \centering
    \begin{tabular}{lcccccc}
    Galaxy Catalog & Sky Coverage & Num. of Galaxies & $z_\mathrm{med}$ & $\frac{\sigma_{z}}{1+z}_\mathrm{med}$ & $f_\mathrm{cov,all}$ & $f_\mathrm{cov,20}$ \\
    & $[\mathrm{deg}^2]$ & & & & & \\
    \hline\hline
    \gladep & 37201 & $9.9 \times 10^5$ & 0.08 & 0.014 & 5\% & 23\% \\
    \des & 5169 & $3.5 \times 10^8$ & 0.85 & 0.096 & 7\% & 17\% \\
    \hline
    \end{tabular}
    \caption{\label{tab:data_catalogs} Summary and comparison of the galaxy redshift catalogs used in this analysis after all relevant cuts have been applied. Here, $f_\mathrm{cov,all}$ and $f_\mathrm{cov,20}$ the percentage of localization volume of all \ac{GW} events and the twenty events with the smallest volume, respectively (see Eq.~\ref{eq:cat_fcov}).
    }
\end{table*}

\begin{figure*}[ht!]
    \centering
    \begin{minipage}[c]{0.56\textwidth}
        \includegraphics[width=\linewidth]{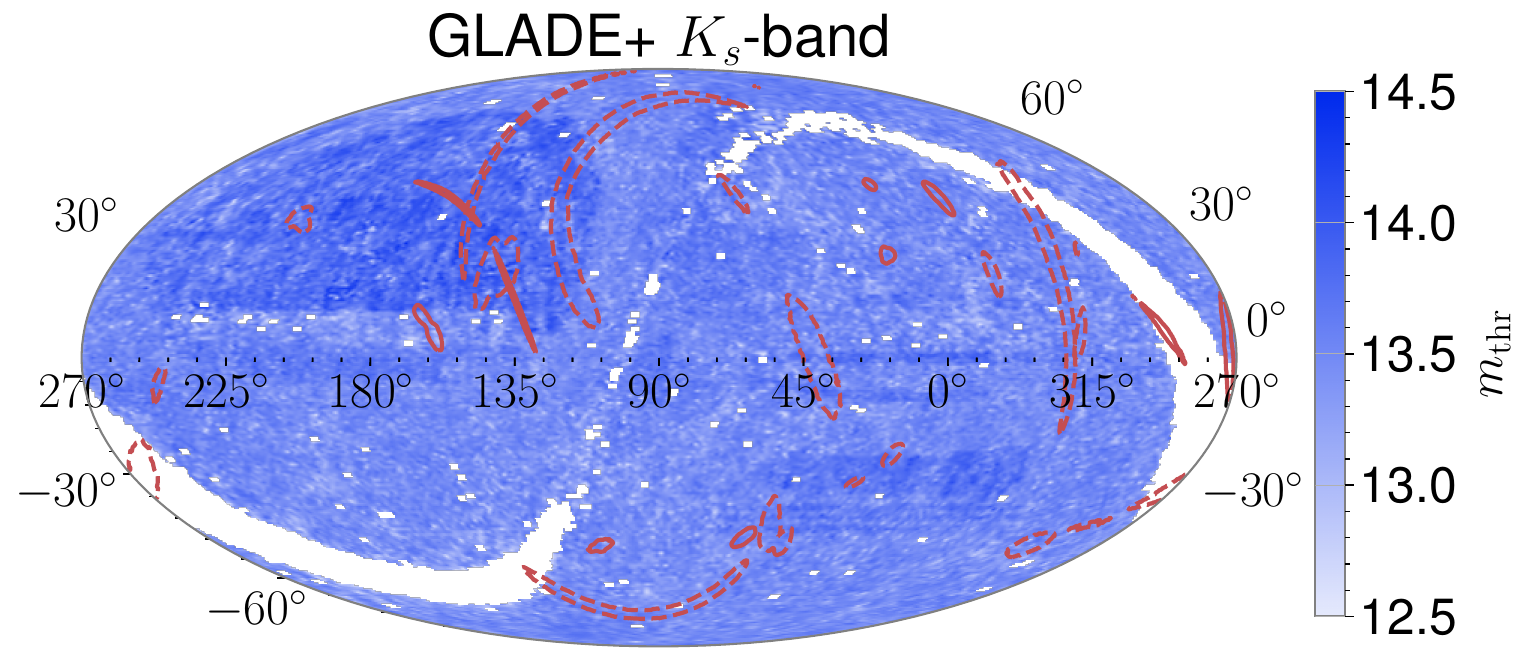}\\[0.2cm]
        \includegraphics[width=\linewidth]{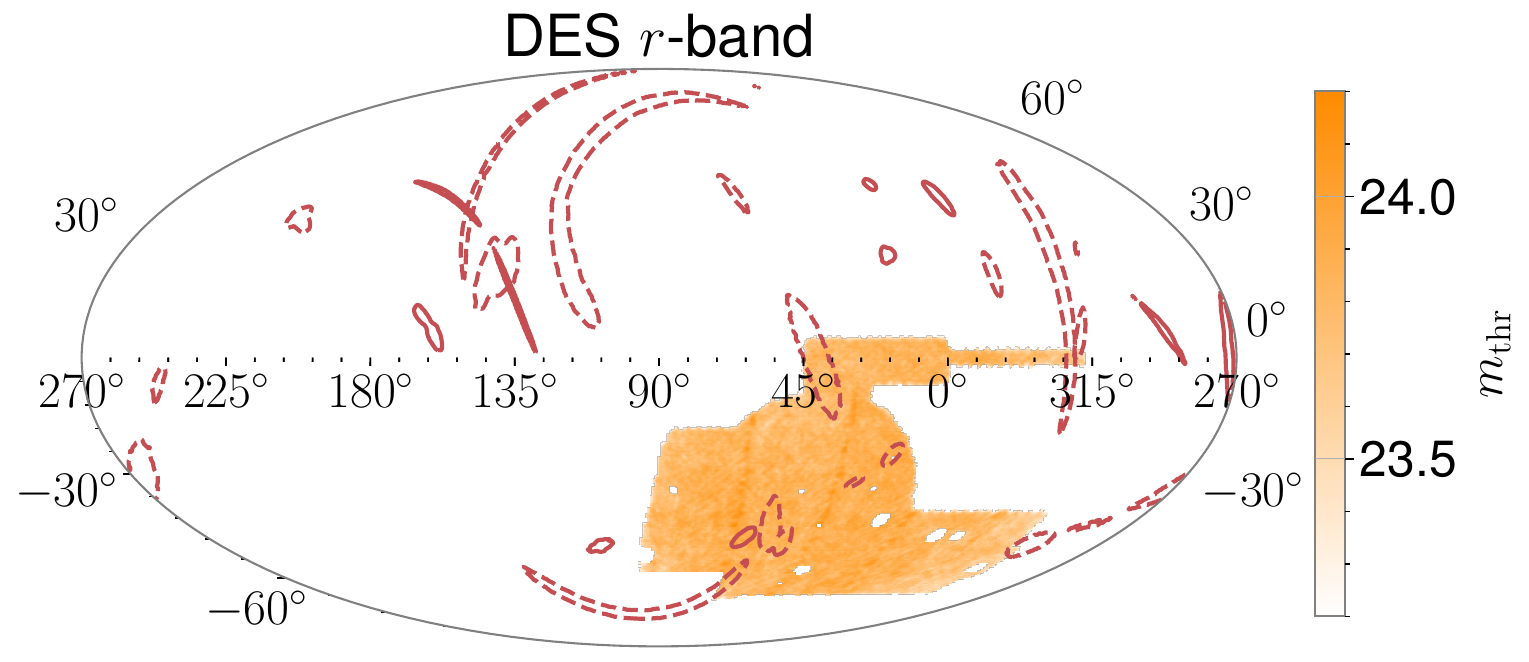}
    \end{minipage}%
    \hfill
    \begin{minipage}[c]{0.42\textwidth}
        \includegraphics[width=\linewidth]{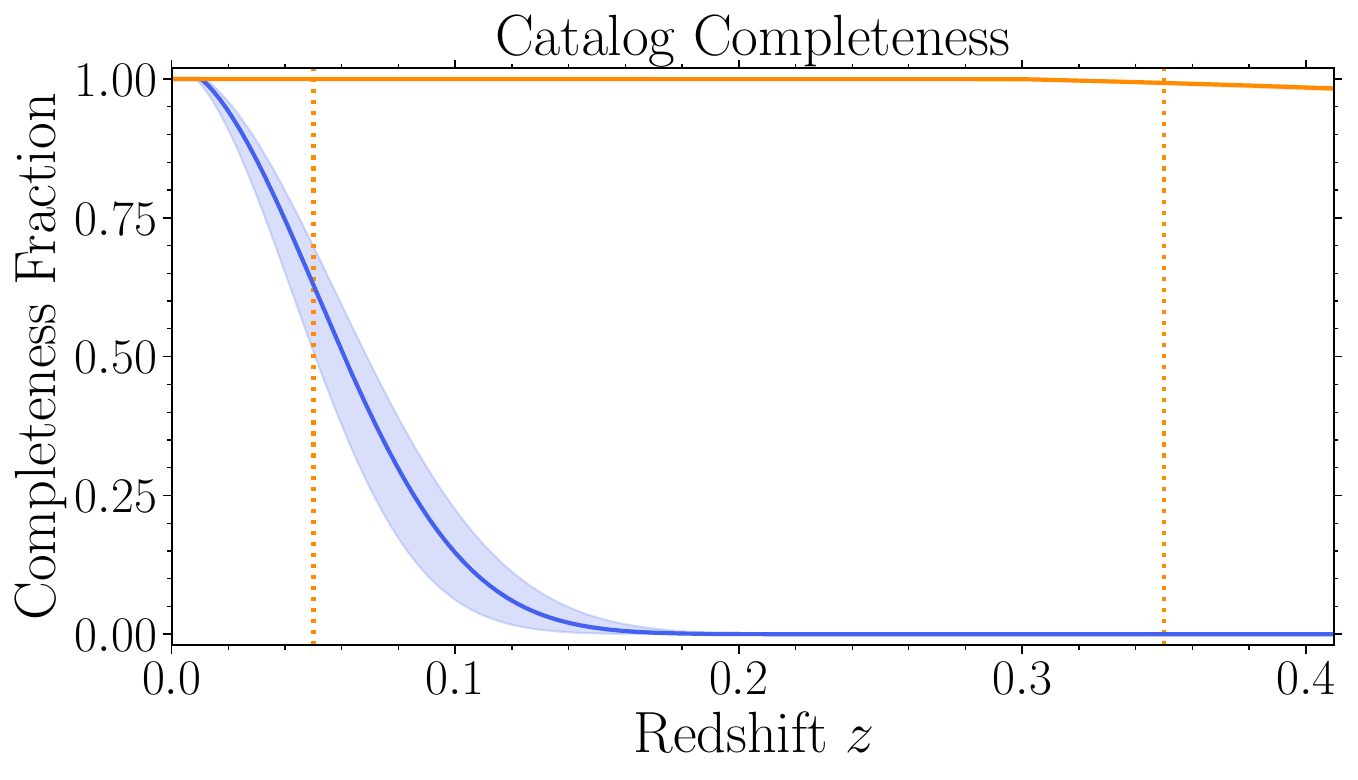}\\[0.2cm]
        \includegraphics[width=\linewidth]{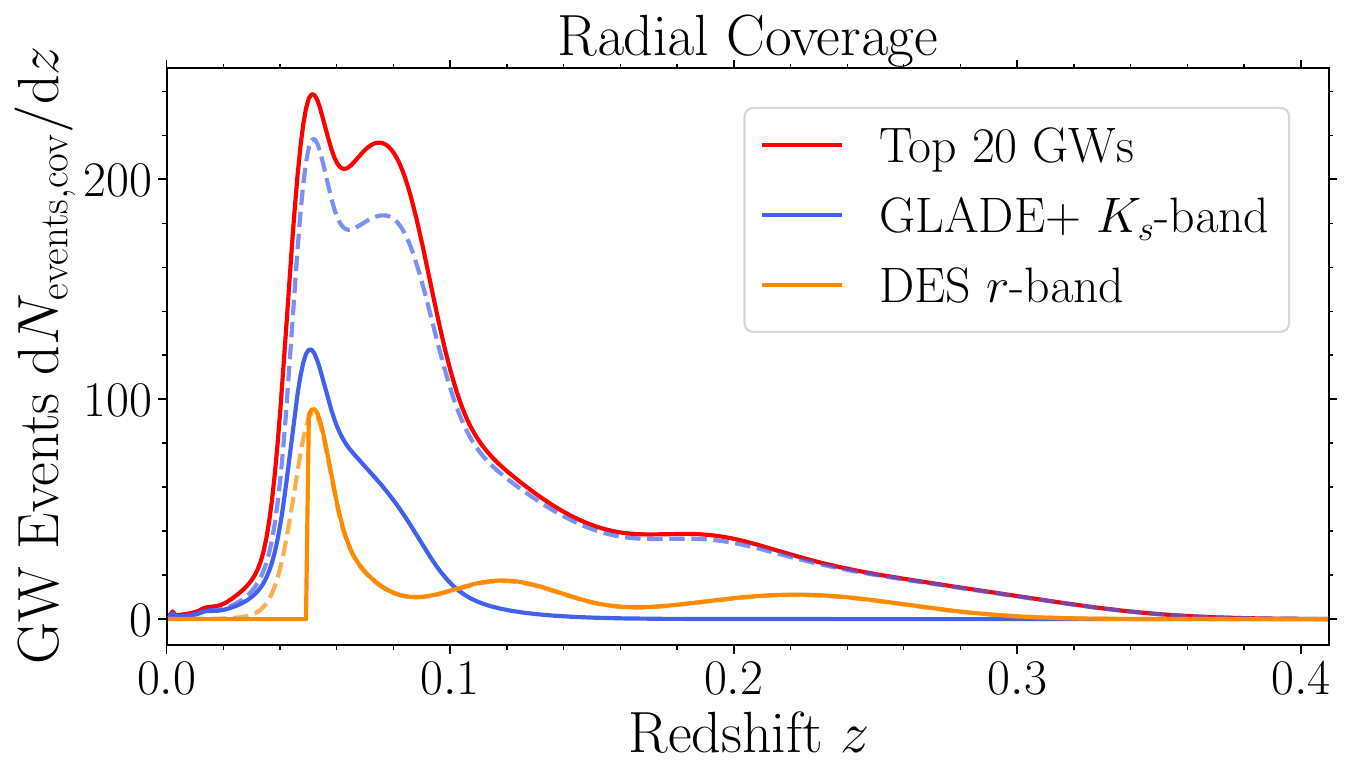}
    \end{minipage}
    \vspace{0.3cm}
    \caption{Left column: survey footprints of the \gladep (top) and \des (bottom) galaxy catalogs. 
    Contours of the 90\% credible areas for the twenty \ac{GW} events with the smallest localization volume are overlaid in red. O4b (pre-O4b) events are shown with solid (dashed) contours.
    Right column: (top) The completeness fraction as a function of redshift of each catalog as defined as the difference between a uniform in comoving volume distribution of galaxies and the out-of-catalog term defined in Section~\ref{subsec:redshift_prior}. 
    The completeness fraction computed with the median magnitude threshold $m_\mathrm{thr}$ over all pixels is shown with solid lines, with $1\sigma$ credible regions overlaid. 
    The redshift cuts of $z_\mathrm{min}=0.05$ and $z_\mathrm{max}=0.35$ on the in-catalog term used for \des are shown with dotted lines (\gladep does not cut on redshift).
    (Bottom) Radial coverage of both catalogs (blue and orange) with respect to the skymaps of the twenty \ac{GW} events with the smallest localization volume (red). The dashed lines correspond to the overlap of areas between the events and the catalogs, and the solid lines correspond to the overlap of volumes (see Appendix~\ref{sec:appendix_fractional_coverage}). 
    }
    \label{fig:galaxy_catalog_comparison}
\end{figure*}

The second catalog we employ is the Dark Energy Survey \citep[DES,][]{DES:2005dhi} Year 6 Gold catalog \citep{DES:2025key}, a cosmology-focused photometric survey of 4923 deg$^2$ in the southern sky, which gathered 669 million objects using the Dark Energy Camera (DECam) at the 4-m Blanco telescope at Cerro Tololo Inter-American Observatory in Chile. Typical photometric redshift errors for each galaxy are of the order $\sigma_z / (1+z) < 10\%$. We choose to use $r$-band photometry because the average photometric magnitude errors in this filter are smaller than in the other three bands ($giz$). The $10\sigma$ magnitude depth of the survey in $r$-band is 23.9. We use magnitudes which are corrected for Galactic extinction, apply photometric quality cuts, and use the XGBoost object classification score included in the catalog to mitigate stellar contamination \citep{Chen:2016btl}. The number of galaxies remaining after these quality cuts is 350 million. We use K-corrections from \citet{2010MNRAS.405.1409C}. Further details of all choices and methodology for using the DES Year 6 catalog can be found in \citet{McMahon:2026nhi}. We adopt the measured $r$-band Schechter parameters from that analysis ($M^*_r=-20.9$, $\alpha=-1.01$, $M_\mathrm{max}=-24.29$, $M_\mathrm{min}=-16.33$). We use a coarse $\mathtt{nside}=128$ to compute the $m_\mathrm{thr}$ and normalization maps.

As bright nearby objects saturate in the camera CCDs, these are not present in the catalog. For this reason, we apply a low-end redshift cut at $z_\mathrm{min}=0.05$ for the in-catalog part of the \des redshift prior (Eq. \ref{eq:total}). This cut also removes the need for peculiar velocity corrections at low redshift. At the large redshift end, we adopt a conservative cut $z_\mathrm{max}=0.35$ on the in-catalog part of the prior, as this is the highest value at which the galaxy redshift distribution is consistent with the comoving volume distribution assumption within less than $1\sigma$. The out-of-catalog term compensates for catalog incompleteness outside of the redshift cuts. For our faint-end Schechter absolute magnitude limit of $M_\mathrm{min}=-16.33$ and average apparent magnitude threshold $m_\mathrm{thr}=23.8$, \des is $99.9\%$ complete at $z=0.35$.

A summary of the characteristics of the galaxy catalogs is given in Table~\ref{tab:data_catalogs}. The left-hand side panels of Figure~\ref{fig:galaxy_catalog_comparison} present the catalog footprints, overlaid with the sky localizations of the twenty best-localized GWTC-5.0 events included in our analysis. These GW events have the smallest overlap with possible potential host galaxies and therefore are expected to be the most cosmologically informative. 
We propose a useful statistic, $f_\mathrm{cov}$, for the \ac{GW} coverage of each of the catalogs, defined as the percentage of the total localization volume of the \ac{GW} events that overlap with the three-dimensional completeness fraction of the catalog. Further details about the derivation of $f_\mathrm{cov}$ can be found in Appendix~\ref{sec:appendix_fractional_coverage}.

The redshift distribution of the twenty best-localized events and their overlap with the catalogs is shown in the right-hand side panel of Fig.~\ref{fig:galaxy_catalog_comparison}. 
While the majority of the \ac{GW} sample distribution is not covered by either catalog, the best-localized events are better covered by both catalogs compared to the rest of the events (see Table~\ref{tab:data_catalogs}).
While \gladep is wide and shallow in coverage and \des is comparatively narrow and deep, they both cover a similar fraction of the \ac{GW} distribution and therefore are expected to provide similar constraints on $H_0$.

\section{Results}\label{sec:results}
\noindent In this section, we present our cosmological results based on the spectral and dark siren analyses, which deliver joint inference of cosmological and population hyperparameters.
These include parameters that describe the assumed mass and spin distributions, and merger rate models, as well as different parametrizations for deviations of GR on cosmological scales. For the $H_0$ results, we will also combine our dark siren constraints with those from the bright siren GW170817, similarly to the analysis presented by \citet{LIGOScientific:2025jau}.

We sample the posterior in Equation~\eqref{eqn:sec2_marginal_lk} with the normalizing-flows-enhanced nested-sampling package \texttt{nessai}~\citep{Williams:2021qyt,nessai} via the \texttt{bilby} package~\citep{Ashton:2018jfp}.
For our fiducial dark and spectral siren analyses we present combined results from \icarogw and \gwcosmo as posterior distributions built from an equal-weighted mixture of samples (50\% from each pipeline), except for the analyses considering alternative spin parametrizations, and modified gravitational wave propagation results, which rely on \icarogw samples only.

Section~\ref{subsec:lcdm_results} focuses on the measurement of $H_0$ in a flat-$\Lambda$CDM model, obtained with either spectral sirens or dark sirens with galaxy catalog data from \gladep or \des, as well as exploring the impact of a number of modeling assumptions.
Section~\ref{MG results} presents constraints on modified \ac{GW} propagation. When quoting results, we report the median value plus its 68\% symmetric \ac{CI}. We use the relative decrease in average uncertainty, computed from the 68\% \ac{CI}, as a metric to measure the improvement of our results.

\begin{table*}[t]
\centering
\begin{ruledtabular}
\begin{tabular}{l c cc cc l c}
\hline

\multicolumn{6}{c}{\textbf{Mass Models \& Luminosity Weighting}} 
& \multicolumn{2}{c}{\textbf{Spin Models}} \\
\cline{1-6} \cline{7-8}

& 
& \multicolumn{2}{c}{\textbf{\gladep}} 
& \multicolumn{2}{c}{\textbf{\des}} 
& & \\

\cline{3-4} \cline{5-6}
\textbf{Mass Model}
& \textbf{Spectral} 
& $\epsilon=0$ & $\epsilon=1$ 
& $\epsilon=0$ & $\epsilon=1$ 
& \textbf{Spin Model} 
&  \\

\hline\hline

\fullpop 
& \BFHzeroMassFullPop 
& \BFHzeroDarkGladePlusUni
& \BFHzeroDarkGladePlusLum 
& \BFHzeroDarkDESUni
& \BFHzeroDarkDESLum 
& \textsc{Gaussian} 
& \BFHzeroSpinGaussian \\

\fullpopthreepeak 
& \BFHzeroMassFullPopBPLThreeP 
& \BFHzeroDarkGladePlusUnithreeP
& \BFHzeroDarkGladePlusLumthreeP
& \BFHzeroDarkDESUnithreeP
& \BFHzeroDarkDESLumthreeP 
& \textsc{Transition} 
& \BFHzeroSpinTransition \\

\hline
\end{tabular}
\end{ruledtabular}
\caption{Bayesian evidence comparison: logarithmic Bayes factors (BF) using base-10 logarithms for different mass and spin population models, as well as galaxy catalog choices.
The mass model and luminosity weighting comparison (left) is restricted to full CBC models only (i.e.~excluding the \ac{MLTP} mass model) and all values are relative to the \fullpop model with no catalog information (i.e.~spectral), while the spin model comparison (right) assumes a \ac{MLTP} spectral analysis, normalizing the spin model BFs to the \textsc{Gaussian} model.
In all cases a positive value indicates a preference for a given model over the reference model. The full CBC models used in the mass model and luminosity weighting columns implicitly assume a uniform spin distribution.
Overall for the mass and galaxy catalog comparison, the \fullpopthreepeak dark-siren analysis with \des assuming that host weighting is uniform with respect to luminosity is the most preferred, while for the spin model comparison the \textsc{Transition} model is preferred over the \textsc{Gaussian} one. }\label{tab:bayes_factors}
\end{table*}

\subsection{$\Lambda$CDM Cosmology}
\label{subsec:lcdm_results}
\noindent 
Table~\ref{tab:bayes_factors} reports the Bayes factors obtained by fitting the data with different population and galaxy catalog model assumptions, assuming a flat-$\Lambda$CDM cosmology.
These Bayes factors are computed using the evidence from each nested sampling process, which is a direct product of the hierarchical inference.
	
According to this table, the spectral siren analyses show a mild preference for the \fullpopthreepeak mass model over the \fullpop. Despite this, we choose to adopt the \preferredmassmodel as our fiducial model in the rest of our analyses, as this means our analyses are more directly comparable to those in \citet{LIGOScientific:2025jau}, which also used the \preferredmassmodel mass model as a fiducial choice. 
The inclusion of galaxy catalog information does not yield a significant preference for any catalog when using the \preferredmassmodel. In the case of the \fullpopthreepeak model, the inclusion of galaxy catalog information is mildly disfavored relative to the \fullpopthreepeak spectral analysis in all cases and with both catalogs, except for \des with $\epsilon=0$ luminosity weights (no-weighting), where it is mildly favored.
However, an important consideration for inclusion of galaxy catalog Bayes factors is that a number of choices, such as those detailed in Section~\ref{subsec:gal_cat}, are made in the construction of the galaxy catalog prior which may impact the overall Bayes factors. In effect, we are not just comparing the inclusion of galaxy catalog information against the spectral result, but also the specific choices made in the construction of the galaxy catalog prior.

When considering only BBHs - which we parametrize with the \ac{MLTP} mass model - we compare spin models. The preferred spin distribution is the \textsc{Transition} model, which is favored significantly over the \textsc{Gaussian} spin model. 
However, since the scope of the \textsc{Transition} and \textsc{Gaussian} models is restricted to BBHs, we do not use these spin models as fiducial models, as the fiducial analyses assume the \fullpop and \fullpopthreepeak mass models that probe the mass distribution of the full population of CBCs.

In the following, we choose as our fiducial model the \preferredmassmodel mass distribution, augmented by the \des galaxy catalog using $\epsilon=1$ luminosity weights to infer cosmological parameters. We justify this choice as no catalog is significantly favored or disfavored by the data, and we see a clear improvement in the $H_0$ constraint when including the \des catalog with luminosity weighting over the other catalogs.

\begin{figure*}[ht!]
    \centering
    \includegraphics[width=0.9\textwidth]{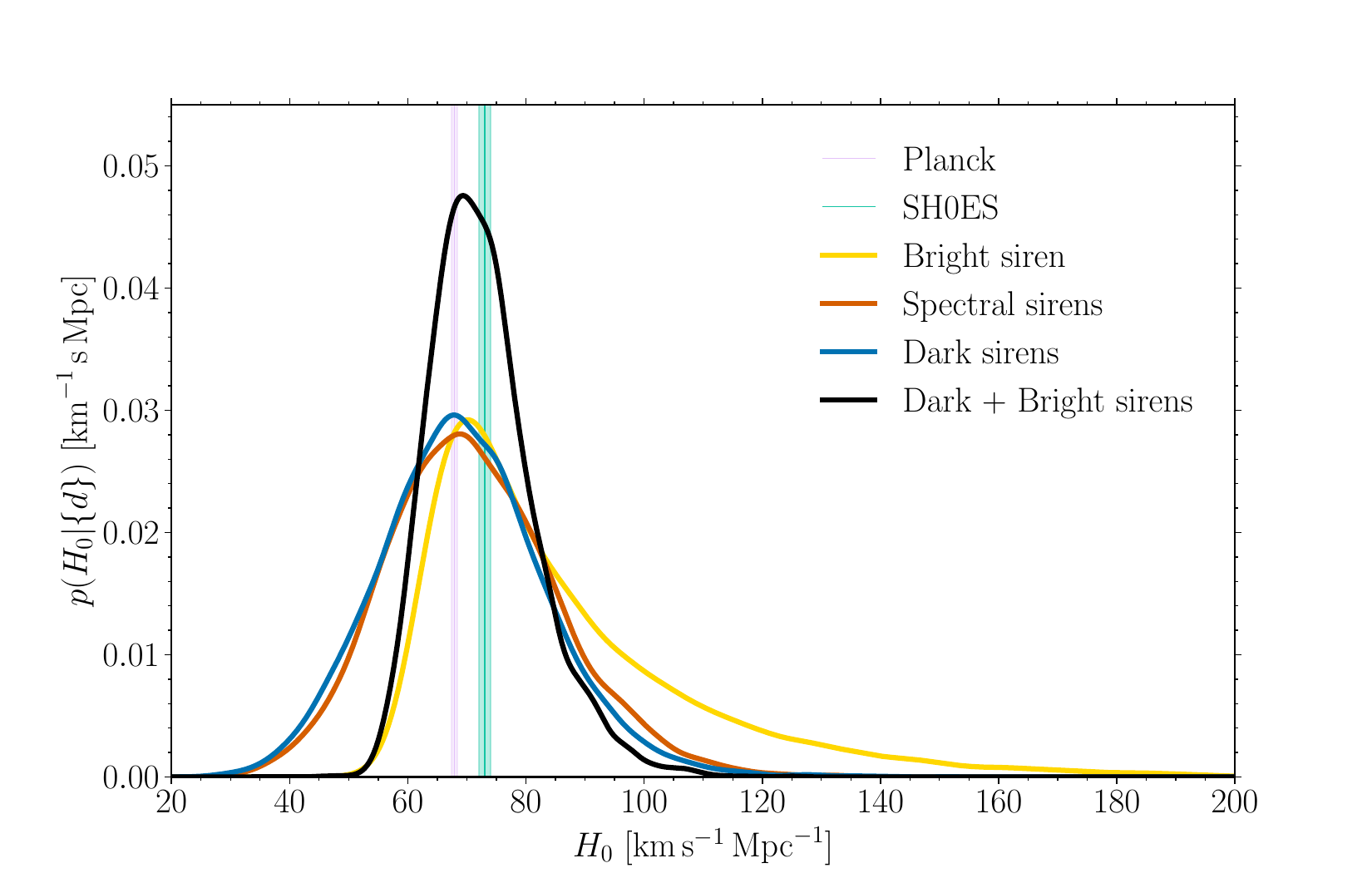}
    \caption{Hubble constant posterior for different cases. 
    Yellow curve: posterior obtained from the bright siren GW170817 and its \ac{EM} counterpart.
    Orange curve: posterior obtained with the spectral siren method and the \preferredmassmodel\ mass model.
    Blue curve: posterior obtained using all dark sirens with \des in the luminosity-weighting case ($\epsilon=1$) and spectral siren information from the \preferredmassmodel \ mass model.
    Black curve: posterior after combining the dark and bright siren results.
    The pink and green shaded areas identify the 68\% \ac{CI} constraints on $H_0$ inferred from CMB anisotropies \citep{Planck:2015fie} and in the local Universe from SH0ES \citep{Riess:2021jrx}, respectively. 
    Our fiducial result using both bright and dark sirens produced a constraint of $H_0 = \Hzerofiducial \, \Hunit$, while the spectral and dark analyses produced constraints of $H_0 = \HzeroEmptycatFullpopsixtygwtcfive \, \Hunit$ and $H_0 = \HzeroFullpopDarkDESepsOnesixty \, \Hunit$, respectively.}
    \label{fig:H0 combined catalog}
\end{figure*}

Figure~\ref{fig:H0 combined catalog} presents the marginalized posterior distributions of the Hubble constant for the different standard siren methodologies, assuming our fiducial model for spectral and dark siren analyses.
When combined with the posteriors of the bright siren result of GW170817, we find $ H_0 = \HzeroCatFullpopCombinedsixtygwtcfive  \, \Hunit$.
Spectral and dark siren constraints alone, i.e.~when not combined with GW170817, provide $H_0 = \HzeroEmptycatFullpopsixtygwtcfive \, \Hunit$ and $ H_0 = \HzeroCatFullpopsixtygwtcfive \, \Hunit$, respectively.
These constraints are now tighter than
the constraint obtained with the only confidently identified bright siren available, namely GW170817, which gives $H_0 = \Hzerobrightonlysixtygwtcfive \, \Hunit$ (Figure~\ref{fig:H0 combined catalog}, yellow curve). In particular, the dark sirens analysis excludes the tail of high $H_0$ values present in the bright siren result. 
We compute our GW170817 posterior analogously to \citet{LIGOScientific:2025jau}, by taking the inferred parameters of a spectral siren run with the fiducial \preferredmassmodel mass model and using these as fixed in order to perform a bright siren analysis.

With the current set of data, the main driver of the Hubble constant measurement remains the presence of features in the mass distribution.
We assess this by comparing the case where galaxy-catalog information is not included and constraints on $H_0$ are solely driven by our population assumptions, which corresponds to the spectral siren result (Figure~\ref{fig:H0 combined catalog}, orange curve), to our fiducial dark siren scenario (Figure~\ref{fig:H0 combined catalog}, blue curve).
From Figure~\ref{fig:H0 combined catalog} we see an improvement of \HzeroImprovementCatAndEmptyfiducialsixty{} in the $H_0$ constraint when including the DES $r$-band galaxy catalog information compared to the spectral siren result.
The two pipelines find different improvements: while \icarogw finds an improvement of \HzeroImprovementCatAndEmptyicarogwsixty{}, \gwcosmo finds an improvement of \HzeroImprovementCatAndEmptygwcosmosixty{} which is partly due to the different ways they treat the galaxy catalog completeness correction (as elaborated in Sec.~\ref{subsec:redshift_prior}).
The spectral siren analysis is further discussed in Appendix~\ref{sec:appendix_pop_results}. 

Figure~\ref{fig:h0_summary_plot} summarizes our findings alongside previous measurements from the \ac{LVK}. We adopt the same methodology used in the previous LVK measurement~\citep{LIGOScientific:2025jau}, fully marginalizing over the \ac{CBC} mass distribution and merger rate parameters. 
This approach is more statistically robust than assuming fixed population parameters, but complicates direct comparison with previous studies which constrain $H_0$ with standard sirens. 

\begin{figure*}
    \centering
    \includegraphics[width=0.95\textwidth]{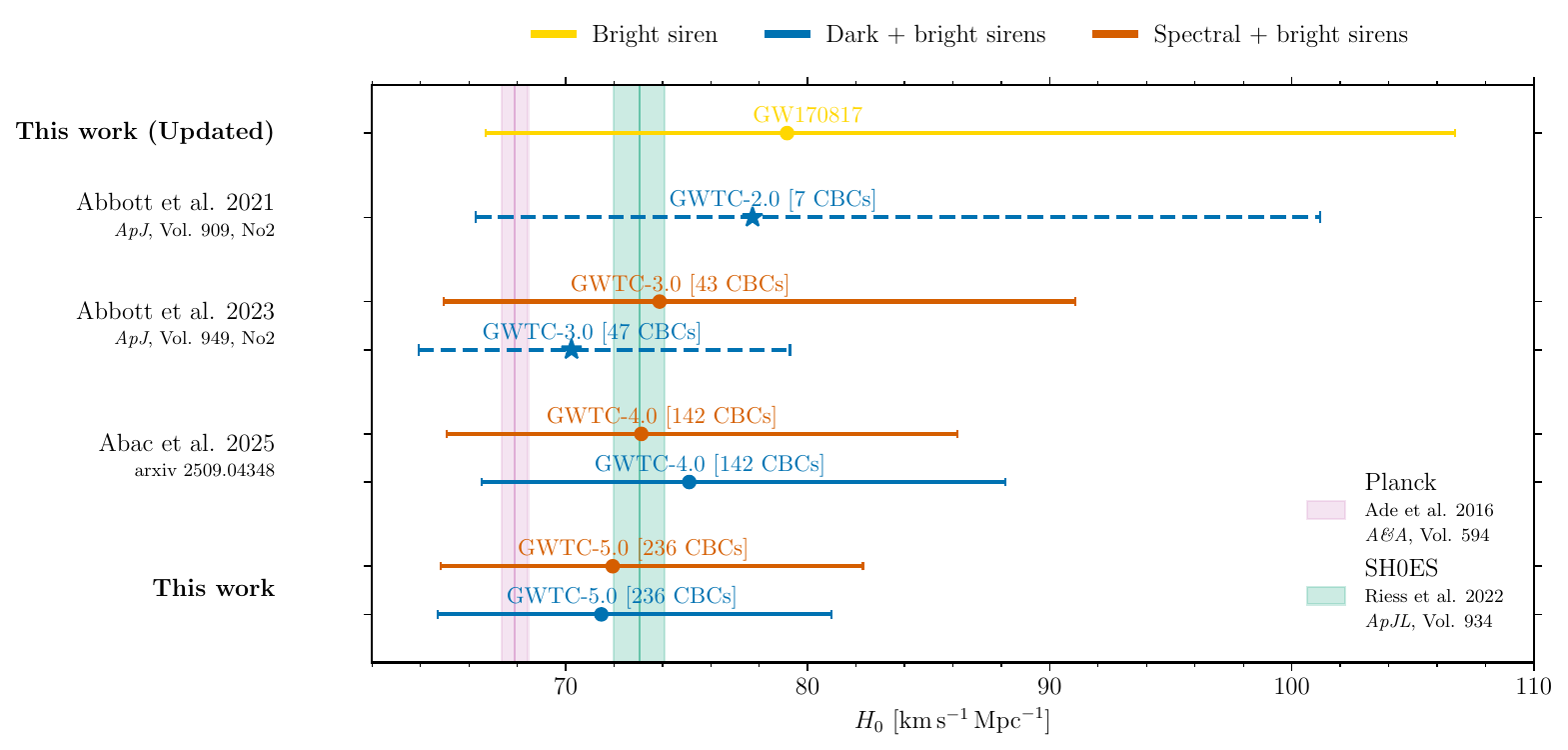}
    \caption{Summary of $H_0$ measurements from \ac{GW} detections, combining bright with dark or spectral siren analyses conducted by \ac{LVK} pipelines from \ac{O1} up to \ac{O4b}. In yellow, we report the bright siren result that was recalculated in \ac{O4a} and combined with the results from \ac{O4a} and this analysis.
    All other previous combined results use the bright siren samples from~\citet{LIGOScientific:2019zcs}. We report the dark siren results in blue and spectral siren results in orange, including the bright siren in both cases. The capped errorbar covers the symmetric 68.3\% \ac{CI}. Studies that assumed a fixed population model are marked with a dashed line style, and a star as a marker for the median value. 
    The total number of \ac{CBC} events used in the analysis is indicated in square brackets on top of each result. For details on the analysis settings, see the respective publications. The pink and green vertical bands indicate the Planck~\citep{Planck:2015fie} and SH0ES~\citep{Riess:2021jrx} median and $1\sigma$ values, respectively. The error bars obtained in this work are based on our fiducial mass model \preferredmassmodel.}
    \label{fig:h0_summary_plot}
\end{figure*}

\begin{table*}[t]
	\centering
	\begin{tabular}{lccc}
		\multicolumn{4}{c}{} \\
		\hline
		\multicolumn{4}{c}{$\Lambda$CDM -- Dark sirens} \\
		\hline
		Population model & \ac{GW} candidates & $H_0$ (Dark sirens) & $H_0$ (Dark + bright sirens) \\
		& & $[\Hunit]$  & $[\Hunit]$ \\
		\hline\hline
		\textsc{Multi Peak} & \NbrBBHgwtcfive & $\HzeroCatMLTPsixtygwtcfive$ $(\HzeroCatMLTPninetygwtcfive)$ & $\HzeroCatMLTPbrightsixtygwtcfive$ $(\HzeroCatMLTPbrightninetygwtcfive)$ \\
		\textsc{FullPop}-4.0 & \NbrCBCgwtcfive&  $\HzeroCatFullpopsixtygwtcfive$ $(\HzeroCatFullpopninetygwtcfive)$ & $\HzeroCatFullpopCombinedsixtygwtcfive$ $(\HzeroCatFullpopCombinedninetygwtcfive)$ \\
		\fullpopthreepeak & \NbrCBCgwtcfive &  $\HzeroBPLthreePdarksixty$ $(\HzeroBPLthreePdarkninety)$ & $\HzeroBPLthreePdarkbrightsixty$ ($\HzeroBPLthreePdarkbrightninety$) \\
		\hline
	\end{tabular}
	\caption{\label{tab:results_summary_LCDM}
	Constraints on the Hubble constant obtained in this work under the $\Lambda$CDM cosmological model, assuming a uniform prior $H_0 \in \mathrm{U}(10,200)\,\Hunit$. 
	The first column lists the mass model adopted in the analysis. 
	The second column gives the number of \ac{GW} events included in the dark-siren analysis. 
	The third column reports the $H_0$ measurement obtained using only dark sirens and the fiducial \des galaxy catalog with luminosity weighting, quoted as the median together with the symmetric 68.3\% and 90\% \acp{CI}, with the latter in parentheses. 
	The fourth column shows the corresponding constraints after combining the dark-siren analysis with the bright-siren posterior of GW170817.}
\end{table*}

\begin{figure*}[t]
	\centering
	\includegraphics[width=\textwidth]{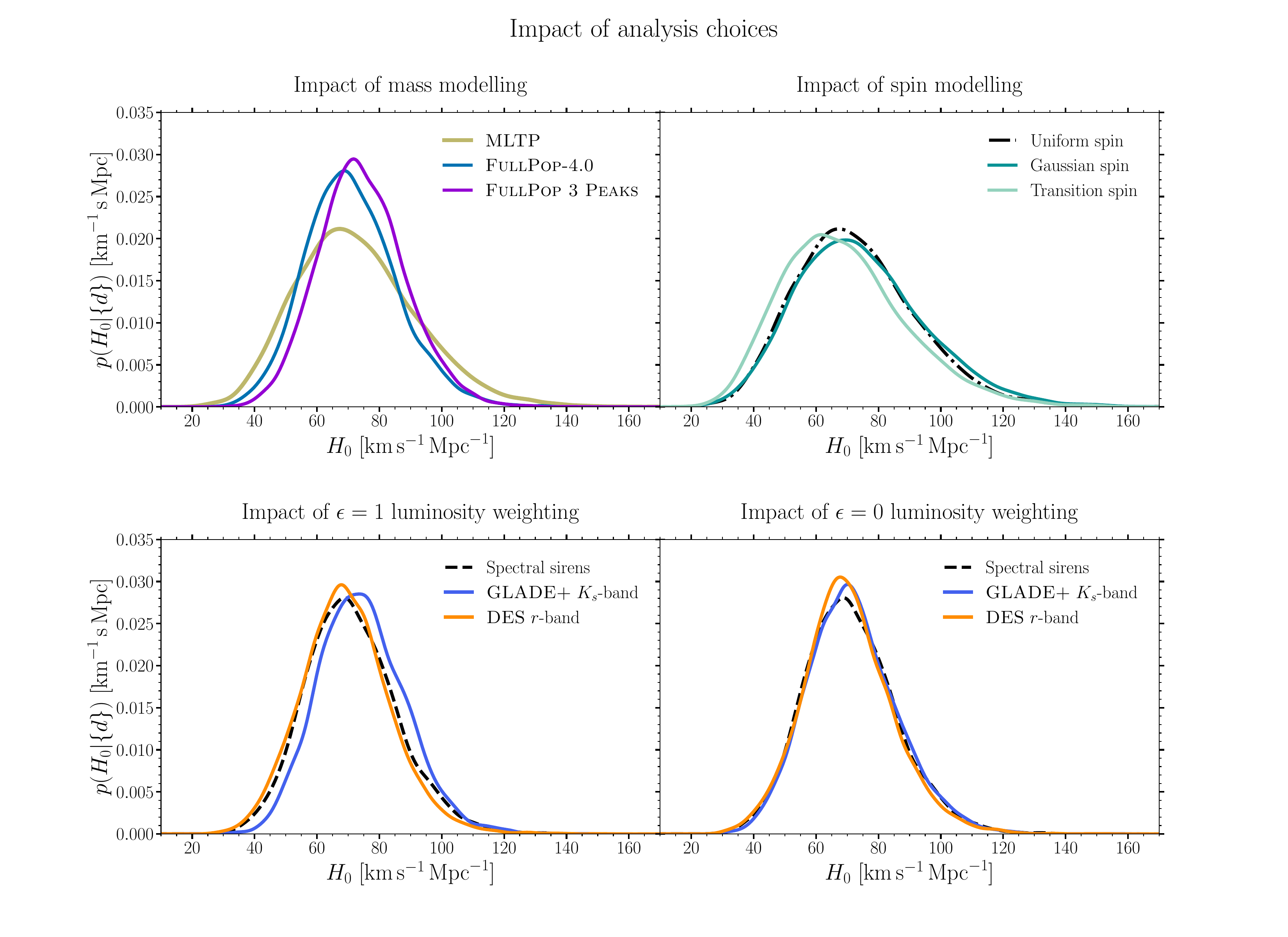} 
	\caption{Left top panel: Hubble constant posteriors with the spectral siren method, assuming three different mass models, one \acp{BBH}-only and two for all \acp{CBC}. See Section~\ref{subsubsec:population_models} and Appendix~\ref{sec:appendix_pop} for definitions of these models. 
		Right top panel: Hubble constant posteriors comparing the effects of two different assumed spin models, compared to a case where no spin information is taken into account. All analyses in this panel assume the \ac{MLTP} mass model and are \acp{BBH}-only. Left bottom panel: Hubble constant posteriors for dark siren analysis with two different galaxy catalogs, assuming that host weighting is proportional to luminosity. 
		The bottom two panels both assume the \fullpop mass model. 
		Right bottom panel: Hubble constant posteriors for dark siren analysis with different galaxy catalogs, assuming that host weighting is uniform with respect to luminosity.
	}
	\label{fig:H0 lumi weights and mass models}
\end{figure*}

In Table~\ref{tab:results_summary_LCDM} and Figure~\ref{fig:H0 lumi weights and mass models} we illustrate the impact of population models, both mass and spin distributions, and galaxy weighting on the marginalized posteriors of $H_0$. All curves in the top (bottom) panels of Figure~\ref{fig:H0 lumi weights and mass models} are from spectral (dark) siren analyses, while the event GW170817 is excluded from the dark and spectral siren inferences, as it is treated solely as a bright siren in this paper (this choice is validated and discussed in detail in Section~\ref{sec:perspective}).
The top-left panel of Figure~\ref{fig:H0 lumi weights and mass models} presents results based on the three different source mass models in the galaxy luminosity-weighting case: the \textsc{MLTP}, the \fullpop and the \fullpopthreepeak models.
The top-right panel shows the impact of spin modeling between an implicit uniform spin-magnitude prior, the \textsc{Gaussian} spin model and the \textsc{Transition} spin model.
The bottom panels explore the difference between the no-weighting (right) and luminosity-weighting (left) cases for the galaxy catalogs considered in our analysis, namely \gladep and \des (cf.~Section~\ref{subsec:gal_cat}), while keeping the source mass model fixed to our fiducial mass model \preferredmassmodel. 

From the top-left panel of Figure~\ref{fig:H0 lumi weights and mass models}, we find some differences in the measurements of $H_0$ due to assumptions about the shape of the mass spectrum.
The \textsc{MLTP} model provides a broader posterior with respect to the other two mass models as it cannot properly account for all the substructures of the mass function which appear preferred by the data (see Table~\ref{tab:bayes_factors}). 
The \fullpopthreepeak model instead shows a very mild preference for higher $H_0$ values.
Such a behavior is connected to the appearance of an additional reconstructed feature in the \fullpopthreepeak mass distribution with respect to the \fullpop one.
The Hubble constant is strongly degenerate with population parameters determining the position of sharp features in the mass function.
As a consequence, a shift of the inferred value of $H_0$ can lead to a shift of the value of some of these parameters, in particular the position of features and the low and high end cuts of the mass functions.
A mass model will compensate for any missing features by adjusting the other parameters of the model. If the inferred source frame distribution has a dearth of higher black hole masses, the analysis will compensate by adjusting the redshifts of the sources upwards, and hence shift the recovered $H_0$ to higher values.
As shown in the left panel of Figure~\ref{fig:mass spectra dark} in the Appendix, the \fullpopthreepeak shifts the inferred minimum and maximum masses of the mass function, thus affecting the retrieved value of $H_0$.

The top-right panel of Figure~\ref{fig:H0 lumi weights and mass models} shows that there are almost negligible differences on the inference of $H_0$ between the two spin models, with the \textsc{Gaussian} model producing a constraint of $H_0 = \HzeroMLTPspectralGaussianSpinsixty\,\Hunit$, while the \textsc{Transition} model prefers slightly lower values, $H_0 = \HzeroMLTPspectralTransitionSpinsixty\,\Hunit$.
However, the \textsc{Gaussian} model is strongly disfavored by the data when compared to the \textsc{Transition} model (see Table~\ref{tab:bayes_factors}). In contrast with our other analyses, the spin results are evaluated using a threshold on the log-likelihood variance of $<\,1$, rather than a cut on the effective number of posterior samples. This is a more robust threshold for analyses including spin parameters, as discussed in~\citet{Talbot:2023pex} and~\citet{LIGOScientific:2025jau}.

The bottom panels of Figure~\ref{fig:H0 lumi weights and mass models} show the effect of the choice of different luminosity weights---either $\epsilon = 0$ or $\epsilon = 1$ (see Equation~\eqref{eq:gal_weights}), which is demonstrated for both galaxy catalogs considered.
The two choices of luminosity weighting balance computational cost and avoid inaccuracies that may arise when large choices of $\epsilon$ cause a very small number of the most luminous galaxies to dominate enough to cause numerical precision issues in the evaluation of the likelihood.
Although fixing these weights introduces a potential systematic uncertainty~\citep{Perna:2024lod,Hanselman:2024hqy},
the results for the no-weighting and luminosity-weighting cases are in good agreement, with differences well within statistical error. 
The ability to constrain luminosity weights would be of astrophysical value, but we find no strong evidence based on Bayes factors to favor uniform weighting over luminosity-based weighting (see Table~\ref{tab:bayes_factors}). 
This outcome reflects the relatively limited impact of the galaxy catalogs on the inference with the datasets used here. 
The bottom panels of Figure~\ref{fig:H0 lumi weights and mass models} indicate that although small differences appear between the use of different galaxy catalogs, the resulting $H_0$ posteriors are still in agreement with one another.
We expect that these differences will become significant with larger datasets and better-localized events, in which case marginalizing over the weighting power-law index may be more robust.

The inclusion of \gladep information improves over the spectral siren constraints on the Hubble constant by approximately \HzeroImprovementGladePepsOneSixty{}, while the inclusion of \des information provides a \HzeroImprovementCatAndEmptyfiducialsixty{} improvement.

All constraints obtained in this section assume a fixed value of $\Omega_{\rm m}= 0.3065$ as well as a fixed dark energy equation-of-state parameter $w_0=-1$.
Inferring the values of these parameters independently is not possible at present using our dark siren methods, due to the computational cost of constructing redshift priors with varying $\Omega_{\rm m}$ and $w_0$.
However, in Appendix~\ref{app: additional systematics} we performed a spectral siren analysis with varying $\Omega_{\rm m}$ and $w_0$ and found that the posterior distributions of these parameters are consistent with the priors, due to the limited constraining power of our data at high redshift, while the uncertainties on other parameters of interest are only marginally affected. This confirms that allowing these parameters to vary does not influence our main results.

\subsection{Modified Gravity\label{MG results}}
\label{subsec:mg_results}

In this section we present the results obtained by introducing parameterized deviations from \ac{GR} that affect the luminosity distance ratio, $\DLGW/\DLEM$, as described in Section~\ref{subsubsec: gravity models}. These results are based on spectral siren analyses carried out using our fiducial mass model (\fullpop) without explicitly modeling the spin distribution, equivalent to a uniform prior on the spin magnitudes.

For each modified gravity (MG) model, we consider two different uniform priors for the Hubble constant: a wide prior, $H_0 \in \text{U}(10, 120)\ \Hunit$, and a narrow prior, $H_0 \in \text{U}(65, 77)\ \Hunit$.
This choice is motivated by the following considerations. In general, $H_0$ and any parameter governing modified \ac{GW} propagation are correlated to some extent, as both affect the luminosity distance--redshift relation. 
Consequently, the most agnostic approach to constraining deviations from GR involves marginalizing over $H_0$ using a broad enough prior---hence the adoption of the wider range. 
The broad prior adopted for $H_0$ in this section is narrower than the one used for the $\Lambda$CDM case. This is because, for certain extreme combinations of $H_0$ and $\Xi_0$, a wider prior on $H_0$ would lead to assigning very high redshifts---beyond $z \gtrsim 10$---to the \ac{GW} sources in the sample. Our redshift priors, by construction, do not cover these redshifts as we assume this case to be highly improbable. Furthermore, this would cause instability in our treatment of selection effects because at these very high redshifts the stability criterion for \ac{MC} integration could fail (see Appendix~A of \citet{LIGOScientific:2025jau}).
To avoid these issues, we restrict the $H_0$ prior accordingly. 
Conversely, it is also valuable to explore constraints on GR under the assumption of prior knowledge of other cosmological parameters, which motivates our second choice of a narrower prior which encompasses the region of the current Hubble tension at approximately $4\sigma$~\citep{Planck:2018vyg,Riess:2021jrx,DiValentino:2024yew}.

First, we discuss results for the MG parametrizations, see Equations~\eqref{eq: def Xi parametrization dl}, \eqref{eq:aM_param} and \eqref{eq:aM_ansatz}. The uniform priors used in the $\Xi_0$--$n$ analysis are $\Xi_0 \in \text{U}(0.435, 10)$ and $n \in \text{U}(0.1,10)$. Adopting a wide $H_0$-prior, we find $\Xi_0 = \MGxidarksixtygwtcfive \,$, while with the narrow $H_0$-prior we obtain $\Xi_0 = \MGxidarknarrowsixtygwtcfive \,$.
This result is consistent with \ac{GR}, recovered in the limit $\Xi_0=1$.

For the $\alpha_M$ parametrization we use the prior $c_M \in \text{U}(-2.95, 10)$. 
We find $c_M = \MGcMdarksixtygwtcfive\,$ with a wide $H_0$-prior, and $c_M = \MGcMdarknarrowsixtygwtcfive \,$ with a narrow $H_0$-prior.
This is also consistent with \ac{GR}, recovered in the limit $c_M=0$. In the case of both the wide and narrow $H_0$ priors, we find that additionally inferring MG parameters slightly worsens the constraints on $H_0$ with respect to the $\Lambda$CDM case, as expected due to the correlation between $H_0$ and MG parameters.

Both MG analyses display a strong correlation between the parameters describing deviations from \ac{GR} and the merger rate model parameter $\gamma$ (see Equation~\ref{eq:MDrate}), consistent with the findings of \citet{Mancarella:2021ecn,Leyde:2022orh,Chen:2023wpj}.
This correlation can be seen in Figure~\ref{fig:mg_cornerplots_simple} of Appendix~\ref{sec:appendix_MG}. The correlation occurs because the MG parameters modify the relationship between $\DLGW$ and $z$, therefore affecting the observability of \ac{GW} sources as a function of redshift. A similar change could be reproduced by adjusting the merger rate of \acp{CBC} as a function of redshift, which is what $\gamma$ controls, leading to degeneracy.

\begin{figure*}[t]
    \centering
    \includegraphics[width=0.8\textwidth]{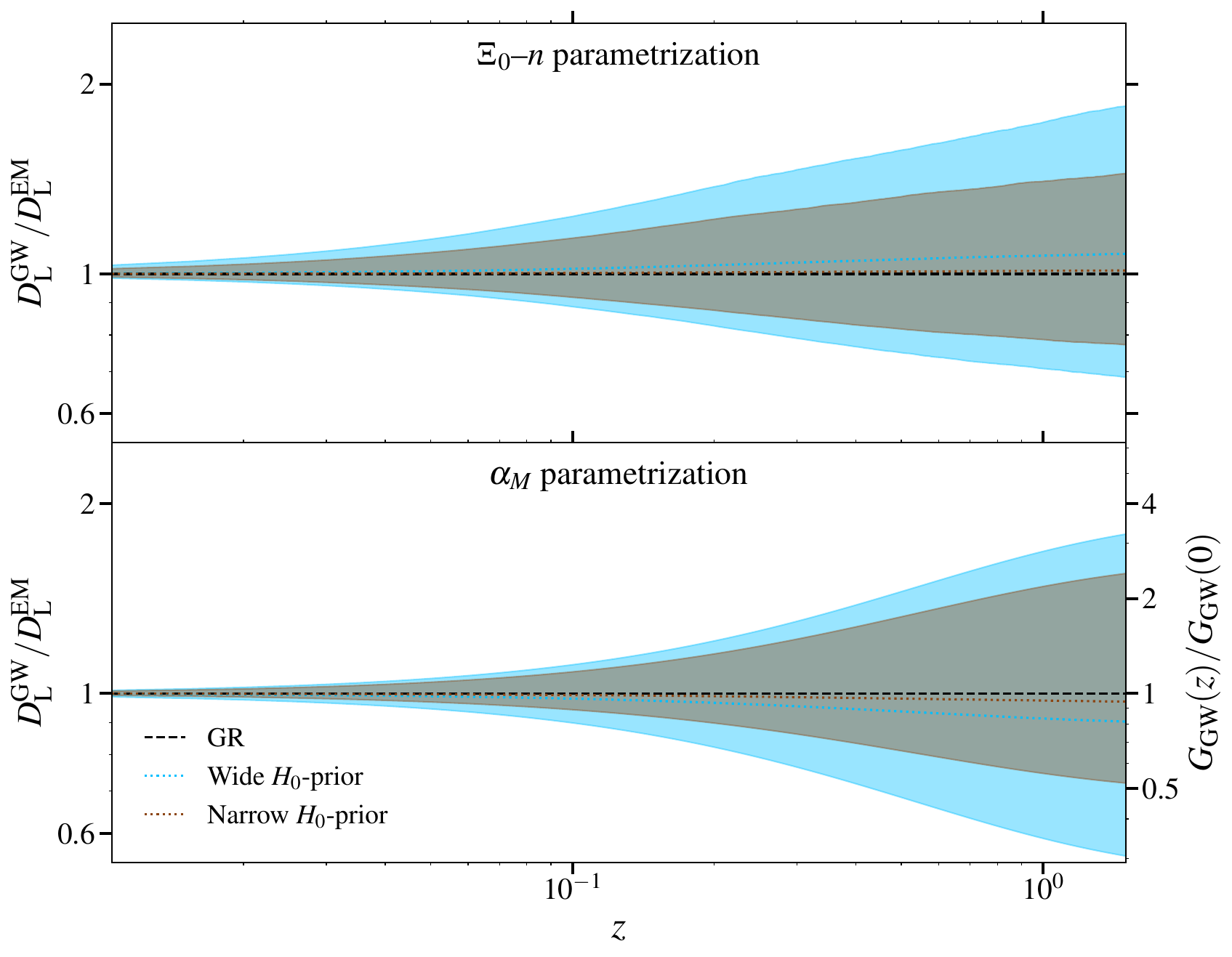}
    \caption{Reconstructed ratio $\DLGW/\DLEM$ as a function of cosmological redshift $z$, for the two modified gravity parametrizations considered, $\Xi_0$--$n$ and $\alpha_M$. 
	The blue (brown) bands indicate wide and narrow $H_0$-prior results, respectively (see Table~\ref{tab:results_summary_MGdark} for details).
	In all cases the contours show the 90\% \ac{CI} with median (dotted curve) reconstructed from spectral siren analyses with the \fullpop mass model. The black dashed curve represents the \ac{GR} limit. Note that the reconstructed distance ratio is asymmetric at higher redshifts. The additional righthand $y$-axis on the lower panel converts the constraints on the $\alpha_M$ parametrization into an effective gravitational coupling, which is discussed in section~\ref{subsubsec: gravity models} and Appendix~\ref{sec:appendix_MG}. This interpretation is not applicable to the $\Xi_0-n$ parametrization.
    }
    \label{fig:dL_z_PPC}
\end{figure*}

Figure~\ref{fig:dL_z_PPC} presents the reconstructed relation between redshift and \ac{GW} luminosity distance ratio $\DLGW / \DLEM$ for the two modified gravity models, obtained with our fiducial spectral siren analysis. The reconstructions for each analysis are derived from their respective posterior distributions provided in Appendix~\ref{sec:appendix_MG}. In \ac{GR}, the distance ratio plotted in Figure~\ref{fig:dL_z_PPC} is always one. 
The slight asymmetry of the contours around $D_{\rm L}^{\rm GW}/D_{\rm L}^{\rm EM}=1$ is inherited from the asymmetry of the marginalized posteriors on $\Xi_0$ and $c_M$ visible in Figure~\ref{fig:mg_cornerplots_simple}.
The additional right-hand $y$-axis for the $\alpha_M$ results enables their interpretation in terms of an effective gravitational coupling that can be derived from GW propagation (see Section~\ref{subsubsec: gravity models}). At low redshifts our results correspond to a few percent constraint on the ratio of this coupling at the source and the observer; the constraints inflate to order unity at $z\sim1$ due to the lack of events at higher redshifts.

\section{Discussion and perspectives}\label{sec:discussion}

\noindent In this section, we compare our results to the literature, and discuss possible improvements and future developments which constitute negligible systematics at present.

\subsection{Comparison with Existing Results}
\noindent We begin by discussing our constraints on the Hubble constant. 
All our results remain statistically consistent with the values reported by the Planck~\citep{Planck:2015fie,Planck:2018vyg} and SH0ES~\citep{Riess:2020fzl,Riess:2021jrx} collaborations at the 68\% \ac{CI}. Our new spectral siren measurement yields a \HzeroImprovementspectralsixtyfromgwtcfour{} improvement with respect to the equivalent analysis in~\citet{LIGOScientific:2025jau}, driven by the increased number of events. Including our best dark siren result from the galaxy catalogs, we find a \HzeroImprovementCombinedsixtyfromgwtcfour{} improvement from \ac{O4a} dark sirens when using the \des catalog. 
For comparison, the \gladep dark siren constraint shows an improvement of \HzeroImprovementCombinedsixtyfromgwtcfourgladep{} from \ac{O4a} dark sirens.

We also report constraints on \ac{GW} propagation parameters for GWTC-5.0 with the \fullpop mass model, which is comparable with the results from \citet{LIGOScientific:2025jau}.
Our new measurement yields a \ImprovementXizerowidesixty{} improvement in $\Xi_0$ over GWTC-4.0 with a wide $H_0$ prior, and a \ImprovementXizeronarrowsixty{} improvement in constraint with a narrow $H_0$ prior. We find a \ImprovementcMwidesixty{} improvement for $c_M$ compared to \ac{O4a} with a wide $H_0$ prior and a \ImprovementcMnarrowsixty{} improvement with a narrow $H_0$ prior. 
Previously, modified gravity parameters were also constrained using GWTC-2.0 \citep{Ezquiaga:2021ayr} and GWTC-3.0 \citep{Mancarella:2021ecn,Leyde:2022orh,Mastrogiovanni:2023emh,Chen:2023wpj}.

Assuming $\alpha_M$ is sourced by a scalar degree of freedom within the \ac{EFT} framework, and in the class of Horndeski-type theories, our constraints from \ac{GW} observations can be compared to those from \ac{LSS} and the \ac{CMB}. 
When analyzing \ac{LSS} and \ac{CMB} data, it is essential to ensure that the scalar sector remains free from ghost and gradient instabilities. These theoretical consistency requirements further restrict the allowed parameter space. Recent \ac{LSS} analyses that assume luminal tensor propagation~\citep{Noller:2018wyv,Baker:2020apq,Seraille:2024beb,Ishak:2024jhs} impose such constraints. In a \ac{GW}-only analysis, we assume that stability can be enforced by appropriate choices of additional \ac{EFT} operators---particularly the braiding parameter $\alpha_B$---which influence only the scalar sector. For theories with $\alpha_B = 0$, regions with $\alpha_M < 0$ are typically ruled out by stability arguments.

The latest available LSS bounds correspond to the clustering measurements from DESI 2024~\citep{Ishak:2024jhs}. This work finds the bound $c_M < 1.14$ (95\% \ac{CI}), assuming vanishing braiding and a $\Lambda$CDM background. Relaxing the braiding assumption and marginalizing over it yields a constraint of $c_M = 1.05 \pm 0.96$ at 68\% \ac{CI}.  
More stringent constraints can be obtained by combining different LSS observables. In particular, the \ac{ISW} effect from galaxy--\ac{CMB} cross-correlations has been shown to provide significant improvements~\citep{Renk:2017rzu,Seraille:2024beb}. Combining \ac{LSS} and \ac{CMB} observables to \ac{ISW},~\citet{Seraille:2024beb} find $c_M =0.54^{+0.90}_{-0.60}$ at 95\% \ac{CI} after marginalization over the braiding parameter. 
While consistent with these bounds, our best result is approximately ${\sim}6\%$ stronger and ${\sim}20\%$ weaker relative to the latter two, respectively.
Despite this, our results are based on an entirely independent dataset with different systematics. 

\subsection{Perspectives \label{sec:perspective}}
\paragraph{Considerations on the Spectral Siren Analysis}
Spectral siren information is dependent on the shape of the mass distribution of compact objects, and thus can be sensitive to choices of priors for the hyperparameters. We do not change our priors from previous studies \citep{LIGOScientific:2021aug,KAGRA:2021duu,LIGOScientific:2025jau}. Even if the analytical formulation of the underlying mass model is the same, significantly extending the prior range of the population parameters could alter the reconstructed mass spectra~\citep{Gennari:2025nho}. We do not consider this possibility here. 

The choice of mass model formulation and corresponding priors alters the resulting cosmological constraints. Using a mass model which includes all \ac{CBC}s improves the $H_0$ constraint due to the additional mass features in the \fullpop and \fullpopthreepeak models when compared to the \ac{BBH}-only \textsc{MLTP} model. From Table~\ref{tab:bayes_factors}, we find a slight preference for the \fullpopthreepeak model, however we continue to use the \fullpop model for our fiducial result, so our results remain comparable to those in~\citet{LIGOScientific:2025jau}.

We consider the impact of modeling mass--spin evolution \citep{GWTC:AstroDist,Biscoveanu:2022qac,Li:2023yyt,Li:2024rmi,Pierra:2024fbl,Tong:2025xvd}.
While this analysis is performed only for the \textsc{MLTP} model, we find that including spin information increases the $H_0$ uncertainty by \HzeroImprovementspectraMLTPGaussspinvsNospinsixtyAbs{} and \HzeroImprovementspectraMLTPTransspinvsNospinsixtyAbs{} for the \textsc{Gaussian} and \textsc{Transition} models, respectively, relative to the default spectral \textsc{MLTP} analysis.
Additionally, we find that the \textsc{Transition} model is significantly favored by Bayes factor ($\mathrm{log}_{10}\mathcal{B} = \BFHzeroSpinTransition$) relative to the \textsc{Gaussian} model, motivating the inclusion of more complex spin modeling in the future. 

Similarly, while current data do not robustly support evolution of the mass distribution with redshift~\citep{Heinzel:2024hva,Lalleman:2025xcs,Gennari:2025nho,GWTC:AstroDist}, considering this effect may become important as \ac{GW} detector sensitivity improves. Such evolution could introduce biases if not properly modeled~\citep{Pierra:2023deu,Agarwal:2024hld,Roy:2024oxh}. 
Nevertheless, because cosmological effects imprint a coherent and predictable modulation on the mass spectrum observed across different redshifts, it is expected that appropriate modeling should allow disentanglement of these from astrophysical evolution~\citep{Ezquiaga:2022zkx,Chen:2024gdn,Mali:2024wpq}. 
In future studies, it would be valuable to incorporate comprehensive correlation modeling or adopt data-driven approaches~\citep{Farah:2024xub}, which offer increased flexibility and robustness by reconstructing features directly from the observations, without strong parametric assumptions.

\paragraph{Combination with Bright Sirens} When combining dark siren events with bright sirens such as GW170817, particularly within a sample that includes both \ac{BNS} mergers with and without \ac{EM} counterparts, the correct approach would be to model the joint \ac{GW} and \ac{EM} detection probabilities and perform a unified hierarchical inference. At present, while our pipelines fully account for \ac{GW} selection effects, they do not yet model the \ac{EM} detection probability (potential systematics related to \ac{EM} selection effects, and related mitigation strategies, can be found in ~\citealt{Chen:2020dyt,Chen:2023dgw,Mancarella:2024qle,Muller:2024wzl,Salvarese:2024jpq}). Consequently, we exclude GW170817 from the dark siren inference and instead combine its posterior with that of the dark sirens a posteriori. \citet{LIGOScientific:2025jau} verified that this choice does not introduce any bias in the inferred mass spectrum. This effect will need to be included in expectation of more bright siren events.

\paragraph{Considerations on the Analysis with Galaxy Catalogs} Our reported results suggest that the information gained from galaxy catalogs is subdominant to spectral siren constraints. Many of the events which have significant overlap with either galaxy catalog contain $\mathcal{O}(10^{3-5})$ galaxies within their volume (See Table~\ref{tab:search_setup_parameter}). Improvements to the \ac{GW} detector sensitivity will improve the distance precision and future spectroscopic surveys will improve the galaxy redshift precision, both of which will have a large impact on the constraint from galaxy catalog dark sirens \citep{Borghi:2023opd,Cross-Parkin:2025xwf}.

Our galaxy catalogs have different optical filters and unique systematic considerations (see Section~\ref{subsec:gal_cat}). We model the expected number density with a redshift-independent Schechter function and assume that missing galaxies are uniformly distributed in comoving volume and isotropically in sky position. While the latter is the most conservative choice, viable alternative assumptions include having them trace the distribution of cataloged galaxies~\citep{Finke:2021aom} or follow prior knowledge of large-scale structure~\citep{Dalang:2023ehp,Leyde:2024tov,Dalang:2024gfk, Leyde:2025rzk}. With deeper catalogs such as \des, redshift evolution of the Schechter function may have an effect on the expected number of galaxies, but we do not model this effect in this analysis.

Furthermore, in this work, we model the uncertainty on galaxy redshift using a Gaussian distribution. However, this assumption likely represents an oversimplification, as photometric redshift error distributions can be more complex and even vary on a galaxy-by-galaxy basis. More comprehensive approaches, such as the use of full photo-\textit{z} PDFs, have been explored in the literature (e.g.,~\citealt{DES:2020nay,Bom:2024afj,Alfradique:2023giv,Alfradique:2026dsq}). 
Redshift uncertainties can propagate into derived quantities that depend on redshift, such as K-corrections and absolute magnitudes (or luminosities).~\citet{Turski:2023lxq} investigated two common error models (Gaussian and modified Lorentzian) and found that, under current levels of uncertainty, the choice of redshift error model does not significantly affect constraints on the Hubble constant. Nonetheless, this conclusion may not hold as future catalogs become more complete and systematic uncertainties are reduced, potentially making the choice of redshift uncertainty model more consequential.

Additionally, both catalogs present unique assumptions and challenges which have a large effect on the resulting measurement. \gladep is a combination of independent catalogs from different instruments and incongruent magnitude limits, mixing photometric and spectroscopic redshifts. The effect of combining catalogs in this way is not yet fully understood. Further issues can arise if the galaxy catalog deviates from the given Schechter function parametrization, which can be caused by residual stellar contamination, for example.

\des, while it avoids many of the above issues by consisting of a single photometric catalog, does show overdensities in the redshift distribution of galaxies which are persistent across the entire footprint. While some of these features can be explained by real astrophysical clustering and overdensities \citep{McMahon:2026nhi}, they may also originate from biased spectroscopic training data which may project spurious features across the entire galaxy distribution. This effect will be mitigated in the future with better photometric observations across more filters and wavelengths.

\paragraph{Considerations on Modified Gravity}

The considerations in the previous paragraphs apply also to modified-gravity analyses. 
The possible evolution of mass features with redshift could be potentially more impactful in this case, due to the redshift dependence of modified \ac{GW} propagation.

A specific point to address is the flexibility of parametrizations used here. Our parameterized forms in Equations~(\ref{eq: def Xi parametrization dl}) and (\ref{eq:aM_param}) enforce upon the distance ratio $\DLGW/\DLEM$ a limited range of redshift-dependent shapes. If these are a poor match for the behavior of an underlying modified gravity theory, constraints on $\Xi_0$ and $c_M$ may not sufficiently capture a mismodelled deviation from GR. 
There remains scope for further model-independent methods to agnostically constrain the distance ratio.

We note that in full Horndeski gravity, the function $\alpha_M$ can also be constrained through its effects on the \ac{CMB}. However, in this work we have fixed $\Omega_{\rm m}$ to a value inferred in a flat-$\Lambda$CDM analysis of Planck data~\citep{Planck:2015fie}. This may introduce a small bias in constraints on $c_M$, which we do not expect to be significant given the order-of-magnitude of the constraints obtained here. A fully correct approach would be to jointly analyze the Planck data alongside our \ac{GW} events, which is beyond the scope of the present work; a related discussion is found in \cite{Lagos:2019kds}.

\section{Conclusions}\label{sec:conclusion}

\noindent We have presented cosmological constraints obtained from the \gwtc-\thisgwtcversionfull{} catalog of \ac{GW} events identified by the \ac{LVK} detectors. 
Our headline results are updated bounds on the Hubble constant: $H_0 =\Hzerofiducial\, \Hunit$, combining \NbrCBCgwtcfive{} dark siren events with the bright siren GW170817. 
This fiducial result uses the \textsc{FullPop-4.0} mass model and redshift priors constructed from the DES Year 6 Gold galaxy catalog, with the GW host probability proportional to the galaxy luminosity. A summary of the different $H_0$ values obtained using various data sets and model assumptions can be found in Table~\ref{tab:results_summary_LCDM}.

The $H_0$ bounds obtained from our fiducial dark siren analysis are, for the first time, tighter than the single bright siren constraint ($ H_0=\HzeroFullpopDarkDESepsOnesixty\, \Hunit$ vs.\ $ H_0=\Hzerobrightonlysixtygwtcfive\, \Hunit$).  This represents a turning point in the chronology of dark sirens cosmology; it has taken $\sim{\cal O}(200)$ dark sirens to stand favorably beside one bright siren.
The change between the headline results of GWTC-\thisgwtcversionfull{} and GWTC-4.0 is \HzeroImprovementCombinedsixtyfromgwtcfour{} in terms of reduced uncertainty, though we highlight that these use different galaxy catalogs (\des and \gladep respectively). Note that dark siren constraints from GWTC-2.0 and GWTC-3.0 shown in Figure~\ref{fig:h0_summary_plot} are not marginalized over mass distribution or merger rate parameters, due to methodological restrictions affecting analyses prior to GWTC-4.0. The bounds indicated by a star in Figure~\ref{fig:h0_summary_plot} should be considered as artificially tight for this reason. It remains true that our dark siren analyses are dominated by their spectral siren components, with galaxy catalog information improving the constraint on $H_0$ by $\HzeroImprovementCatAndEmptyfiducialsixty$ in the fiducial case. This is to be compared to the improvement of $8\%$ reported in~\cite{LIGOScientific:2025jau}.

We have considered here a range of parameterized models for the mass distribution of compact objects. The tightest constraints on $H_0$ are obtained using the \fullpop and \fullpopthreepeak models, which enable the \ac{NS} and \ac{BH} distributions to be jointly analyzed. For the first time we have also considered the impact of different spin distributions on cosmological constraints. We find strong evidence favoring the \textsc{Transition} spin model over the \textsc{Gaussian} spin model, although the impact of spin modeling on $H_0$ is relatively modest at this time. In a similar vein, we investigate the choice of luminosity weight applied to the galaxy catalog. We find that weighting of $\epsilon=1$ significantly widens the variation in the $H_0$ posterior between the different galaxy catalogs. This is understandable physically, as luminosity weighting increases the effective catalog coverage of the event. This upweights the catalog's constraining power relative to the spectral sirens contribution. 

For the first time, this work has investigated how galaxy catalogs of different sky area, completeness and redshift error distribution impact cosmological constraints. 
Although the DES Year 6 Gold galaxy catalog has only a moderate footprint at $\sim\!5000$ deg$^2$, and hence overlaps with only a fraction of events ($\sim\!6\%$, see Table~\ref{tab:data_catalogs}), this is compensated for by its substantially higher completeness at redshifts up to $z\!\sim\!0.8$ than in the GLADE+ dataset (see Figure~\ref{fig:galaxy_catalog_comparison}). 
As a result, the constraints obtained on $H_0$ with \des are very similar to those using the $K_s$-band of the GLADE+ catalog, which is nearly full-sky but highly incomplete above $z=0.1$. 
This finding bodes well for future analyses with Stage IV survey data, which will have depths broadly similar to DES Year 6 Gold but over larger fractions of the sky. 
For the present, however, many events are only weakly informed by the distribution of potential galaxy hosts; instead, features in the mass distribution of \acp{CBC} dominate the constraints. 
The DES catalog was chosen for our fiducial analysis due to a lack of strong preference for any catalog against the \fullpop spectral analysis in our Bayesian model comparison (see Table~\ref{tab:bayes_factors}), however it does mildly improve the $H_0$ constraint when including it, relative to the spectral case.

In addition to placing constraints on the Hubble constant, we have presented bounds on parameterized deviations from \ac{GR} affecting the \ac{GW} luminosity distance. A summary of these constraints can be seen in Table~\ref{tab:results_summary_MGdark}. 
Using two commonly-employed parametrizations, we obtain the dark siren bounds of $\Xi_0 = \MGxidarksixtygwtcfive \,$, $n = \MGndarksixtygwtcfive \,$ and $c_M = \MGcMdarksixtygwtcfive \,$, where the \ac{GR} limit is recovered in the cases $\Xi_0=1$ and $c_M=0$, respectively.
Hence, our results show good consistency with GR on cosmological distance scales. The improvement in constraints on these parameters vary between ${\sim}\ImprovementcMnarrowsixty$ and ${\sim}\ImprovementXizeronarrowsixty$ (when using a narrow $H_0$ prior) relative to previous \ac{GW} analyses. 
This is because these constraints utilize higher-redshift \ac{GW} events and benefit from the ${\sim}1.7$-fold increase in \ac{GW} event number in GWTC-\thisgwtcversionfull{} over GWTC-4.0.
For the $c_M$ parameter, the constraints presented here using a narrow $H_0$ prior are comparable to those found via analyses of electromagnetic data, e.g. \cite{Ishak:2024jhs}.

\begin{table*}[t]
\centering

\begin{tabular}{lcc}
\hline
\multicolumn{3}{c}{Modified gravity -- Spectral sirens} \\
\hline
Parametrization $\Xi_0$--$n$ & $\Xi_0$ & $n$ \\
\hline\hline

Wide $H_0$-prior &
$\MGxidarksixtygwtcfive$ $(\MGxidarkninetygwtcfive)$ &
$\MGndarksixtygwtcfive$ $(\MGndarkninetygwtcfive)$
\\

Narrow $H_0$-prior &
$\MGxidarknarrowsixtygwtcfive$ $(\MGxidarknarrowninetygwtcfive)$ &
$\MGndarknarrowsixtygwtcfive$ $(\MGndarknarrowninetygwtcfive)$
\\

\hline
\end{tabular}

\vspace{0.2cm}

\begin{tabular}{lc}
\hline
Parametrization $\alpha_M$ & $c_M$ \\
\hline\hline

Wide $H_0$-prior &
$\MGcMdarksixtygwtcfive$ $(\MGcMdarkninetygwtcfive)$
\\

Narrow $H_0$-prior &
$\MGcMdarknarrowsixtygwtcfive$ $(\MGcMdarknarrowninetygwtcfive)$
\\

\hline
\end{tabular}
\caption{\label{tab:results_summary_MGdark}
Values of the modified-gravity parameters $\Xi_0$, $n$ and $c_M$ constrained assuming two different models of modified \ac{GW} propagation. All analyses are carried out assuming our fiducial population model \fullpop ($\NbrCBCgwtcfive$ \ac{GW} candidates) in a spectral siren analysis. The prior on $H_0$ is $H_0 \in \text{U}(10, 120)\, \Hunit$ in the wide case and $H_0 \in \text{U}(65, 77)\, \Hunit$ in the narrow case. We adopt uniform priors for $\Xi_0 \in \text{U}(0.435, 10)$ and $n \in \text{U}(0.1,10)$, and a uniform prior for $c_M \in \text{U}(-10, 50)$. 
Columns are: $H_0$ prior chosen for the analysis (first column), modified gravity parameter measurement reported as a median with 68.3\% (second column, first value) and 90\% (second column, second value) symmetric \ac{CI}. Note that, in contrast to Table~\ref{tab:results_summary_LCDM}, the bright siren GW170817 is not used as it is uninformative in this analysis. }
\end{table*}

The next few years will also see further data releases from Stage IV galaxy surveys such as the Dark Energy Spectroscopic Instrument~\citep{DESI:2016fyo}, Euclid~\citep{EUCLID:2011zbd,Euclid:2024yrr}, and the start of observations by the Vera Rubin Observatory~\citep{LSST:2008ijt}.
Using data from these EM campaigns is expected to strengthen the informativeness of the galaxy catalog component of the dark siren method. Forecasts using simulations of the 100 highest \ac{SNR} events in O5 with a spectroscopic galaxy catalog are presented in \cite{Borghi:2023opd,Borghi:2025pav}; these indicate bounds on $H_0$ of  $\sim2\%$ are achievable for catalogs with an average of $50\%$ completeness or higher inside the \ac{GW} horizon. Having this kind of galaxy data in hand will accelerate the progress towards competitive \ac{GW} constraints on the Hubble constant presented in Figure~\ref{fig:h0_summary_plot}.

Future runs of the LVK detectors presumably will yield further bright siren detections, although these are rare. Such an event would likely give \ac{GW} measurements of $H_0$ a rapid boost in constraining power. However, with or without such events, the methods and analyses of this paper demonstrate that dark and spectral sirens can provide steady progress towards the goals of precision \ac{GW} cosmology.

\emph{Data Availability:} 
All strain data analyzed as part of \gwtc[\thisgwtcversion] are publicly available through \ac{GWOSC}.
The details of this data release and information about the digital version of the \gwtc{} are described in detail in~\citet{OpenData}.
The data products generated by the methods described within this work are available from Zenodo~\citep{lvk_cosmo_data_release_gwtc5}.

\section*{Acknowledgements}
This material is based upon work supported by NSF's LIGO Laboratory, which is a
major facility fully funded by the National Science Foundation.
The authors also gratefully acknowledge the support of
the Science and Technology Facilities Council (STFC) of the
United Kingdom, the Max-Planck-Society (MPS), and the State of
Niedersachsen/Germany for support of the construction of Advanced LIGO 
and construction and operation of the GEO\,600 detector. 
Additional support for Advanced LIGO was provided by the Australian Research Council.
The authors gratefully acknowledge the Italian Istituto Nazionale di Fisica Nucleare (INFN),  
the French Centre National de la Recherche Scientifique (CNRS) and
the Netherlands Organization for Scientific Research (NWO)
for the construction and operation of the Virgo detector
and the creation and support  of the EGO consortium. 
The authors also gratefully acknowledge research support from these agencies as well as by 
the Council of Scientific and Industrial Research of India, 
the Department of Science and Technology, India,
the Science \& Engineering Research Board (SERB), India,
the Ministry of Human Resource Development, India,
the Spanish Agencia Estatal de Investigaci\'on (AEI),
the Spanish Ministerio de Ciencia, Innovaci\'on y Universidades,
the European Union NextGenerationEU/PRTR (PRTR-C17.I1),
the ICSC - CentroNazionale di Ricerca in High Performance Computing, Big Data
and Quantum Computing, funded by the European Union NextGenerationEU,
the Comunitat Auton\`oma de les Illes Balears through the Conselleria d'Educaci\'o i Universitats,
the Conselleria d'Innovaci\'o, Universitats, Ci\`encia i Societat Digital de la Generalitat Valenciana and
the CERCA Programme Generalitat de Catalunya, Spain,
the Polish National Agency for Academic Exchange,
the National Science Centre of Poland and the European Union - European Regional
Development Fund;
the Foundation for Polish Science (FNP),
the Polish Ministry of Science and Higher Education,
the Swiss National Science Foundation (SNSF),
the Russian Science Foundation,
the European Commission,
the European Social Funds (ESF),
the European Regional Development Funds (ERDF),
the Royal Society, 
the Scottish Funding Council, 
the Scottish Universities Physics Alliance, 
the Hungarian Scientific Research Fund (OTKA),
the French Lyon Institute of Origins (LIO),
the Belgian Fonds de la Recherche Scientifique (FRS-FNRS), 
Actions de Recherche Concert\'ees (ARC) and
Fonds Wetenschappelijk Onderzoek - Vlaanderen (FWO), Belgium,
the Paris \^{I}le-de-France Region, 
the National Research, Development and Innovation Office of Hungary (NKFIH), 
the National Research Foundation of Korea,
the Natural Sciences and Engineering Research Council of Canada (NSERC),
the Canadian Foundation for Innovation (CFI),
the Brazilian Ministry of Science, Technology, and Innovations,
the International Center for Theoretical Physics South American Institute for Fundamental Research (ICTP-SAIFR), 
the Research Grants Council of Hong Kong,
the National Natural Science Foundation of China (NSFC),
the Israel Science Foundation (ISF),
the US-Israel Binational Science Fund (BSF),
the Leverhulme Trust, 
the Research Corporation,
the National Science and Technology Council (NSTC), Taiwan,
the United States Department of Energy,
and
the Kavli Foundation.
The authors gratefully acknowledge the support of the NSF, STFC, INFN and CNRS for provision of computational resources.

This work was supported by MEXT,
the JSPS Leading-edge Research Infrastructure Program,
JSPS Grant-in-Aid for Specially Promoted Research 26000005,
JSPS Grant-in-Aid for Scientific Research on Innovative Areas 2402: 24103006,
24103005, and 2905: JP17H06358, JP17H06361 and JP17H06364,
JSPS Core-to-Core Program A.\ Advanced Research Networks,
JSPS Grants-in-Aid for Scientific Research (S) 17H06133 and 20H05639,
JSPS Grant-in-Aid for Transformative Research Areas (A) 20A203: JP20H05854,
the joint research program of the Institute for Cosmic Ray Research,
University of Tokyo,
the National Research Foundation (NRF),
the Computing Infrastructure Project of the Global Science experimental Data hub
Center (GSDC) at KISTI,
the Korea Astronomy and Space Science Institute (KASI),
the Ministry of Science and ICT (MSIT) in Korea,
Academia Sinica (AS),
the AS Grid Center (ASGC) and the National Science and Technology Council (NSTC)
in Taiwan under grants including the Science Vanguard Research Program,
the Advanced Technology Center (ATC) of NAOJ, 
the Mechanical Engineering Center of KEK
and Vietnam National Foundation for Science and Technology Development 
(NAFOSTED) 103.01-2025.147.

Additional acknowledgements for support of individual authors may be found in the following document: \\
\url{https://dcc.ligo.org/LIGO-M2300033/public}.
For the purpose of open access, the authors have applied a Creative Commons Attribution (CC BY)
license to any Author Accepted Manuscript version arising.
We request that citations to this article use 'A. G. Abac {\it et al.} (LIGO-Virgo-KAGRA Collaboration), ...' or similar phrasing, depending on journal convention.

\software{Calibration of the \ac{LIGO} strain data was performed with \GSTLAL{}-based
calibration software pipeline~\citep{Viets:2017yvy}.
Data-quality products and event-validation results were computed using the
\soft{DMT}{}~\citep{DMTdocumentation}, \soft{DQR}{}~\citep{DQRdocumentation},
\soft{DQSEGDB}{}~\citep{Fisher:2020pnr}, \soft{gwdetchar}{}~\citep{gwdetchar-software},
\soft{hveto}{}~\citep{Smith:2011an}, \soft{iDQ}{}~\citep{Essick:2020qpo},
\soft{Omicron}{}~\citep{Robinet:2020lbf} and
\soft{PythonVirgoTools}{}~\citep{pythonvirgotools} software packages and contributing
software tools.  Analyses in this catalog relied upon the \LALSUITE{} software
library~\citep{lalsuite, Wette:2020air}.  The detection of the signals and subsequent
significance evaluations in this catalog were performed with the
\GSTLAL{}-based inspiral software
pipeline~\citep{Messick:2016aqy,Sachdev:2019vvd,Hanna:2019ezx,Cannon:2020qnf},
with the \MBTA{} pipeline~\citep{Adams:2015ulm,Aubin:2020goo}, and with the
\PYCBC{}~\citep{Usman:2015kfa,Nitz:2017svb,Davies:2020tsx} and the
\CWB{}~\citep{Klimenko:2004qh,Klimenko:2011hz,Klimenko:2015ypf} packages.
Estimates of the noise spectra and glitch models were obtained using
\BAYESWAVE{}~\citep{Cornish:2014kda,Littenberg:2015kpb,Cornish:2020dwh}.
Source-parameter estimation was performed
with the \BILBY{} library~\citep{Ashton:2018jfp,Romero-Shaw:2020owr} using the
\DYNESTY{} nested sampling package~\citep{Speagle:2019ivv}. 
\PESUMMARY{} was used to postprocess and collate parameter-estimation
results~\citep{Hoy:2020vys}.  The various stages of the parameter-estimation
analysis were managed with the \ASIMOV{} library~\citep{Williams:2022pgn}.
Plots were prepared with \MATPLOTLIB{}~\citep{Hunter:2007ouj},
\SEABORN{}~\citep{Waskom:2021psk} and \GWPY{}~\citep{gwpy-software}.
\NUMPY{}~\citep{Harris:2020xlr} and \SCIPY{}~\citep{Virtanen:2019joe} were used
in the preparation of the manuscript.
We made use of the software packages \gwcosmo, see \url{https://git.ligo.org/lscsoft/gwcosmo} and \icarogw, see \url{https://github.com/simone-mastrogiovanni/icarogw}.}

\appendix
\numberwithin{equation}{section}

\section{Mass, merger rate and spin models}\label{sec:appendix_pop}
\noindent In this appendix, we describe the population models that we have considered in this paper, both in terms of mass and merger rate of \acp{CBC}. 
All the adopted population models are composed of various simple mathematical functions which are described in the GWTC-4.0 cosmology paper \citep{LIGOScientific:2025jau}.

The \ac{MLTP} model~\citep{LIGOScientific:2020kqk} is the direct extension of the \ac{PLP} model used in the GWTC-4.0 cosmology paper. The equations describing the primary and secondary mass distributions for this model are given in Appendix C of the GWTC-4.0 cosmology paper.
We report the parameter priors of the \ac{MLTP} model in Table~\ref{tab:priors_MLTP}.

The \fullpop model~\citep{LIGOScientific:2025pvj} spans the full mass distribution of \acp{CBC} and therefore includes \acp{BNS}, \acp{NSBH}, and \acp{BBH}.
It consists of a broken power-law continuum, Gaussian peaks, and smoothing at the edges of the distribution.
It additionally includes notch filters to allow for both lower and upper mass gaps~\citep{Ozel:2010su,Farr:2010tu,Fryer:2011cx,Belczynski:2012yt,Mali:2025sac}.
The depth of these mass gaps is a free parameter: the data can determine whether the rate goes to zero within the gap or if the gap is partially or totally filled.
This model is an extension of the \textsc{PowerLaw--Dip--Break} model described in \citet{Fishbach:2020ryj,Farah:2021qom}, and is the same as the \fullpop model described in \citet{LIGOScientific:2025pvj}.
The equations describing the primary and secondary mass distributions of \fullpop can be found also in Appendix C of the GWTC-4.0 cosmology paper.
The full set of parameter priors, descriptions and notations, are shown in Table~\ref{tab:priors_fullpop}.

Additionally, we include an extended version of the \fullpop model, named the \fullpopthreepeak, which incorporates an additional Gausssian peak (bringing the total number of peaks to three) to better capture potential features in the mass distribution. 
The primary and secondary mass distributions of the \fullpopthreepeak model are of the form
\begin{equation}
\label{eq: bpl3p mass distribution m1}
\begin{split}
p(m|\PEhyparameter) &= \Biggl[ (1-\lambda_{\rm g})\mathcal{B}(m|m_{\rm min},m_{\rm max},\alpha_1,\alpha_2,b) \\
&\quad + \lambda_{\rm g}\lambda_{\rm g}^{(1)} \mathcal{G}(m|\mu_{\rm g}^{(1)},\sigma_{\rm g}^{(1)},m_{\rm min},m_{\rm max}) \\
&\quad + \lambda_{\rm g}(1-\lambda_{\rm g}^{(1)})\lambda_{\rm g}^{(2)} \mathcal{G}(m|\mu_{\rm g}^{(2)},\sigma_{\rm g}^{(2)},m_{\rm min},m_{\rm max}) \\
&\quad + \lambda_{\rm g}(1-\lambda_{\rm g}^{(1)})(1-\lambda_{\rm g}^{(2)}) \mathcal{G}(m|\mu_{\rm g}^{(3)},\sigma_{\rm g}^{(3)},m_{\rm min},m_{\rm max}) \Biggr]\,,
\end{split}
\end{equation}
which also retains the smoothing functions at the edges of the distribution and the notch filter as in the \fullpop model.

The notation for the Gaussian components has been updated to reflect the presence of three peaks, with means $\mu_{\rm g}^{(1,2,3)}$ and standard deviations $\sigma_{\rm g}^{(1,2,3)}$. 
The fractions of events in each peak are governed by $\lambda_{\rm g}$, $\lambda_{\rm g}^{(1)}$, and $\lambda_{\rm g}^{(2)}$. The rest of the model components, including the broken power law $\mathcal{B}$ and the smoothing functions, remain unchanged from the \fullpop model.
In this construction, to have an equal prior weight on each peak, we employ Beta priors on $\lambda_{\rm g}^{(1)}$ and $\lambda_{\rm g}^{(2)}$ such that its equivalent to a symmetric Dirichlet prior. Table~\ref{tab:priors_bpl3p} summarizes hyperparameters and prior distributions for the \fullpopthreepeak model where they differ from the \fullpop model.

\begin{table}[t]
\centering
\begin{tabular}{ccc}
\multicolumn{3}{c}{} \\
\multicolumn{3}{c}{\textsc{Multi Peak}} \\
\hline
\textbf{Parameter} & \textbf{Description} & \textbf{Prior} \\
\hline
\hline
$\alpha$ &  Spectral index of primary-mass power law & $\text{U}(1.5,12)$\\ 
$\beta$& Spectral index of secondary-mass power law & $\text{U}(-4,12)$\\ 
$m_{\rm min}$&  Minimum primary mass [$M_\odot$]& $\text{U}(2,10)$\\ 
$m_{\rm max}$& Maximum primary mass [$M_\odot$]& $\text{U}(50,200)$\\ 
$\delta_{\rm m}$& Smoothing parameter [$M_\odot$]& $\text{U}(10^{-3},10)$\\ 
$\mu_{\rm g}^{\rm low}$& Location of the first peak [$M_\odot$]& $\text{U}(5,100)$\\ 
$\sigma_{\rm g}^{\rm low}$& Width of the first peak [$M_\odot$] & $\text{U}(0.4,10)$\\ 
$\mu_{\rm g}^{\rm high}$& Location of the second peak [$M_\odot$]& $\text{U}(5,100)$\\ 
$\sigma_{\rm g}^{\rm high}$ & Width of the second peak [$M_\odot$] & $\text{U}(0.4,15)$\\ 
$\lambda_{\rm g}$& Fraction of sources in the peaks & $\text{U}(0,1)$\\ 
$\lambda_{\rm g}^{\rm low}$& Fraction of sources in the first peak & $\text{U}(0,1)$\\
\hline
\end{tabular}
\caption{\label{tab:priors_MLTP}
 Summary of the hyperparameters priors used for the \ac{MLTP} model. \text{U} stands for uniform prior.}
\end{table}

\begin{table}[t]
\centering
\begin{tabular}{ccc}
\multicolumn{3}{c}{} \\
\multicolumn{3}{c}{\fullpop} \\
\hline
\textbf{Parameter} & \textbf{Description} & \textbf{Prior} \\
\hline
\hline
$\alpha_{1}$ & Spectral index of the power law before $b$ & $\text{U}(-4,12)$\\
$\alpha_{2}$ & Spectral index of the power law  after $b$ & $\text{U}(-4,12)$\\
$\beta_{1}$ & Spectral index of the pairing function before $m_{\rm break}$ & $\text{U}(-4,12)$\\
$\beta_{2}$ & Spectral index of the pairing function after $m_{\rm break}$ & $\text{U}(-4,12)$\\
$m_{\rm min}$&  Minimum primary and secondary mass [$M_\odot$]& $\text{U}(0.4,1.4)$\\ 
$m_{\rm max}$& Maximum primary and secondary mass [$M_\odot$]& $\text{U}(50,200)$\\ 
$\delta_{\rm m}^{\rm min}$&  1st smoothing parameter of the low mass [$M_\odot$]& $\rm \text{LU}(10^{-2},1)$\\ 
$\delta_{\rm m}^{\rm max}$& 2nd smoothing parameter of the low mass [$M_\odot$] & $\rm \text{LU}(10^{-3},10)$\\ 
$\mu_{\rm g}^{\rm low}$& Location of the first peak [$M_\odot$]& $\text{U}(5,150)$\\ 
$\sigma_{\rm g}^{\rm low}$& Width of the first peak [$M_\odot$] & $\text{U}(0.4,10)$\\ 
$\mu_{\rm g}^{\rm high}$& Location of the second peak [$M_\odot$] & $\text{U}(5,150)$\\ 
$\sigma_{\rm g}^{\rm high}$ & Width of the second peak [$M_\odot$] & $\text{U}(0.4,15)$\\
$\lambda_{\rm g}$& Fraction of sources in peaks & $\text{U}(0,1)$\\ 
$\lambda_{\rm g}^{\rm low}$& Fraction of sources in the first peak & $\text{U}(0,1)$\\ 
$m_{\rm d}^{\rm low}$ & Left side of the dip  [$M_\odot$]& $\text{U}(1.5,3)$\\
$m_{\rm d}^{\rm high}$ & Right side of the dip [$M_\odot$]& $\text{U}(5,9)$\\ 
$\delta_{\rm d}^{\rm min}$  & Smoothing of the left side of the dip [$M_\odot$]& $\rm \text{LU}(0.01,2)$\\ 
$\delta_{\rm d}^{\rm max}$  & Smoothing of the right side of the dip [$M_\odot$]& $\rm \text{LU}(0.01,2)$\\ 
$\rm A$  & Amplitude of the dip ($A=1$; completely empty gap)& $\text{U}(0,1)$\\ 
\hline
\end{tabular}
\caption{\label{tab:priors_fullpop}
 Summary of the hyperparameters priors used for the \fullpop mass model. \text{U} (\text{LU}) stands for uniform (log-uniform) prior.}
\end{table}

\begin{table}[t]
\centering
\begin{tabular}{ccc}
\multicolumn{3}{c}{} \\
\multicolumn{3}{c}{\fullpopthreepeak} \\
\hline
\textbf{Parameter} & \textbf{Description} & \textbf{Prior} \\
\hline
\hline
$\lambda_{\rm g}^{(1)}$ & Gaussian peak fraction one & $\text{Beta}(1,2)$ \\ 
$\lambda_{\rm g}^{(2)}$ & Gaussian peak fraction two & $\text{Beta}(1,1)$ \\ 
$\mu_{\rm g}^{(1)}$ & Location of the first peak [$M_\odot$] & $\text{U}(5,150)$ \\ 
$\sigma_{\rm g}^{(1)}$ & Width of the first peak [$M_\odot$] & $\text{U}(0.4,10)$ \\ 
$\mu_{\rm g}^{(2)}$ & Location of the second peak [$M_\odot$] & $\text{U}(5,150)$ \\ 
$\sigma_{\rm g}^{(2)}$ & Width of the second peak [$M_\odot$] & $\text{U}(0.4,15)$ \\ 
$\mu_{\rm g}^{(3)}$ & Location of the third peak [$M_\odot$] & $\text{U}(5,150)$ \\ 
$\sigma_{\rm g}^{(3)}$ & Width of the third peak [$M_\odot$] & $\text{U}(0.4,15)$ \\ 
\hline
\end{tabular}
\caption{\label{tab:priors_bpl3p}
 Summary of the new hyperparameters priors used in the \fullpopthreepeak model. \text{U} (\text{LU}) stands for uniform (log-uniform) prior.}
\end{table}

Moreover, we describe the merger rate evolution as a function of the redshift, modeled with a Madau--Dickinson parametrization~\citep{Madau:2014bja}, which is characterized by parameters $\{ \gamma, \kappa, z_{\mathrm{p}}\}$, where $\gamma$ and $\kappa$ are the power-law slopes respectively before and after the redshift turning point between the two power-law regimes, $z_{\mathrm{p}}$. Explicitly, 
\begin{equation}
    \psi \left(z | \gamma, \kappa, z_{\mathrm{p}}\right) = \left[1+\left(1+z_{\mathrm{p}}\right)^{-\gamma-\kappa}\right] \frac{(1+z)^\gamma}{1+\left[(1+z) /\left(1+z_{\mathrm{p}}\right)\right]^{\gamma+\kappa}} \, ,
    \label{eq:MDrate}
\end{equation}
with the parameter priors shown in Table~\ref{tab:rate_model_params}.
This parametrization is more complex than the one adopted in studies that focus solely on \ac{GW} population properties, where usually it takes the form of simple power\textendash laws, $\psi(z) \propto (1+z)^{\gamma}$~\citep{KAGRA:2021duu,GWTC:AstroDist}. 
Our choice is motivated by the fact that, when varying the cosmology, a \ac{GW} event at given distance can be associated with a redshift which is significantly higher than the one corresponding to the fiducial cosmology.  The model in Equation~\eqref{eq:MDrate} ensures that the merger rate decays after a peak at $z=z_{\mathrm{p}}$, consistently with astrophysical expectations.
The Madau–Dickinson distribution is typically used to describe the cosmic star formation rate, while the \ac{CBC} merger rate is then obtained by convolving this with a time-delay distribution. In practice, this is equivalent to using the same functional form with different values of $\gamma$ and $\kappa$, and by adopting wide priors on these parameters we effectively account for a broad range of possible delay times.

\begin{table*}[t]
\centering
\begin{tabular}{ccc}
\multicolumn{3}{c}{Merger rate model} \\
\toprule
\textbf{Parameter} & \textbf{Description} & \textbf{Prior} \\
\midrule\midrule
$\gamma$ & Slope of the power law before the point $z_{\rm p}$ & $\text{U}(0,12)$ \\
$\kappa$ & Slope of the power law after the point $z_{\mathrm{p}}$ & $\text{U}(0,6)$ \\
$z_{\rm p}$ & Redshift turning point between the power laws & $\text{U}(0,4)$\\ 
\bottomrule
\end{tabular}
\caption{\label{tab:rate_model_params}
 Summary of the hyperpriors used in the merger rate evolution model. \text{U}stands for uniform prior.}
\end{table*}

Finally, the hyperparameters prior ranges used for the \textsc{Gaussian} and \textsc{Transition} spin model, described in Section~\ref{subsubsec:population_models} are shown in Tables ~\ref{tab:gaussian_spin_params} and~\ref{tab:spin_model_params}.

\begin{table*}[ht]
\centering
\begin{tabular}{ccc}
\multicolumn{3}{c}{Gaussian Spin model} \\
\toprule
\textbf{Parameter} & \textbf{Description} & \textbf{Prior} \\
\midrule\midrule
$\mu_{\chi}$ & Mean of the spin magnitude distribution & U(0, 1) \\
$\sigma_{\chi}$ & Variance of the spin magnitude distribution & U(0.05, 5) \\
$\sigma_t$ & Width of the Gaussian component of the $\cos \theta$ distribution & U(0.1, 0.5) \\
$\zeta_{\text{spin}}$ & Mixing fraction for the spin components & U(0, 1) \\ 
\bottomrule
\end{tabular}
\caption{\label{tab:gaussian_spin_params}
 Summary of the hyperpriors used in the Gaussian spin model.}
\end{table*}

\begin{table*}[t]
\centering
\begin{tabular}{ccc}
\multicolumn{3}{c}{Spin Transition model} \\
\toprule
\textbf{Parameter} & \textbf{Description} & \textbf{Prior} \\
\midrule\midrule
$\mu_{\chi_{1}}$ & Mean of the spin magnitude distribution before the transition. & $\text{U}(0,1)$ \\
$\mu_{\chi_{2}}$ & Mean of the spin magnitude distribution after the transition. & $\text{U}(0,1)$ \\
$\sigma_{\chi_{1}}$ & Variance of the spin magnitude distribution before the transition. & $\text{U}(0.05,5)$ \\
$\sigma_{\chi_{2}}$ & Variance of the spin magnitude distribution after the transition. & $\text{U}(0.05,5)$ \\
$m_{\rm t}$ & Transition mass between the first and second spin distribution [$M_\odot$]. & $\text{U}(10,120)$ \\
$\delta_{m_{\rm t}}$ & Steepness of the spin transition [$M_\odot$]. & $\text{U}(1,25)$ \\
$\lambda_{\rm f}$ & Mixing fraction at between first and second spin distributions at mass=0.  & $\text{U}(0.8,1)$ \\
\bottomrule
\end{tabular}
\caption{\label{tab:spin_model_params}
 Summary of the hyperpriors used in the spin transition model.}
\end{table*}

\section{Catalog diagnostic derivation}\label{sec:appendix_fractional_coverage}

\noindent
We propose a useful diagnostic statistic for galaxy catalogs $f_\mathrm{cov}$, defined as the percentage of the total localization volume of $N_\mathrm{events}$ \ac{GW} events that overlap with the three-dimensional completeness fraction of the catalog:

\begin{equation}\label{eq:cat_fcov}
    f_\mathrm{cov} = N_\mathrm{events}^{-1} \int \mathrm{d}D_L\,\frac{\mathrm{d}N_\mathrm{events,cov}}{\mathrm{d}D_L}.
\end{equation}
The number of \ac{GW} events, denoted by $\mathrm{d}N_\mathrm{events,cov}/\mathrm{d}D_L$, covered by a galaxy catalog per luminosity distance is given by

\begin{equation}\label{eq:cat_cov}
    \frac{\mathrm{d}N_\mathrm{events,cov}}{\mathrm{d}D_L} = \sum^{N_\mathrm{events}}_{j=1} \sum_{i=1}^{N_\mathrm{pix}} \;p_\mathrm{cat}(D_L,\,\Omega_{i,j}) \;p_\mathrm{pix}(\Omega_{i,j}) \;\mathcal{N}(D_L,\,\mu(\Omega_{i,j}),\,\sigma(\Omega_{i,j})) \,,
\end{equation}
where $N_\mathrm{events}$ is the number of events, $N_\mathrm{pix}$ is the number of pixels in the event skymap, and $p_\mathrm{cat}$ is the fraction of galaxies which are observable at the threshold magnitude of a given pixel, as a function of distance. $\mathcal{N}$ is a normalized Gaussian function of the mean $\mu$ and standard deviation $\sigma$ distance given in each pixel of an event's skymap with sky location $\Omega$, and $p_\mathrm{pix}$ is the localization probability assigned to that pixel. 
The variable $p_\mathrm{cat}$ is an adaptation of the out-of-catalog term in Eq.~\ref{eq:total}:

\begin{equation}\label{eq:pcat}
    p_\mathrm{cat}(D_L,\,\Omega_{i,j}) = 1 - \Big( \int_{x_\mathrm{thr}(D_L,\,\Omega_{i,j})}^{x_\mathrm{max}} \mathrm{d}x \;x^{\alpha+\epsilon} \;e^{-x} \Big) / \Big( \int_{x_\mathrm{min}}^{x_\mathrm{max}} \mathrm{d}x \;x^{\alpha+\epsilon} \;e^{-x} \Big) \,,
\end{equation}
where

\begin{equation}\label{eq:pcat_xthr}
    x_\mathrm{thr}(D_L,\,\Omega_{i,j}) = 10^{0.4[M^* - M_\mathrm{thr}(D_L,\,m_\mathrm{thr}(\Omega_{i,j}))]},
\end{equation}
and
\begin{equation}\label{eq:pcat_xmax_xmin}
    x_\mathrm{max} = 10^{0.4(M^* - M_\mathrm{max})}\,, \quad\quad
    x_\mathrm{min} = 10^{0.4(M^* - M_\mathrm{min})} \,.
\end{equation}
The Schechter parameters $\alpha$, $M^*$, $M_\mathrm{min}$, and $M_\mathrm{max}$ are identical to parameters used when constructing the line-of-sight priors for each catalog. We choose the luminosity weighting factor $\epsilon=1$ for diagnostic purposes. $M_\mathrm{thr}$ is the absolute magnitude at a given distance and apparent magnitude threshold for each pixel $m_\mathrm{thr}$. Further details about the derivation of $p_\mathrm{cat}$ can be found in Appendix B of \citet{LIGOScientific:2025jau}.

\section{Joint population and cosmological inference details}\label{sec:appendix_pop_cosmo_inference}

\begin{figure*}[t]
	\centering
	\includegraphics[width=0.47\textwidth]{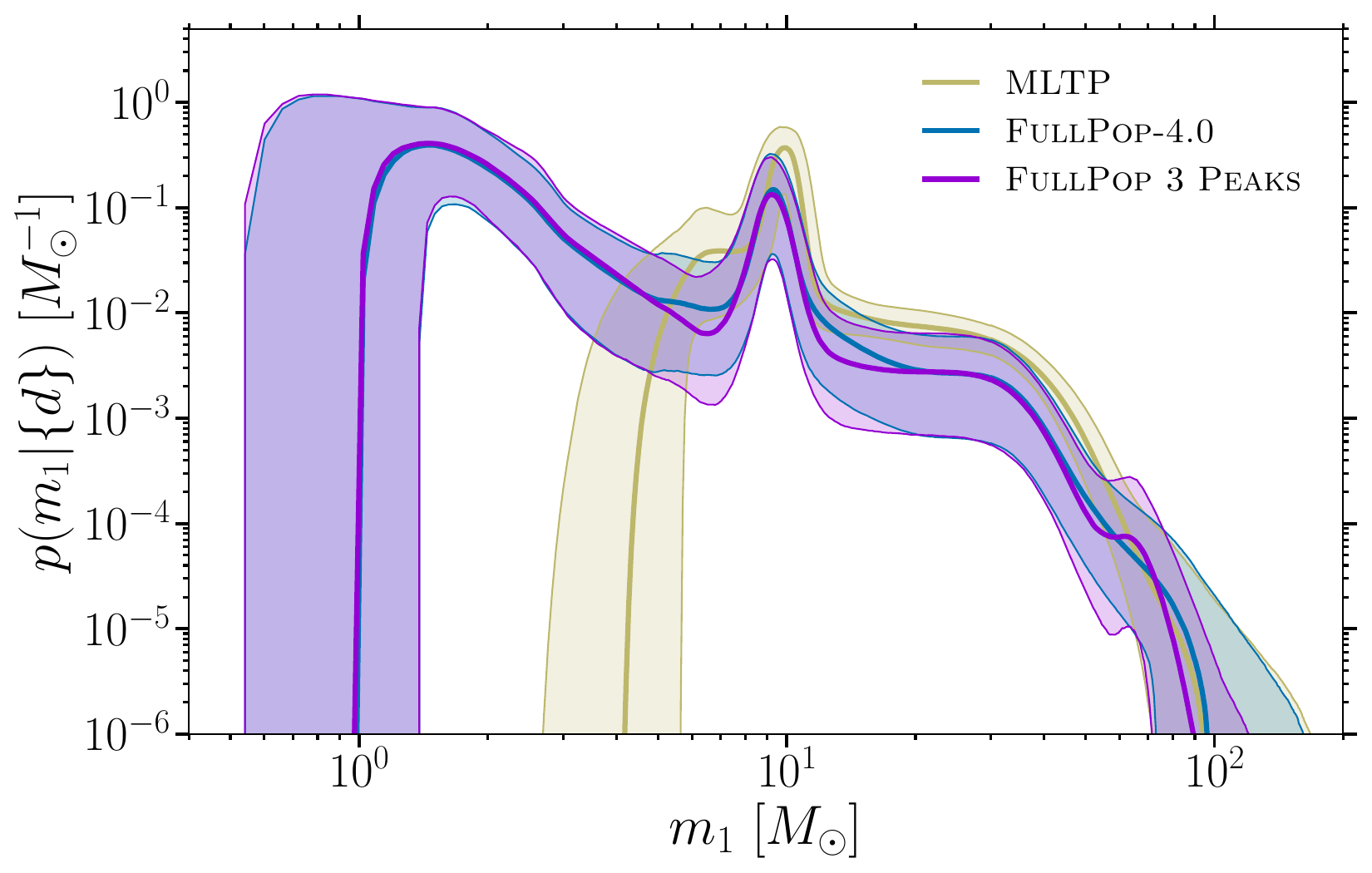}
	\includegraphics[width=0.47\textwidth]{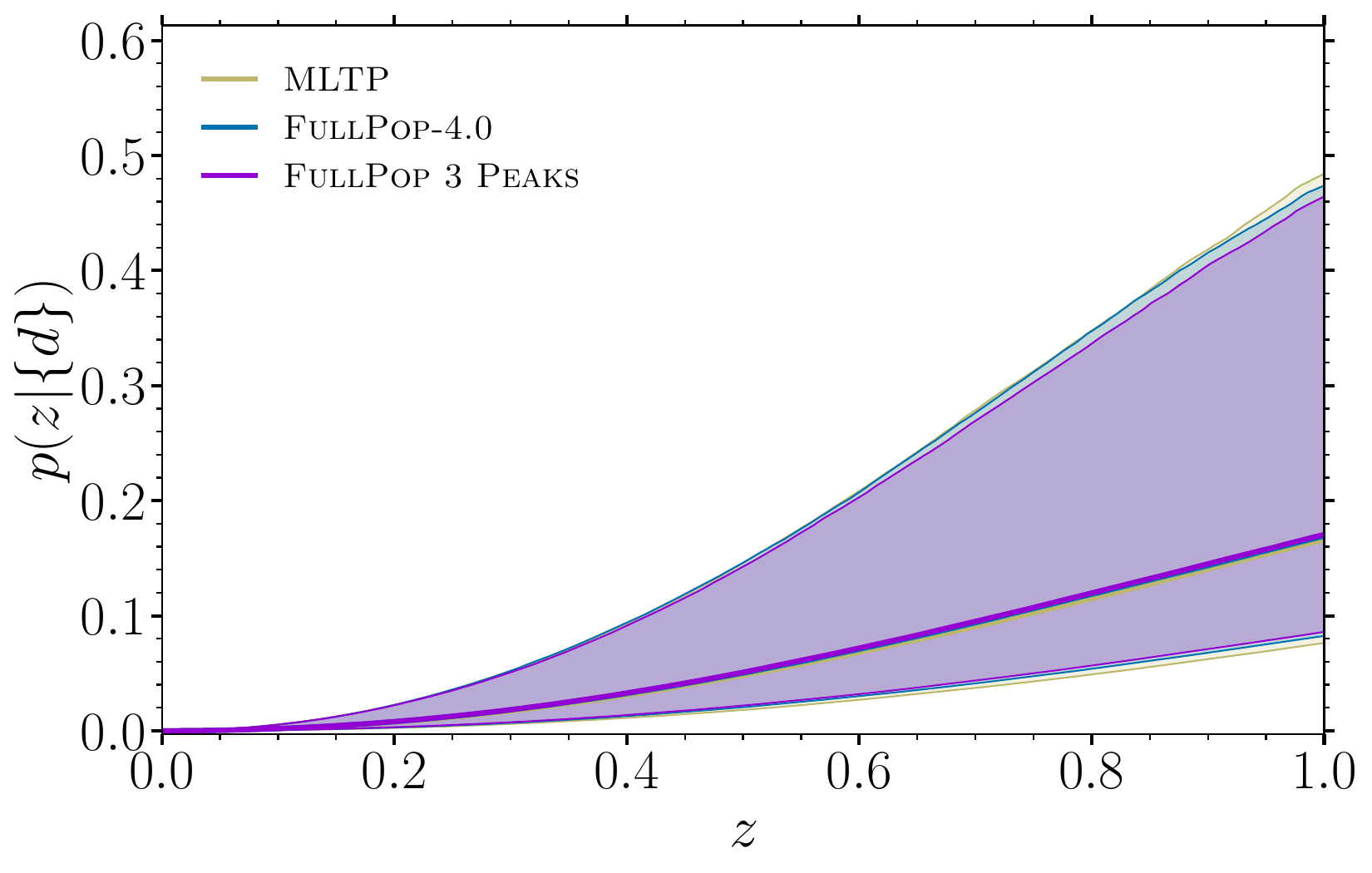}
	\caption{Left panel: Reconstructed source-frame primary-mass distribution (solid curve: median; shaded region: 90\% \ac{CI}). Right panel: reconstructed \ac{CBC} merger rate as defined in the main text. Results in both panels are obtained from spectral siren analyses using the \ac{MLTP},  \fullpop and \fullpopthreepeak mass models.}
	\label{fig:mass spectra dark}
\end{figure*}

This appendix provides supplementary information regarding selected results from the joint astrophysical and cosmological analyses discussed in this paper. 
We show how the source population is reconstructed in the three different population models considered here.
The left panel of Figure~\ref{fig:mass spectra dark} shows the reconstructed primary mass spectrum using the \ac{MLTP}, \fullpop and \fullpopthreepeak mass models in the spectral siren analysis. 
The reconstructed mass distribution clearly presents different features.
A markedly distinct peak around $10\Msun$ as well as an overdensity at around $35\Msun$ is captured by all three mass models, while an additional peak at higher masses is present in the \textsc{BLPL+3P} model.
The high-end cut-off, at around 80-90 $\Msun$ is similar for all three models, while the low-end cut-off of the BBH population is found around 5 $\Msun$, identified by the \textsc{MLTP} model.
With the use of the \fullpop and  \fullpopthreepeak models, we also gain access to the \ac{NS} mass range.
In particular, we find support for a minimum mass value around $1\Msun$, though the \fullpopthreepeak  model prefers a slightly higher value.
Overall, both \fullpop and  \fullpopthreepeak mass model reconstructs features in agreement with our single-population model, namely the \ac{MLTP}.
The right panel of Figure~\ref{fig:mass spectra dark} presents the reconstruction of the \ac{CBC} merger rate, defined as $p(z|\{ d\}) \propto (\mathrm{d} V_{\rm c}/\mathrm{d} z)\psi(z | \PEhyparameter)/(1+z)$, (see Section~\ref{sec:method} for definitions of these quantities) as derived in the same spectral siren scenarios of the left panel.
We find that the reconstructed redshift distributions are consistent across the three mass models considered in our analysis.
We also note that uncertainty on the merger rate grows rapidly in redshift due to the bulk of observations being located at low redshift.
Moreover in the absence of observations falling in the region around or above the expected peak of Madau-Dickinson function, any conclusion about the shape of the redshift distribution at the corresponding redshifts ($z\gtrsim1$) is driven by the assumed parametric form of the merger rate and by the prior range of the associated parameters.

\begin{figure}[t]
	\centering
	\includegraphics[width=0.55\columnwidth]{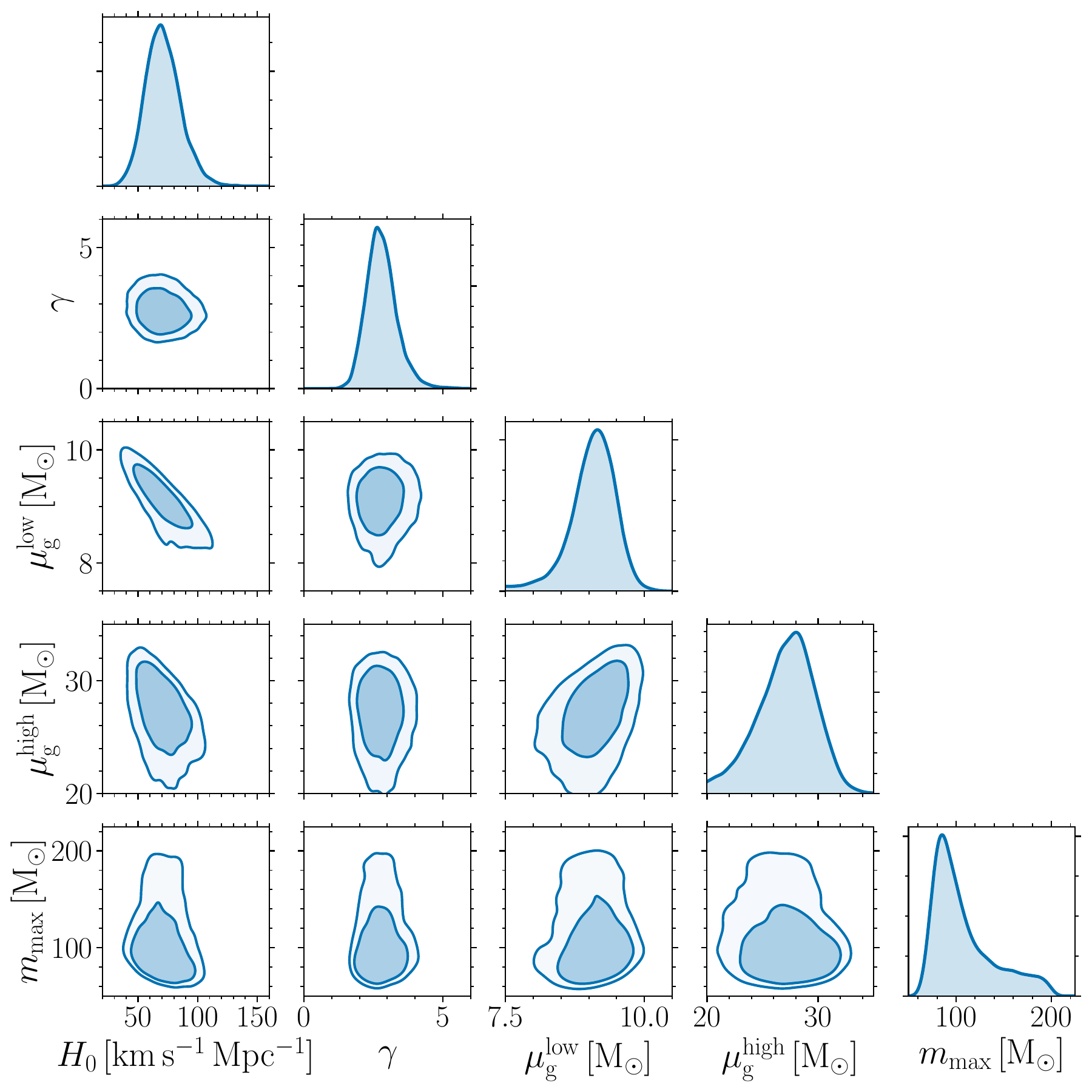}
	\caption{Corner plot showing $H_0$ and a subset of population parameters assuming the \fullpop mass model. The parameter $\gamma$ is the low redfshift index of the Madau--Dickinson distribution, $\mu_{\rm g}^{\rm low}$ and $\mu_{\rm g}^{\rm high}$ are the central locations of the two peaks in the mass model, while $m_{\rm max}$ is the maximum allowed mass for either binary component. The solid contours indicate the 68.3\% and 90\% \ac{CR}.
	}
	\label{fig:corner plot catalog}
\end{figure}

Finally, Figure~\ref{fig:corner plot catalog} shows a reduced corner plot highlighting a subset of the population and cosmological hyperparameters inferred using our fiducial mass model in the dark siren analysis and luminosity-weighting case.
We observe a correlation between $H_0$ and the locations of the two \ac{BH} mass peaks, $\mu_{\rm g}^{\rm low}$ and $\mu_{\rm g}^{\rm high}$ (see Table~\ref{tab:priors_fullpop}), consistent with trends seen in our previous analysis~\citep{LIGOScientific:2021aug,LIGOScientific:2025jau}.
Changing $H_0$ shifts the inferred redshift of the sources, which in turn rescales their intrinsic masses, so the mass spectrum shifts alongside $H_0$ to match the observed signals. In contrast, the maximum mass parameter $m_{\rm max}$ shows only a marginal correlation with $H_0$.

Overall, the Hubble constant appears to correlate only with certain mass scales, showing no significant correlation with merger-rate parameters such as the power-law index $\gamma$. 

\section{Spectral siren results}
\label{sec:appendix_pop_results}

\begin{figure}[t!]
    \centering
    \includegraphics[width=0.9\textwidth]{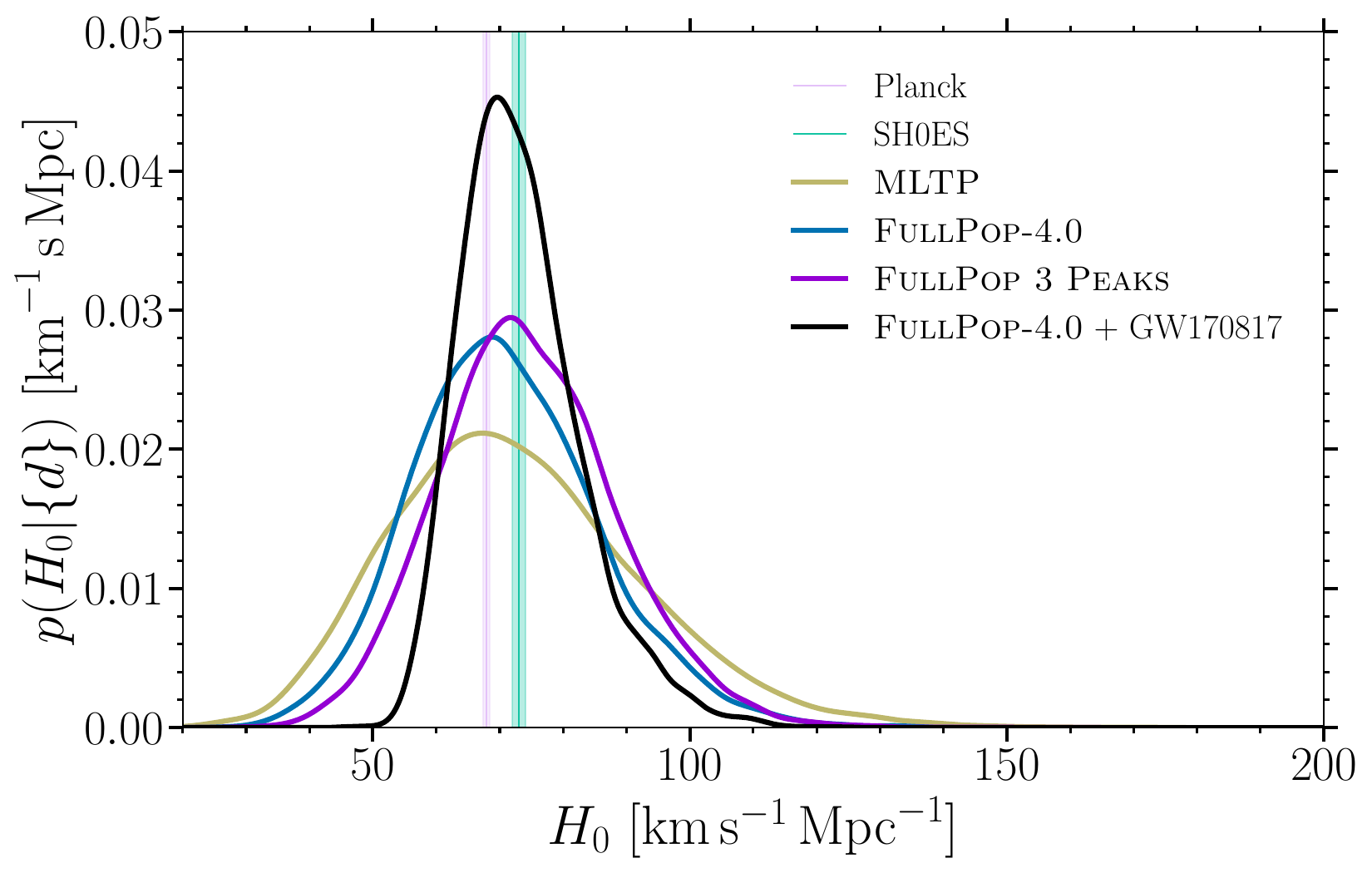}
    \caption{Hubble constant posteriors with the spectral sirens method assuming different population mass models, namely the \ac{MLTP} (gold curve), \fullpop (blue curve) and \fullpopthreepeak (purple curve). The black curve corresponds to the combined posterior between the \fullpop result and the bright siren posterior measured with GW170817. The pink and green shaded areas identify the 68\% \ac{CI} constraints on $H_0$ inferred from CMB anisotropies~\citep{Planck:2015fie} and in the local Universe from SH0ES~\citep{Riess:2021jrx} respectively.}
    \label{fig:H0 combined spectral}
\end{figure}

\noindent In this appendix we report details on results using the spectral sirens method.
Figure~\ref{fig:H0 combined spectral} displays the marginalized posteriors for the Hubble constant estimated with each of the three mass models considered. 
As for the galaxy catalog results (see Figure~\ref{fig:H0 combined catalog}), we show the marginalized posterior for $H_0$ from the spectral siren analysis, with different mass models, as well as the posterior for the \fullpop model combined  with the bright siren GW170817 (blue curve). The analyses using the \ac{MLTP}, \fullpop and \fullpopthreepeak mass models yield $ H_0 = \HzeroMLTPspectralsixtygwtcfive\, \Hunit$, $H_0 = \HzeroEmptycatFullpopsixtygwtcfive\, \Hunit$ and $H_0 = \HzeroBPLthreePspectralsixty\, \Hunit$,  respectively. We observe that, as for the dark siren analysis, the best precision is also achieved using the \fullpop population mass model, which benefits from a larger number of \ac{GW} events and more mass features.
Our most precise estimate is obtained by combining the \fullpop model with GW170817, which leads to a value of $H_0 = \HzeroFullpopspectralbrightsixtygwtcfive \, \Hunit$, similar to the dark sirens results.

\begin{figure}[t]
    \centering
    \includegraphics[width=0.55\textwidth]{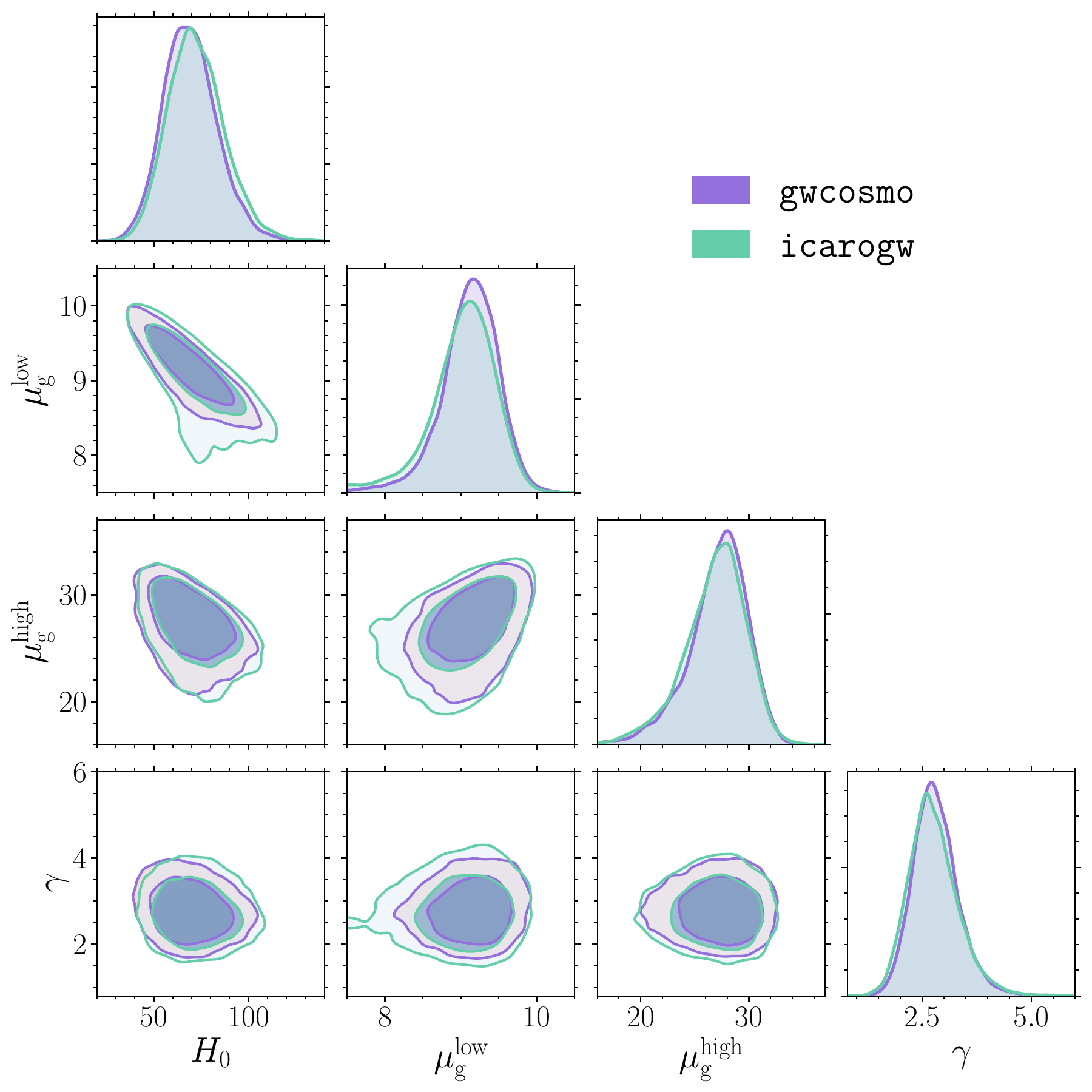}
    \caption{
        Spectral siren reduced corner plot of the Hubble constant and a subset of the \fullpop model mass parameters obtained with \gwcosmo and \icarogw.  The contours indicate the 68.3\% and 90\% \ac{CR}.
    }
    \label{fig:corner_and_mass_spectral}
\end{figure}

Figure~\ref{fig:corner_and_mass_spectral} shows the reconstructed primary mass spectrum from the spectral analysis using the \ac{MLTP}, \fullpop and \fullpopthreepeak mass models. 
As for the dark siren analysis, the \ac{MLTP} and \fullpop models identify two peaks at $\MugLowSpectralsixtygwtcfive M_{\odot}$ and $\MugHighSpectralsixtygwtcfive M_{\odot}$.
For the \ac{NS} region, the results are again consistent with the galaxy catalog analysis, supporting the presence of a shallow dip between $\LeftDipSpectralsixtygwtcfive  M_{\odot}$ and $\RightDipSpectralsixtygwtcfive M_{\odot}$.

Figure~\ref{fig:corner_and_mass_spectral} presents the reduced corner plot showing the most interesting population and cosmological parameters derived from the spectral siren analysis with the \fullpop mass model, as in Figure~\ref{fig:corner plot catalog}. 
In addition, in Figure~\ref{fig:corner_and_mass_spectral} we display results obtained with our two pipelines separately, to show explicitly their consistency.
These results are consistent with those obtained from the dark siren analysis.

Finally, in addition to constraints on the Hubble constant, with the spectral siren approach in principle we are able to infer the present-day matter density of the Universe, $\Omega_{\rm m}$ and the dark energy equation-of-state parameter $w_0$. To facilitate comparison with the results of Section~\ref{subsec:lcdm_results}, the main results of this section keep $\Omega_{\rm m}$ fixed. 
See Section~\ref{subsec:lcdm_results} and Figure~\ref{fig:numerical_stability} for a discussion of the impact of varying $\Omega_{\rm m}$ and $w_0$. 

A summary of the different $H_0$ values obtained  using different data sets and model assumptions can be seen in Table~\ref{tab:results_summary_LCDM_spectral}.

\begin{table*}[t]
\centering
\begin{tabular}{lccc}
\multicolumn{4}{c}{} \\
\hline
\multicolumn{4}{c}{$\Lambda$CDM -- Spectral sirens} \\
\hline
\textbf{Population model} & \textbf{\ac{GW} sources} & $H_0$ (Spectral sirens) & $H_0$ (Spectral + bright sirens) \\
& & $[\Hunit]$  & $[\Hunit]$ \\
\hline\hline
\textsc{Multi Peak} & \NbrBBHgwtcfive \,(\NbrBBHBrightgwtcfive) & $\HzeroMLTPspectralsixtygwtcfive \, (\HzeroMLTPspectralninetygwtcfive)$ & $\HzeroMLTPspectralbrightsixtygwtcfive \, (\HzeroMLTPspectralbrightninetygwtcfive)$ \\
\textsc{FullPop}-4.0 & \NbrCBCgwtcfive\, (\NbrCBCtotgwtcfive) &  $\HzeroEmptycatFullpopsixtygwtcfive \, (\HzeroEmptycatFullpopninetygwtcfive)$ & $\HzeroFullpopspectralbrightsixtygwtcfive \, (\HzeroFullpopspectralbrightninetygwtcfive)$ \\
\fullpopthreepeak & \NbrCBCgwtcfive \,(\NbrCBCtotgwtcfive) &  $\HzeroBPLthreePspectralsixty \, (\HzeroBPLthreePspectralninety)$ & $\HzeroBPLthreePspectralbrightsixty \, (\HzeroBPLthreePspectralbrightninety)$ \\
\hline
\end{tabular}
\caption{\label{tab:results_summary_LCDM_spectral}
Constraints on the Hubble constant obtained in this work under the $\Lambda$CDM cosmological model, assuming a uniform prior $H_0 \in \mathrm{U}(10,200)\,\Hunit$. 
	The first column lists the mass model adopted in the analysis. 
	The second column gives the number of \ac{GW} events included in the dark-siren analysis. 
	The third column reports the $H_0$ measurement obtained using only spectral sirens and the \fullpop mass model, quoted as the median together with the symmetric 68.3\% and 90\% \acp{CI}, with the latter in parentheses. 
	The fourth column shows the corresponding constraints after combining the dark-siren analysis with the bright-siren posterior of GW170817.}
\end{table*}

\section{Robustness checks}\label{app: additional systematics}

\begin{figure*}
    \centering
    \includegraphics[width=0.95\textwidth]{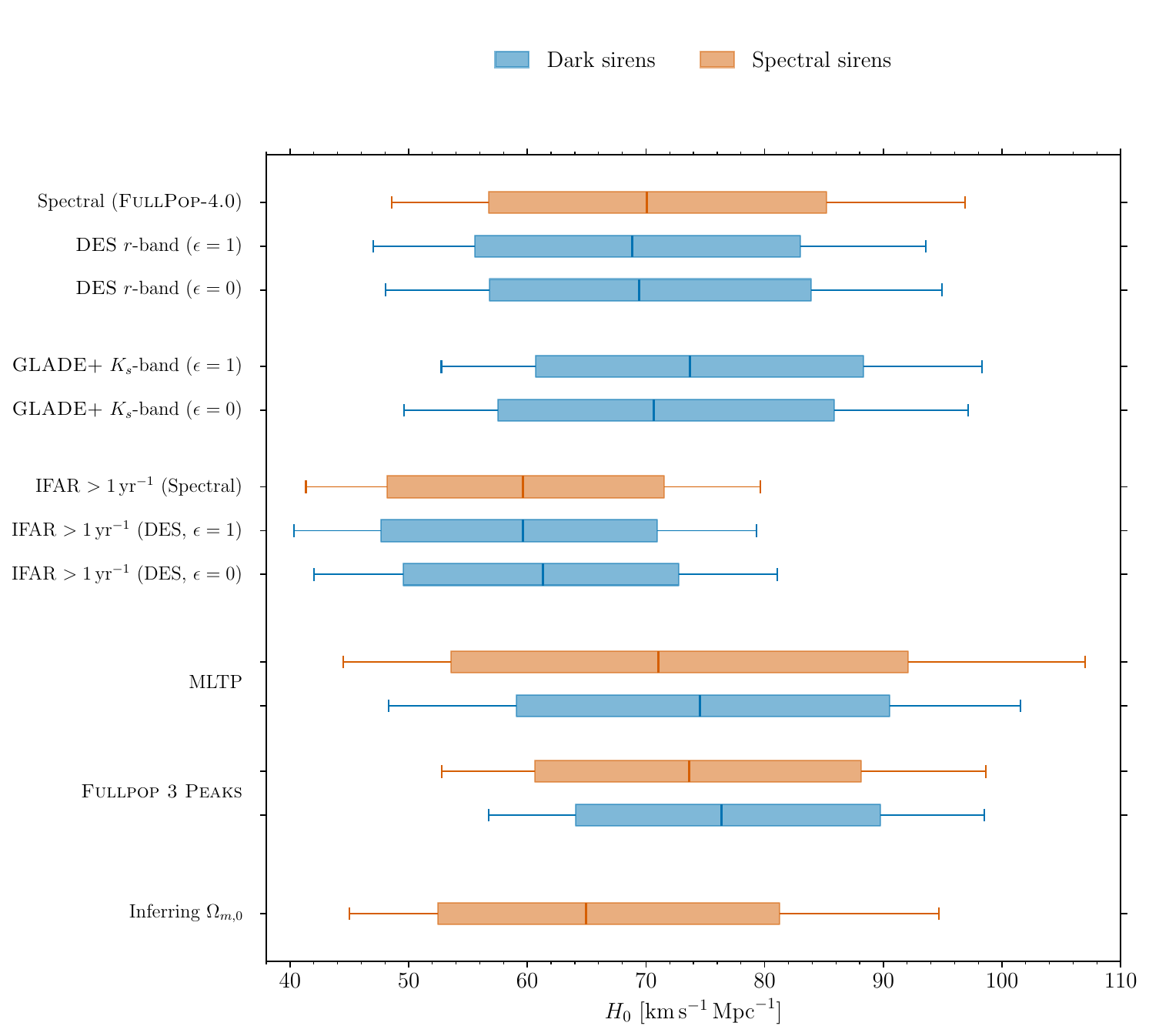}
    \caption{Robustness checks against various systematics discussed in Section~\ref{sec:results}, compared to the fiducial results ($\Lambda$CDM,~\fullpop, and luminosity-weighting case for the dark sirens case). 
    The box-plots show the median value as a vertical segment. The colored boxes stretch to the 68.3\% \ac{CI}, while the whiskers extend to encompass the 90\% \ac{CI}. The labels indicate variations with respect to the fiducial results. The posteriors shown in this plot do \emph{not} include bounds from GW170817.  In the analysis with varying $\Omega_{\rm m}$ the prior used was U$(0,1)$.}
    \label{fig:numerical_stability}
\end{figure*}

\noindent
In this appendix, we summarize checks conducted to ensure robustness of our results.
Figure~\ref{fig:numerical_stability} shows a summary of the constraints on $H_0$ varying several assumptions discussed in Sections~\ref{subsec:lcdm_results} and~\ref{subsec:mg_results}, with additional numerical stability checks that we discuss below. In particular, we display the effects of luminosity weighting, varying mass models and varying other parameters of the cosmic expansion history - specifically $\Omega_{\rm m, 0}$.
For these tests, we used the dark siren approach with the \fullpop mass model and a luminosity-weighting scheme for both the DES and \gladep galaxy catalogs.
For all of the analyses shown in Figure~\ref{fig:numerical_stability}, we find that the medians of each posterior are consistent with the $68\%$ CI of the others, with the exception of the IFAR (inverse false alarm rate) $> 1 \, \text{yr}^{-1}$ case, which is consistent with the others at the $90\%$ CI level.
The posteriors shown in this plot do not include constraints from GW170817. These results were obtained using only \gwcosmo with the exception of the varying $\Omega_{\rm m,0}$ test, which was performed only with \icarogw.

We do not repeat the additional numerical stability tests on the effective number of injections and \ac{PE} samples which were included in~\citep{LIGOScientific:2025jau}, as they were already shown to have a negligible impact on the results.

As mentioned in Section~\ref{Dark Sirens Statistical Framework}, our population models implicitly assume that the \ac{CBC} spin distribution is isotropic with uniform distribution in the spin magnitudes~\citep{GWTC:Results}.
However, we verified that including spin distributions for the \ac{BBH} population using the \textsc{Default} model~\citep{KAGRA:2021duu,GWTC:AstroDist} has no significant impact on the current cosmological constraints (see Section~\ref{sec:discussion}). For the spin-informed tests, we adopted the \textsc{MLTP} mass model, as this model better fits the \ac{BBH} mass spectrum of the \ac{GW} candidates used in our analysis~\citep{GWTC:AstroDist}.

\section{Further comments on modified-gravity analyses}\label{sec:appendix_MG}

\paragraph{Connection between $\alpha_M$ parametrization and gravitational coupling.}
Here we supply further details on the connection between the Horndeski function $\alpha_M(z)$ and effective gravitational coupling strength, referenced in Section~\ref{subsubsec: gravity models}.
In GR, the Planck mass enters via the constant factor appearing in the Einstein-Hilbert Lagrangian density: ${\cal L_{\rm GR}}=(M_P^2/2)\, R$. In Horndeski gravity this factor is generalised to be a function of the scalar field, i.e. ${\cal L_{\rm Hd}} = G_4(\phi) R+\dots$, where the dots indicate additional terms and we restricted the propagation speed of GWs to be luminal. 
$\alpha_M$ is then defined by the derivative:
\begin{align}
\alpha_M(a) = \frac{\rm{d}\ln G_4(\phi[a])}{\rm{d}\ln a}
\label{eq:alphaM}
\end{align}
Using Equation~(\ref{eq:alphaM}) in Equation~(\ref{eq:aM_param}), and restricting to theories where matter is minimally coupled to the metric, one reaches Equation~\eqref{eq:DGW_GGW}.

Here the effective gravitational coupling $G_{\rm GW}$ is associated with the effective Planck mass, that is $G_{\rm GW}=1/(16 \pi G_4)$. As stressed in the main text, $G_{\rm GW}$ is not necessarily the same gravitational coupling strength constrained by large-scale structure surveys (commonly parameterized as $\mu$), although it can be under certain conditions.

In short, by squaring our constraints on the GW\textendash EM distance ratio in the $\alpha_M$ parametrization, we can obtain constraints on redshift evolution of an effective gravitational coupling strength for a large class of modified gravity models. This interpretation is displayed in the additional $y$-axis on the right-hand side of the middle panel in Figure~\ref{fig:dL_z_PPC}. 

Physically, the origin of Equation~(\ref{eq:DGW_GGW}) can be understood as a requirement to ensure the conservation of gravitons, which is respected in most gravity theories -- see \citet{Belgacem:2018lbp} for details. We note that here we considered only the linear, homogeneous part of \ac{GW} propagation; further deviations from GR may arise through propagation on inhomogeneous spacetimes (e.g. \ac{GW} lensing) or in the nonlinear regime.

\begin{figure*}[ht!]
    \centering
    \includegraphics[width=0.48\textwidth]{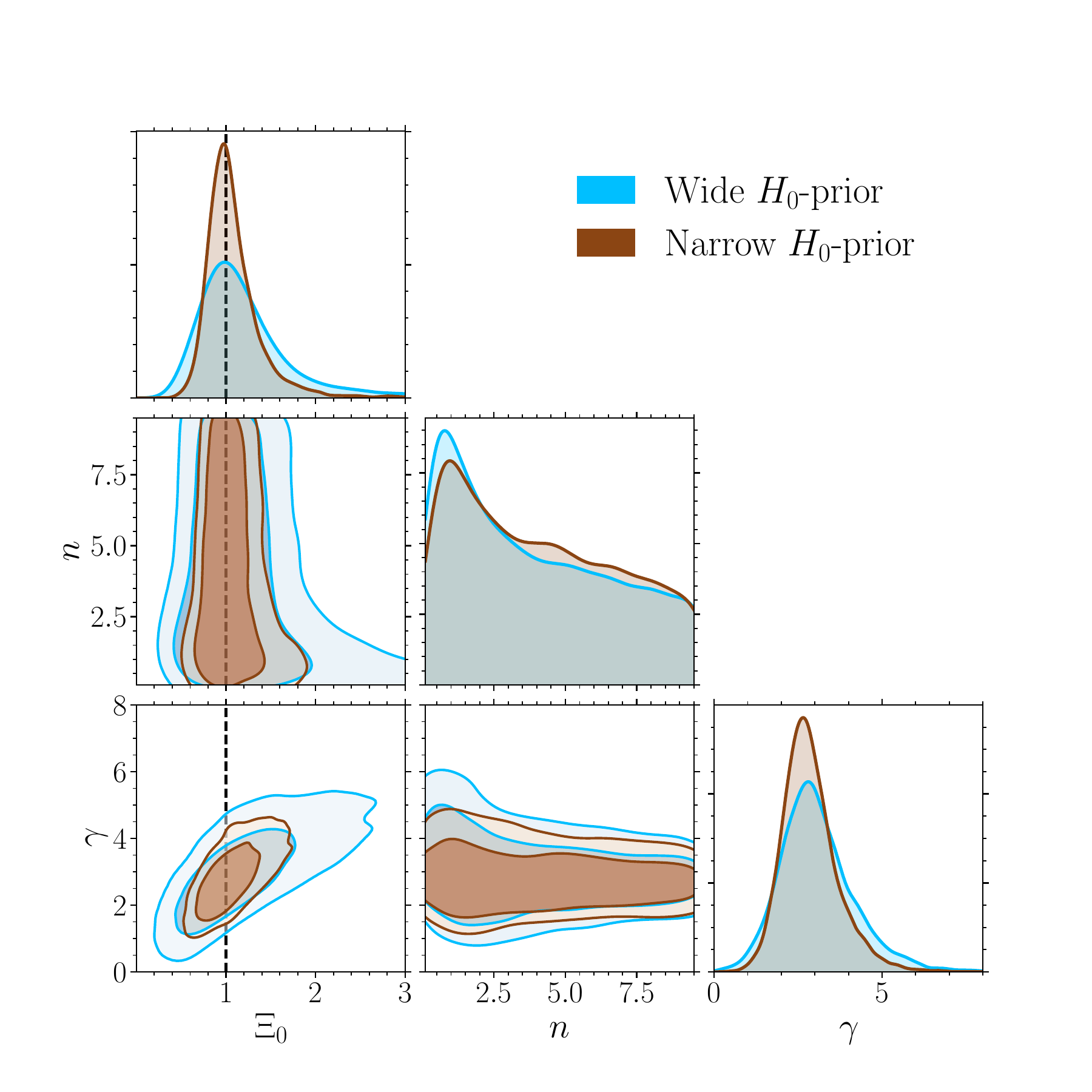} \hfill
    \includegraphics[width=0.48\textwidth]{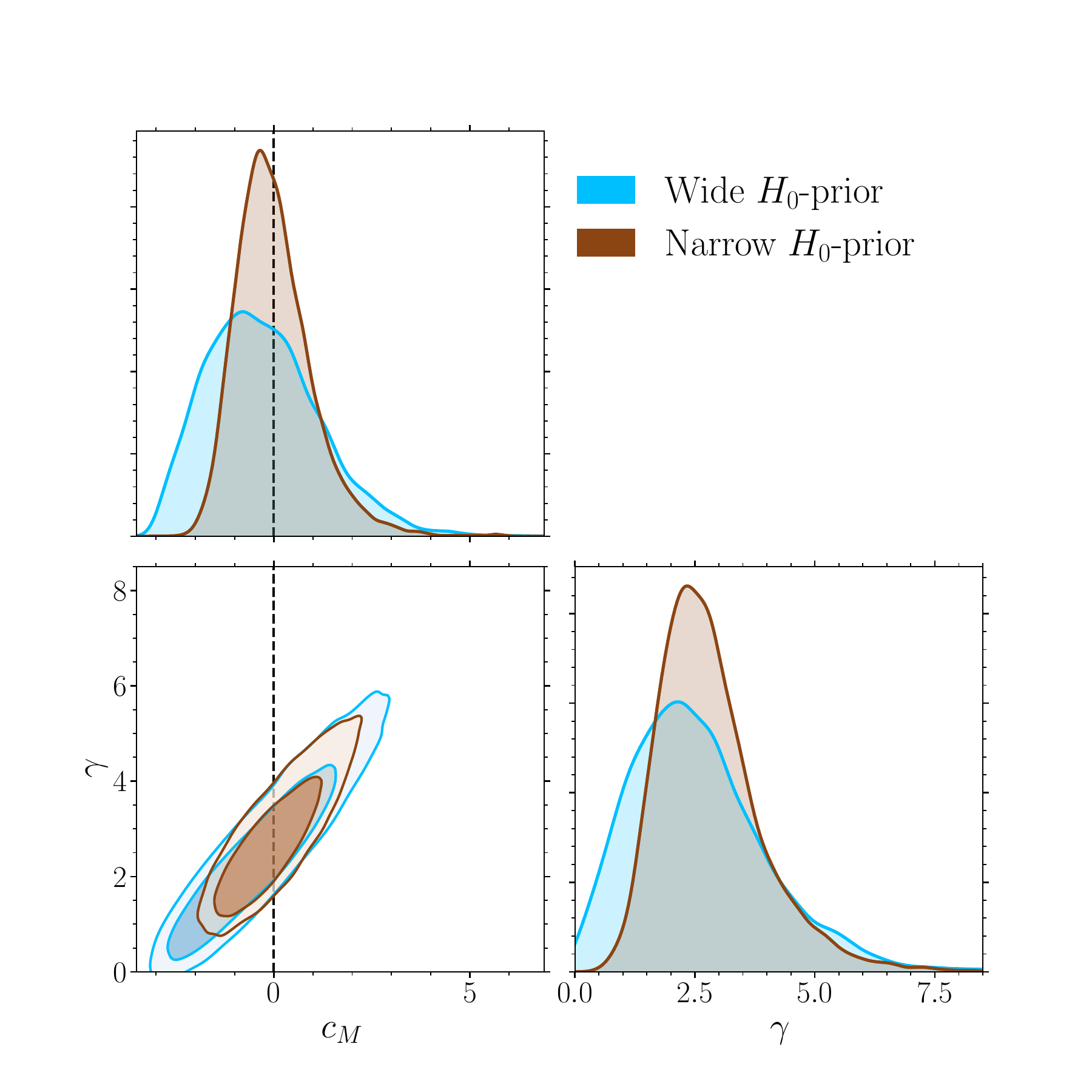}

    \caption{Corner plots of the modified gravity parametrizations $\Xi_0$--$n$ (left), and $c_M$ (right), and the merger rate parameter $\gamma$. These were obtained with the spectral siren method assuming the \fullpop mass model. Vertical dashed lines indicate the \ac{GR} limit. Contours indicate the 68.3\% and 90\% \ac{CR}.}
    \label{fig:mg_cornerplots_simple}
\end{figure*}

\paragraph{Results and corner plots of selected parameters}
Figure~\ref{fig:mg_cornerplots_simple} shows the marginalised posterior contours for the MG parameters and the low-redshift power-law slope of the merger rate, $\gamma$, for each of our three MG analyses. As discussed in the main text, $\gamma$ is the parameter showing greatest degeneracy with MG parameters at present, an effect well-known in the literature \citep{Mancarella:2021ecn,Leyde:2022orh,Chen:2023wpj}. The priors and marginalised constraints on the MG parameters are reported in Table~\ref{tab:results_summary_MGdark}.

All of the results show consistency with \ac{GR}, marked for the MG parameters by the black dashed vertical lines in Figure~\ref{fig:mg_cornerplots_simple}.

As expected, for each alternative cosmological model we find a correlation between modified gravity parameters and the Hubble constant.
$c_M$ is positively correlated with $H_0$, with Pearson correlation coefficient
$\PearsonHcmCataloggwtcfive$, respectively. This explains the narrower error-bars on $c_M$ when restricting $H_0$ with a narrower prior. For the $\Xi_0-n$ parametrization more subtle effects are at play, as remarked upon in Section~\ref{subsec:mg_results}.

\section{Event list}\label{sec:appendix_event_list}
\noindent In this appendix we provide a list of the events used in our analyses with their main properties relevant for our analysis. For the details on the PE and waveform models used, see Sec.~\ref{subsec:gw_events}. For each of the events used in our analyses, Table~\ref{tab:search_setup_parameter} reports  the following properties:
\begin{itemize}
    \item $D_{\rm L}$, and $z$: luminosity distance to the source and the corresponding redshift, calculated from the distance samples assuming Planck-15~\citep{Planck:2015fie} cosmology. We give the median of the samples and the 90\%\,\ac{CI}, cutting away 5\% of samples at the edges of the posterior distribution
    \item Sky localization $\Delta \Omega$: the localization area of the event calculated from the skymap as a fraction of pixels containing the 90\% of the probability
    \item Localization volume $\Delta V$: localization volume of the event at 90\%\,\ac{CI}, calculated as the fraction corresponding to the 90\% of the sky area (see above) of the spherical shell, at the 90\%\,\ac{CI} of the event's redshift distribution
    \item $N_{\rm gal}$, over(under)-density and incompleteness: the number of galaxies inside the 90\% localization volume, the over(under)-density fraction, and the catalog ($K$-band of the GLADE+ catalog) incompleteness percentage.
\end{itemize}

\begin{longrotatetable}
\begin{deluxetable*}{lcccccccccccc}
\tabletypesize{\scriptsize}
\tablecaption{
\scriptsize{
\footnotesize{List of the 236 \ac{CBC} events selected with \ac{FAR} $< \FARcutgwtcfive\,\text{yr}^{-1}$. Columns with a `*' have been computed assuming the reference cosmology in ~\citet{Planck:2015fie}, i.e. $H_0=67.9\,{\rm km\,s}^{-1}\,{\rm Mpc}^{-1}$ and $\Omega_{\rm m}=0.3065$. This table reports all of the \ac{GW} events considered in this work and summarizes some of their properties reported with their median values and 90\% symmetric \ac{CI}~\citep{LIGOScientific:2018mvr,LIGOScientific:2020ibl,LIGOScientific:2021usb,KAGRA:2021vkt,GWTC:Results}. First, second and third columns: \ac{GW} event label, luminosity distance and redshift. 
Fourth and fifth columns: sky localization area, and 3D localization comoving volume. The remaining columns list the number of galaxies inside the localization volume for each event, the under(over)-density of galaxies and the incompleteness fraction of the catalog for each event, for both the \des and \gladep catalogs. The under(over)-density is computed as the fraction between the number of galaxies (in the $K$ and $r$-bands respectively) corrected for incompleteness, and the effective number expected from the Schechter function. We report the minimum and maximum values, respectively.
The lower and upper bounds on the incompleteness probabilities, instead, are derived from the incompleteness fractions at the boundaries of the 90\% localization volume. These columns, number of galaxies, under(over)-density, and incompleteness fraction, do not apply to the GW170817 event. We report these values using a distance prior proportional to $D_\textrm{L}^2$. For events released with GWTC-3.0, we use the first data release associated with \citet{KAGRA:2021vkt}}}.}
\tablehead{
\colhead{Name} &
\colhead{$\DL$} &
\colhead{$z^*$} &
\colhead{\skyareasymbol} &
\colhead{$\Delta V^*$} &
\colhead{$N_{\rm gal}^{\rm DES}$} &
\colhead{$N_{\rm gal}^{\rm GLADE+}$} &
\colhead{Under(over)$^{\rm DES}$} &
\colhead{Under(over)$^{\rm GLADE+}$} &
\colhead{Incompleteness$^{\rm DES}$ (\%)} &
\colhead{Incompleteness$^{\rm GLADE+}$ (\%)}
\\
\colhead{} & \colhead{[Mpc]} & \colhead{} & \colhead{[deg$^2$]} & \colhead{[Gpc$^3$]} &
\colhead{} & \colhead{} & \colhead{} & \colhead{} & \colhead{} & \colhead{}
}
\startdata
GW150914 & $463^{+132}_{-142}$ & $0.10^{+0.03}_{-0.03}$ & 157 & $1.9 \times 10^{-3}$ & 182 & $1.5 \times 10^{3}$ & 0.84\textendash$\sim$1 & 1.05\textendash1.17 & 100\textendash100 & 60\textendash95\\ 
GW151012 & $1056^{+621}_{-493}$ & $0.21^{+0.10}_{-0.09}$ & $1.5 \times 10^{3}$ & 0.30 & $5.6 \times 10^{3}$ & $6.7 \times 10^{3}$ & 0.92\textendash0.96 & $\sim$1\textendash1.05 & 100\textendash100 & 94\textendash100\\ 
GW151226 & $471^{+158}_{-196}$ & $0.10^{+0.03}_{-0.04}$ & 984 & 0.01 & 0 & $9.8 \times 10^{3}$ & $\sim$1\textendash$\sim$1 & 0.96\textendash1.16 & 100\textendash100 & 50\textendash97\\ 
GW170104 & $1126^{+385}_{-466}$ & $0.22^{+0.07}_{-0.09}$ & 950 & 0.14 & $7.7 \times 10^{4}$ & $5.7 \times 10^{3}$ & 0.93\textendash1.04 & $\sim$1\textendash1.07 & 0\textendash0 & 97\textendash100\\ 
GW170608 & $334^{+123}_{-117}$ & $0.07^{+0.03}_{-0.02}$ & 387 & $2.5 \times 10^{-3}$ & 0 & $3.0 \times 10^{3}$ & $\sim$1\textendash$\sim$1 & 0.72\textendash1.11 & 100\textendash100 & 32\textendash81\\ 
GW170729 & $2874^{+1630}_{-1474}$ & $0.49^{+0.22}_{-0.23}$ & $1.1 \times 10^{3}$ & 1.8 & $8.0 \times 10^{4}$ & 100 & 0.85\textendash$\sim$1 & $\sim$1\textendash$\sim$1 & 0\textendash100 & 100\textendash100\\ 
GW170809 & $1117^{+298}_{-352}$ & $0.22^{+0.05}_{-0.06}$ & 269 & 0.03 & $3.8 \times 10^{5}$ & 992 & 0.86\textendash1.04 & $\sim$1\textendash1.07 & 0\textendash0 & 99\textendash100\\ 
GW170814 & $600^{+165}_{-226}$ & $0.12^{+0.03}_{-0.04}$ & 86 & $2.2 \times 10^{-3}$ & $3.2 \times 10^{4}$ & $1.3 \times 10^{3}$ & 1.03\textendash1.43 & 1.03\textendash1.31 & 0\textendash0 & 69\textendash99\\
\enddata
\end{deluxetable*}
\end{longrotatetable}

\bibliography{}

\end{document}